\documentclass[12pt,a4paper]{article}
\usepackage{amsmath}
\usepackage{amssymb}\usepackage{cite}
\usepackage{graphicx}
\usepackage{appendix}

\title {A survey on  uncertainty relations and quantum measurements }

\author{Spiridon Dumitru \\(Retired) ex-Department of Physics, "Transilvania" University, \\
 B-dul Eroilor 29, R-2200 Brasov, Romania,\\
E-mail: s.dumitru42@yahoo.com}

\date{\today}

\begin{document}
\maketitle

\begin{abstract}
  This survey tries to investigate the truths and deficiencies of prevalent philosophy about Uncertainty Relations (UR) and Quantum Measurements (QMS). The respective philosophy,  known as being eclipsed by unfinished controversies, is revealed to be grounded on six basic precepts. But one finds that all the respective precepts are  discredited by insurmountable deficiencies.  So, in regard to UR, the alluded
philosophy discloses oneself to be an unjustified mythology. Then UR appear either as short-lived historical conventions or as simple and limited mathematical formulas, without any essential significance for physics. Such a finding reinforces the Dirac's prediction that UR \textit{"`in their present form will not survive in the physics of future"'}. The noted facets of UR motivate  reconsiderations of associated debates on QMS. Mainly onereveals that, properly, UR have not any essential connection with genuine descriptions of QMS. For such descriptions, it is necessary that, mathematically, the quantum observables to be considered as random variables. The measuring scenarios with a single sampling, such are wave function collapse or Schrodinger's cat thought experiment, are revealed as being useless exercises. We propose to describe QMS as transmission processes for stochastic data. Note that the above-announced revaluation of UR and QMS philosophy does not disturb in any way the practical framework of the usual quantum mechanics. 	
	\end{abstract}
\begin{flushleft}
\end{flushleft}
\tableofcontents
\section{Introduction}
Nearly a century until  nowadays, in the  publications regarding Quantum Mechanics (QM) and even other areas,  have  persisted discussions (debates and controversies) about  the  meaning  of     Uncertainty Relations (UR). Moreover UR in their entirety were ranked to a status of fundamental concept named Uncertainty Principle (UP) (for a bibliography of the better known specific publications see \cite{1,2,3,4,5,6,7,8,9,10,11,12}). Mostly the respective discussions have credited UR/UP with considerable popularity and crucial importance, both in physics and in other domains. The mentioned importance was highlighted by    compliments such as:
\begin{itemize}
	\item 	UR are \textit{"`expression of the most important principle of the twentieth century Physics "`}\cite{13},
\item UP is \textit{"`one of the cornerstones of quantum theory"' }\cite{9} ,
\item UP \textit{"`epitomizes quantum physics, even in the eyes of the scientifically 
           informed public"'} \cite{7}.
\end{itemize}	

       But, as a fact, in spite of such compliments, in scientific literature of our days the essential aspects regarding UR/UP remain as unsolved and misleading questions. Today keeps their topicality many   critiques reported during   last decades, like the next ones:
			
			\begin{itemize}
				\item UR \textit{"`are probably the most controverted  formula in the whole of
 the theoretical  physics"'} \cite{14} .
\item\textit{ "`Still now, 80 years after its inception, there is no general consensus over the scope and    validity of this principle (‘UP’) "`} \cite{7} .

\item \textit{"`Overcoming the early misunderstanding and confusion, the concept}  (notion of   uncertainty – i.e. of UR/UP ) \textit{grew continuously and still remains an active and fertile      research field"'} \cite{8}.
			\end{itemize}
	
       Note that the above reminded appreciations (compliments and critiques) regard mainly the own essence (intrinsic meaning) of UR/UP.  But, within many texts about QM fundamentals, one finds also an adjacent topic which, historically, is a direct 
			sub-sequence of the debates about the mentioned essence. The respective topic refers  to the significance and description of Quantum Measurements (QMS). 

       Marked by the previously noted points, during the decades,   the  discussions  about UR and QMS  meaning and implications  have generated a  true prevalent  philosophy ( i.e. \textit{"`a group of theories and ideas related to the
understanding of a particular subject"'} \cite{15}). For almost a century, the respective philosophy dominates in mainstream physics publications and thinking.
It obstructs (delays) the expected progresses in clarifying some of main aspects regarding the fundamentals/interpretation of QM respectively the essentials of QMS problem. 
	Add here the more alarming observation \cite {16} that : "`\textit{there is still no consensus on ... interpretation and limitations} of QM"'.  Then it becomes of immediate interest to continue searches for finding the truth about   own essence and consecutive topics of the UR/UP and QMS matters.
			
       A search of the alluded type can be done (or  facilitated at least) by a pertinent survey on deficiencies of the mentioned philosophy.  Such a survey (of modest extent) we intend to present in this article. Our survey tries firstly to identify the basic elements of nowadays	prevalent views within UR and QMS  philosophy. Afterward we will investigate truth and value of the respective elements. Within the investigation we promote a number of re-considerations regarding the conventional (and now dominant) views about UR and QMS matters. Mainly we reveal the fact that the alluded views are discredited (and denied) by a whole class of insurmountable deficiencies, overlooked in the mainstream literature. So our survey aims to represent an unconventional analysis of the actual dominant  philosophy about  UR and QMS.The above-announced analysis germinated from some of our preceding investigations (see \cite{17,18,19,20,21}  and references). \textbf{Also, it was stimulated by a number of opinions due to other scientists (usually the respective opinions are  ignored in dominant literature, but here   they are highlighted by specifying the proper bibliographic sources )}. Through the present survey, we try to gather, extend, systematize, improve and consolidate the results of our mentioned investigations in order to present a more argued viewpoints about the approached topics.
							
     In our survey, when it is usefully, we will appeal to the so called 'parsimony principle'. The respective principle (known also as Ockham's razor) will be applied as a heuristic method of simplicity which can be summarized \cite{22} by the next two desiderata:
		
		\begin{itemize}
			\item \textit{"`Of two competing theories, the simpler explanation of an entity is to be preferred"'}.
			\item  \textit{"`Entities are not to be multiplied beyond necessity"'}.
		\end{itemize}
  
The mentioned principle will be   accounted for in order that the text to be easy understood for readers (including students) not highly specialized.

    By the present article-survey, through adequate arguments and details, we try to elucidate what is in fact the true meaning of UR, respectively to evaluate the genuine scientific aspects regarding QMS.
		
      From the conclusions resulting from this survey the most important one is that, in its entirety, the actual prevalent  philosophy about UR must be regarded as a veritable myth without any special or extraordinary status/significance for physics. This because, in reality, the UR reveal themselves to be   nothing but short-lived historical conventions (in empirical, thought-experimental version) or simple and restricted   formulas (in theoretical approach).  But such a conclusion come in consonance, from another perspective, with the Dirac's  guess \cite{23} that: \textit{"`uncertainty relations in their present form will not survive in the physics of future"'}. 
			
			Add here the fact that, essentially, the above mentioned re-evaluation of UR and QMS philosophy does not disturb in any way the basic  framework (principles, concepts, models and working rules) of usual QM. Furthermore, the QMS description remains as a distinct and additional subject comparatively with the elements of QM in itself. The mentioned description requires to  regard quantum observables
			\footnote{\underline{Drafting specifications}: (i) In the next parts of this article, for naming a  physical  quantity,  we shall use the term \textit{"`observable"'}(promoted by the UR and QMS philosophy literature), (ii) Also, according to the mainstream publications, we adopt  the titles \textit{"`commuting"'} or \textit{"`non-commuting"'} observables   for the QM quantities described by operators which\textit{ "`commute"'} respectively \textit{"`do not commute"}',  (iii) For improving fluency of  our text  some of the  corresponding  mathematical notations, formulas and proofs are summarized briefly and unitary in few  Appendices located in the final of the article.}
			as true random variables. Also it must  be dissociated of some fictive  QMS   scenarios  with a unique sampling	(such  scenarios are  schema  with wave function collapse and Schrodinger's cat thought experiment). We recommend  to describe QMS as  transmission processes of stochastic data. 
									
\section{Basic precepts of UR - QMS prevalent  philosophy  }
     Firstly it must be pointed out the fact that, in spite of its prevalence inside of nowadays scientific debates, the actually dominant philosophy about UR and QMS   germinates mainly from an old   doctrine which can be called Conventional Interpretation of UR (CIUR). The mentioned doctrine (or dogma) was initiated by the Copenhagen School founders and, subsequently, during nine decades, it was promoted (or even extrapolated) by the direct as well indirect partisans (conformists ) of the respective school. Currently CIUR enjoys of a considerable acceptance, primarily in QM studies but also in other thinking areas.  Moreover, today, within the normative   (mainstream/authoritarian) physics publications, CIUR dominates the leading debates about foundations and interpretation of QM. 
		
		   	But as a notable fact, in publications, CIUR doctrine, as well as most aspects of UR and QMS philosophy, are presented rather through independent or disparate assertions but not through a complete and systematized set  of clearly defined 	\textit{'precepts'} ( considered as \textit{"` beliefs ...accepted as authoritative by some group or school "`} \cite{24}). That is why, for a fruitful survey of the UR - QMS philosophy, it is of direct interest to   identify such an set of \textit{\textbf{B}}asic \textit{\textbf{P}}recepts ($\textbf{\textit{BP}}$)  
from which the mentioned assertions turn out to be derived or extrapolated.   Note that the aforesaid  set of precepts (i.e. the true   core of CIUR doctrine along with  prevalent philosophy of UR and QMS) can be collected  by means of a careful examination of the today known publications. In its essence  the respective  collection can be presented as follows.				
					
		The history regarding Conventional Interpretation of UR (CIUR) began with two main generative elements which were the following ones:

      (i)  Heisenberg's \textit{“Thought-Experimental”} (\textit{TE}) relation:
\begin{equation}\label{eq:1}
\Delta _{TE} A \cdot \Delta _{TE} B \cong \hbar \quad or\quad \Delta _{TE} A \cdot \Delta _{TE} B \ge \hbar 
\end{equation}	

(ii)	Robertson-Schrodinger  relation of theoretical origin:	
\begin{equation}\label{eq:2}
\Delta _\Psi  A \cdot \Delta _\Psi  B \ge \frac{1}{2}\left| {\left\langle {\left[ {\hat A,\hat B} \right]} \right\rangle _\Psi  } \right|
\end{equation}

For introducing relation \eqref{eq:1} in \cite{25,26}  were imagined some \textit{Thought Experiments} (\textit{TE}) (or 'gedanken' experiments). The respective \textit{TE} referred on simultaneous measurements of two (canonically) conjugate observables A and B regarding a same quantum micro-particle. As such pairs of two observables were considered coordinate \textit{q} and momentum \textit{p} respectively time \textit{t} and energy \textit{E}. Then the quantities 
$ \Delta _{TE} A$  and $ \Delta _{TE} B$  were indicated as corresponding \textit{"`uncertainties"'} of the imagined measurements, while $\hbar$ denotes  the  Planck's constant.

      Relation \eqref{eq:2}  was introduced in \cite{27,28}  and it is depicted as above in terms of traditional QM notations \cite{29,30}. The main features of the respective notations are reminded briefly below in Appendices A and B while some aspects regarding the Dirac's braket QM notations \cite{29,30,31,32} are discussed in Appendix B. 

     Note here the fact that the right-hand side term from \eqref{eq:2}   is dependent on Planck's constant $\hbar$, 
		e.g. $\left| {\left\langle {\left[ {\hat A,\hat B} \right]} \right\rangle _\Psi  } \right| = \hbar $ when A and B 
		are (canonically) conjugate. 
		
		          Starting from the generative elements \eqref{eq:1} and \eqref{eq:2}, CIUR doctrine jointly with UR and QMS philosophy have been evolved around the   following \textit{\textbf{B}}asic \textit{\textbf{P}}recepts 
							($\textit{\textbf{BP}}$):
\begin{itemize}
								\item  $\textit{\textbf{BP}}_1$: Quantities $\Delta_{TE} A$   and $\Delta_{\Psi} A$ from  relations \eqref{eq:1} and \eqref{eq:2}, have similar significances of measuring uncertainties for the observable A. Consequently, the respective relations should be regarded as having a same meaning of Uncertainty Relations  (UR) concerning the simultaneous measurements of observables A and B.  Such a regard is fortified much more   by the fact that 
$ \left| {\left\langle {\left[ {\hat A,\hat B} \right]} \right\rangle _\Psi  } \right| = \hbar $  when A and B are (canonically) conjugate.
     \item  $\textit{\textbf{BP}}_2$:  In case of a solitary observable A, for a micro-particle, the quantities    $\Delta_{TE} A$   or $\Delta_{\Psi} A$ can have always an unbounded small value. Therefore such an observable should be considered as measurable without any uncertainty in all cases of micro-particles (systems) and states.
		   \item  $\textit{\textbf{BP}}_3$:   For two commuting observables A and B (whose operators 
			$\hat A$   and  $\hat B$ commute, i.e. $\left[ {\hat A,\hat B} \right] = 0$ ) relation \eqref{eq:2} allows for the product 
			$\Delta _\Psi  A \cdot \Delta _\Psi  B$  to be no matter how small. Consequently the quantities  $\Delta_{\Psi} A$  and    $\Delta_{\Psi} B$  can be unlimited small at the same time. Such observables have to be regarded  as being compatible, i.e. measurable simultaneously and without interconnected uncertainties, for any micro-particle (system) or state. 
		
			  \item $\textit{\textbf{BP}}_4$:  In case of two non-commuting observables A and B (described by  operators $\hat A$  and $\hat B $  which  do not commute, i.e. $ \left[ {\hat A,\hat B} \right] \ne 0$  )  the relation \eqref{eq:2}   shows that the product $\Delta _\Psi  A \cdot \Delta _\Psi  B$  has as  lower bound a non-null and $\hbar$-dependent quantity.  Then the quantities  $\Delta_{\Psi} A$    and  $\Delta_{\Psi} B$   can be never reduced concomitantly to null values.  For that reason the respective observables must be accounted as measurable simultaneously only with non-null and interconnected uncertainties, for any situation (particle/state). Viewed in a pair such observables are proclaimed as being incompatible, respectively complementary when they are (canonically) conjugate. 
		     \item  $\textit{\textbf{BP}}_5$: The main elements of CIUR doctrine and UR philosophy show quantum particularities of uniqueness comparatively with other non-quantum areas of physics. Such elements are the very existence of relations\eqref{eq:1}  and \eqref{eq:2}, the above asserted measuring features and the discriminating presence of the Planck's constant $\hbar$. 
				   \item  $\textit{\textbf{BP}}_6$: For glorifying the precepts $\textit{\textbf{BP}}_1$ - $\textit{\textbf{BP}}_5$  and adopting the usages of dominant literature, UR philosophy in its entirety should be   ranked to a status of fundamental concept named Uncertainty Principle (UP).
				\end{itemize}
				
				Add here the observation that, in their wholeness, CIUR doctrine conjointly with UR and QMS prevalent philosophy emerge completely from the assertions embedded in basic precepts  $\textit{\textbf{BP}}_1$ - $\textit{\textbf{BP}}_6$.

\section{\textit{\textbf{D}}eficiencies (\textit{\textbf{D}}) of the mentioned precepts }
     The above mentioned emergence conceals a less popularized   fact namely that each of the precepts  
		$\textit{\textbf{BP}}_1$ - $\textit{\textbf{BP}}_6$ is discredited (and denied) by insurmountable deficiencies.  Such a fact can be revealed through a deep analysis of the respective precepts, an analysis which   is of major importance for an authentic and fruitful survey of UR and QMS prevalent philosophy. That is why here below we aim to reveal the most significant ones of the mentioned deficiencies. They will be presented in a meaningful ensemble, able to give an edifying global appreciation regarding the mentioned philosophy. The referred  ensemble includes as distinct pieces the following \textit{\textbf{D}}eficiencies (\textit{\textbf{D}}):

\subsection{$\textbf{\textit{D}}_1$: Provisional character of relation \eqref{eq:1}}
Now it must be noted firstly the aspect that, through an analysis of its origins, relation \eqref{eq:1} shows only a
 provisional (transient) character.  This because it was founded \cite{25,26} on old resolution criterion from optics (introduced by Abe and Rayleigh - see \cite{33}). But the respective criterion was surpassed through the so-called super-resolution techniques worked out in modern experimental physics (see \cite{34,35,36,37,38} and references). Then by means of   of the mentioned techniques can be imagined  some interesting \textit{“Super-Resolution-Thought-Experiments”} (\textit{SRTE}). Through such \textit{SRTE }for two (canonically) conjugate observables A and B, instead of \textit{TE}-uncertainties $\Delta_{TE} A$   and 
 $\Delta_{TE} B$   from \eqref{eq:1}, it becomes possible to discuss situations with some  \textit{SRTE}-uncertainties  denoted as $\Delta_{SRTE} A$   and 
 $\Delta_{SRTE} B$.  For the respective  \textit{SRTE}-uncertainties, instead of  Heisenberg’s   restrictive formula \eqref{eq:1} (first - version), can be suggested some CIUR-discordant relations like as
\begin{equation}\label{eq:3}
\Delta _{SRTE} A \cdot \Delta _{SRTE} B < \hbar 
\end{equation}
Note that an experimental example of discordant relation of \eqref{eq:3}-type  was mentioned in \cite{39} ( where the UR \eqref{eq:1}  \textit{"would be violated by close to two orders of magnitude"}).

      Now one observes that, from the our days scientific perspective, \textit{SRTE} relations like \eqref{eq:3} are suitable to replace the old Heisenberg's formula \eqref{eq:1} (second -  version ). But such suitability invalidates a good part of the precept  $\textit{\textbf{BP}}_1$   and, additionally, it incriminates the CIUR doctrine and UR - QMS philosophy in connection with one of their main (generative) element.

		It is surprising that, after invention of  the super-resolution  techniques, the mainstream  (normative /authoritarian ) publications connected with UR - QMS philosophy  avoided a just and detailed evaluation  of the respective techniques. Particularly, even after eight year after the result reported in \cite{39}, almost all of the dominant  publications omit to discuss the respective result. The surprise is evidenced to a great extent by the fact that   parsimony desiderata noted in Section 1 offer a viable argumentation for completing the evaluations and discussions oft the mentioned kind. 
		
	Another infringement (violation) of Heisenberg's relation \eqref{eq:1}  was reported in 
	\cite{40} as an experimental result. That report  is criticized  vehemently by  CIUR partisans \cite{12} . The respective  criticism  is done  in terms of a few  un-argued 
	(and un-explained) accusatory-sentences. But it is expected  that, if they are justifiable, such kind of  critiques  should be grounded on precise technical details and arguments. This  in order that they to be credible.

		Curiously is also the fact that, over the past decades within the UR philosophy, the debates have neglected the older criticisms of the relation \eqref{eq:1} due to K. Popper \cite{41}.
	
     Taking into account the above revealed aspects one can say that the precept 
		$\textbf{\textit{BP}}_1$ proves oneself to be a 	misleading  (even harmful) basic element for CIUR doctrine and UR - QMS philosophy. But such a proof is a first argument for reporting that the respective doctrine and philosophy cannot be accepted as solid (and credible) scientific constructions.

\subsection{$\textbf{\textit{D}}_2$:  Significance of quantities from  relation \eqref{eq:2}}
The term \textit{"`uncertainty"'} used within CIUR doctrine for quantities  
$\Delta_{\Psi} A$   and  $\Delta_{\Psi} B$ from \eqref{eq:2} is groundlessly because of the following considerations. According the theoretical framework of QM, by their definitions, the respective quantities   signify genuinely the standard deviations of the observables A and B regarded as random variables (see below Appendix A).  With such significances the alluded quantities refer to  intrinsic ( own) properties (known as fluctuations) of the considered particle but not to characteristics of the measurements performed on  respective particle. In fact, on a one hand, for a measured particle in a given state (described by certain wave function $\Psi$) the quantities  $\Delta_{\Psi} A$   and $\Delta_{\Psi} B$  have unique and well definite values. On the other hand for the same particle/state the measuring uncertainties regarding the observables A and B can be changed through the improvements or deterioration  of experimental devices/techniques. 

The above revealed   QM significances for quantities $\Delta_{\Psi} A$   and $\Delta_{\Psi} B$ are genuinely preferable comparatively with the assertions from the precepts 
 $\textit{\textbf{BP}}_1$ -  $\textit{\textbf{BP}}_4$     promoted by CIUR doctrine and UR - QMS philosophy. But such a preference is completely congruent with the previously mentioned desiderata of parsimony principle. 

\subsection{$\textit{\textbf{D}}_3$: Limitations of relation  \eqref{eq:2}}
     Relation \eqref{eq:2}  has only limited validity within the complete theoretical framework of QM. This because, as it is detailed below in Appendix A, for observables A and B, relation \eqref{eq:2} is only a restricted consequence of the generally valid Cauchy-Schwarz    formula, given in  \eqref{eq:A2}.  From such a general formula the relation \eqref{eq:2}  results iff (if and only if) in circumstances when the conditions \eqref{eq:A3} are satisfied. In the respective circumstances in addition to 
		relation \eqref{eq:2}/ \eqref{eq:A7} from \eqref{eq:A2} arises also the formula \eqref{eq:A6}. It is worthy to note    that the mentioned particularities regarding the validity of the relation \eqref{eq:2}   discredit indirectly the precept   $\textit{\textbf{BP}}_1$  of CIUR doctrine and UR - QMS philosophy.  In their essence the specifications recorded here are nothing but concretizations   of parsimony desiderata regarding the respective doctrine and philosophy. 

\subsection{$\textit{\textbf{D}}_4$:  On solitary observables}
      It is surprising to find that, within UR - QMS  philosophy debates, the problem of solitary observables is not discussed carefully. Particularly, were neglected discussions regarding the measurements of such observables. This although the respective discussions can be sub-summed to the question of simultaneous measurements of two observables. Such a sub-summation can be imagined by means of the Thought 
			Experiments (\textit{TE}) which motivated the conventional relation \eqref{eq:1}. Namely, for example, if in the respective \textit{TE} it is of interest only the quantity $\Delta_{TE}A$ , by ignoring completely the quantity $\Delta_{TE}B$ , one can say that  $\Delta_{TE}A$  can be unlimited small. Therefore the observable $A$, regarded as a solitary variable, appears as measurable without any uncertainty in all cases. But, on the other hand, if the same solitary observable A is analyzed in terms of relation \eqref{eq:2}, it cannot be associated with an unlimited small value for the quantity $\Delta_\Psi A$. This because, form a QM perspective, $\Delta_\Psi A$  has a unique and well definite value, evaluated through   the corresponding wave function $\Psi$. Consequently, even in the cases of solitary observables, the CIUR doctrine and the UR - QMS philosophy cannot provide a clear and unequivocal approach as it is suggested by  precept  $\textbf{\textit{BP}}_2$.

\subsection{$\textit{\textbf{D}}_5$:  About commutable observables}
     According to the precept $\textit{\textbf{BP}}_3$  for two observables $A$ and $B$, whose associated operators $\hat A$ and $\hat B$ are commutable,  relation  \eqref{eq:2}, allows for the product $\Delta_\Psi A \cdot \Delta_\Psi B$ to be however small. Then the quantities $\Delta_\Psi A$ and $\Delta_\Psi B$ can be unlimited small at the same time. Such observables are supposed compatible, they being measurable simultaneously and without interconnected uncertainties for any micro-particle (system) or state.
		
       But, as it was shown above in deficiency $\textit{\textbf{D}}_2$, the mentioned assertions from 
			$\textit{\textbf{BP}}_3$, conflict with the genuine significance of the quantities  $\Delta_\Psi A$ and $\Delta_\Psi B$. This because both  $\Delta_\Psi A$ and $\Delta_\Psi B$  have unique values,   determined theoretically by the wave function $\Psi$ which describe the considered state of particle. Or it is possible to have some 'rebellious situations' in which the respective values of  $\Delta_\Psi A$ and $\Delta_\Psi B$ to be simultaneously non-zero but finite entities, even the corresponding observables are commutable.
		
  Such a  'rebellious situation' can be found \cite{20} for the observables 
				$P_x$ and $P_y$ (Cartesian moments) regarding a micro-particle situated in a potential well of   a rectangular 2D configuration. If the well walls are inclined towards the $X$ and $Y$ axes, the both the quantities  $\Delta_\Psi {P}_x$ and 	$\Delta_\Psi {P}_y$  have non-zero but finite values. In that situation for $P_x$ and $P_y$, besides the relation \eqref{eq:2}, it is satisfied however the formula 
				\eqref{eq:A2} with  $\left| {\left( {\delta _\Psi  \hat P_x \Psi ,\delta _\Psi  \hat P_x \Psi } \right)} \right|$ as a non-null quantity.
				
     The above remarks show that, in fact, the cases of commutable   observables require to repudiate   firmly the precept 	$\textit{\textbf{BP}}_3$. Additionally we think that the same cases should be regarded in the spirit of parsimony principle desiderata, by their consideration in QM terms reminded briefly  in \\ Appendices A and B.

\subsection{$\textit{\textbf{D}}_6$: Cases of  angular observables $L_z$ and $\varphi$ }
The precept $\textit{\textbf{BP}}_4$ stipulates that, as a principle, two non-commutable observables $A$ and $B$ cannot be measured simultaneously because the product 
 $\Delta_\Psi A \cdot \Delta_\Psi B$  
has a non-null lower bound. But the respective stipulation is contradicted by some rebellious pairs of observables. Such a pair, widely discussed, is $L_z$ - $\varphi$ (angular momentum - azimuthal angle), regarded in certain particular situations. The respective contradiction was probably the most inciting subject of debates during the history of CIUR doctrine and UR - QMS philosophy (see \cite{5,17,18,19,20, 42,43,44,45,46,47,48,49,50,51,52,53,54,55}). The mentioned debates regarded mainly the quantum rotations which can be called
"`$L_z$ -\textit{ non - degenerate - circular - rotations}"' ($L_z$ -\textit{ndcr}).  But, besides  of that  situations, in QM framework can be discussed also other kinds  of  rotations,  of direct  significance 
 for  $L_z$ -  $\varphi$   pair. Such kinds are   the ones regarding the rotational eigenstates of a Quantum Torsion Pendulum (QTP) and respectively the "`$L_z$ - \textit{degenerate - spatial - rotations}"' ($L_z$ - \textit{dsr}).  The true situations of the $L_z$ - $\varphi$ pair in relation with all kinds of the mentioned rotations will be discussed below in more details. 

\subsubsection{$\textit{\textbf{D}}_{6a}$: About non-degenerate circular rotations}
Let us discuss now the cases of $L_z$ - \textit{non - degenerate - circular \\- rotations} ($L_z$  - \textit{ndcr}). As systems of with  $L_z$ - \textit{ndcr}   can be quoted the following ones: (i) a particle (bead) on a circle, (ii) an 1D rotator and (iii) non-degenerate spatial rotations of   a particle on a sphere or of an electron in a hydrogen atom respectively. The mentioned spatial  rotations are considered as  $L_z$-non-degenerate if the magnetic quantum number $m$ (associated with $L_z$) has a unique value 
			(while, of course, all other specific (orbital) quantum numbers have well-defined values). The rotations of respective systems are  described through  the wave functions given by  
\begin{equation}\label{eq:4}
\Psi \left( \varphi  \right) = \Psi _m \left( \varphi  \right) = \left( {2\pi } \right)^{ - \;\frac{1}{2}}  \cdot \exp \left( {im\varphi } \right)
\end{equation}
Here $\varphi$ is an ordinary polar coordinate (angle) with the corresponding mathematical  characteristics \cite{56}  i.e.  $\varphi\in[0,2\pi)$    and  number $m$ gets only  one value from the set $m= 0,\pm1,\pm2, ...$ . Also in \eqref{eq:4}  the wave function $\Psi(\varphi)=\Psi_m (\varphi)$  has the property $\Psi \left( 0 \right) = \Psi _m \left( {2\pi  - 0} \right): = \mathop {\lim }\limits_{\varphi  \to 2\pi  - \;0} \;\Psi _m \left( \varphi  \right)$ .

In the same context, according to the known QM framework \cite{29} , $L_z$ and $\varphi$ should be regarded as polar observables, described by the conjugated operators and commutator represented as follows  
\begin{equation}\label{eq:5}
\hat L_z  =  - i\hbar \frac{\partial }{{\partial \varphi }}\quad ,\quad \hat \varphi  = \varphi  \cdot \quad ,\quad \left[ {\hat L_z ,\hat \varphi } \right] =  - i\hbar 
\end{equation}
Therefore the conventional relation \eqref{eq:2} motivates as a direct consequence the next formula
\begin{equation}\label{eq:6}
\Delta _\Psi  L_z  \cdot \Delta _\Psi  \varphi  \ge \frac{\hbar }{2}
\end{equation}
Now it is easy to observe that this last formula is explicitly inapplicable in cases described by wave functions \eqref{eq:4}. This because in such cases, for the quantities $\Delta_\Psi L_z$  and
$\Delta_\Psi \varphi$   associated with the pair $L_z$ - $\varphi$, one obtains the following values
\begin{equation}\label{eq:7}
\Delta _\Psi  L_z  = 0\quad ,\quad \Delta _\Psi  \varphi  = \pi  \cdot \left( 3 \right)^{ - \;\frac{1}{2}} 
\end{equation} 
But such values for  $\Delta_{\Psi}L_z$  and $\Delta_{\Psi}\varphi$  are evidently incompatible with the     conventional relation  \eqref{eq:2} /  \eqref{eq:6} .

     In order to avoid the above revealed incompatibility in many mainstream publications the CIUR partisans promoted some unusual ideas such are:
		
\begin{itemize}
	\item  For $L_z$ and $\varphi$ operators and commutator, instead of current expressions \eqref{eq:5}, it is conveniently to adopt other new denotations (definitions).
	\item  The formula \eqref{eq:6}   must be abandoned/proscribed and replaced by one (or more) '\textit{modified} $L_z$ - $\varphi$ \textit{UR} ' able to mime the conventional relation \eqref{eq:2}  for the  $L_z$ - $\varphi$  pair.
\end{itemize}
  
The alluded ideas were   promoted through the conception of 
\textit{'impossibility of distinguishing … between 
two states of angle differing by 2$\pi$'}.  But such a conception has not  any realistic sense in cases of circular rotations. This because in such cases 
the angle $\varphi$  has as physical range the interval $[0, 2\pi)$.  Moreover in the respective cases the wave functions \eqref{eq:4} are normalized on the same interval but not on other strange domains. 

      As regards the '\textit{modified} $L_z$ - $\varphi$ \textit{UR}', along the years, by means of some circumstantial (and more or less fictitious) considerations, were proposed a lot of such relations. In terms of usual QM notations (summarized below in Appendix A), the alluded 
			'\textit{modified} $L_z$ - $\varphi$ \textit{UR}' can be written generically as follows
\begin{equation}\label{eq:8}
f\left( {\Delta _\Psi  L,\Delta _\Psi  g\left( \varphi  \right)} \right) \ge \hbar  \cdot \left\langle
{s\left( \varphi  \right)} \right\rangle _\Psi  
\end{equation}
Here $f\left( {\Delta _\Psi  L,\Delta _\Psi  g\left( \varphi  \right)} \right)$, $g\left( \varphi  \right)$ and $s\left( \varphi  \right)$  denote some specially invented functions depending on the corresponding arguments. Note that some of the mostly known concrete examples of relations \eqref{eq:8} can be found collected in \cite{55}.

     Now it should be noted the fact that the '\textit{modified $L_z$ - $\varphi$ UR}' such are \eqref{eq:8} show some troubling features like the following ones:
		
\begin{itemize}
	\item   Regarded comparatively, the mentioned '\textit{modified} $L_z$ - $\varphi$ \textit{UR}' are not mutually equivalent. This despite of the fact that they were invented in order to substitute the same proscribed formula \eqref{eq:6}. Consequently, none of that modified relations, is agreed unanimously as a suitable model able to give such a substitution.
		\item   Relations \eqref{eq:8} are in fact ad hoc artifices without any  source in mathematical framework of QM. Then, if one wants to preserve QM as a unitary theory, like it is accredited in our days, the relations \eqref{eq:8} must be regarded as unconvincing and inconvenient (or even prejudicial) inventions.
\item	 In fact in relations \eqref{eq:8}  the relevant angular quantities
 $\Delta_{\Psi}L_z$   and 
 $\Delta_{\Psi}\varphi$     are substituted more or less factitious with the adjusting functions \\
 $f\left( {\Delta _\Psi  L_z,\Delta _\Psi  g\left( \varphi  \right)} \right)$ , 
$g\left( \varphi  \right)  $ 
and $s\left( \varphi  \right)$. But, from a genuine perspective, such substitutions, and consequently the corresponding relations, are only mathematical constructs but not elements with useful physical significance.
 Of course that such constructs overload (or even impede) the scientific discussions by additions of extraneous entities which are not associated with true information about the real world.
\end{itemize}

         Then, for a correct evaluation of the facts, all the aspects regarding relations \eqref{eq:8} versus \eqref{eq:6} ought to be judged by taking into consideration the parsimony principle desideratum: \textit{"`Entities are not to be multiplied beyond necessity"'}. Such an evaluation can be started by clarifying firstly   the origin and validity conditions of the formula \eqref{eq:6}  regarded as descendant of conventional relation \eqref{eq:2}. For the respective clarification it is usefully to see some QM elements briefly summarized in Appendix A.
				
	      So it can be observed easy that, in its essence, the relations \eqref{eq:2} follow from the generally valid formulas \eqref{eq:A2}  pertaining to the mathematical framework of QM. But, attention, \eqref{eq:2} results correctly from \eqref{eq:A2}   iff  (if and only if) when it is satisfied the condition \eqref{eq:A3}. In other cases \eqref{eq:2}  are not valid at all. Such an invalidity is completely specific for the cases of $L_z$ - $\varphi$ pair in relations with situations described by the wave functions 
				\eqref{eq:4}. This because in respective cases instead of conditions 
				\eqref{eq:A3}  it is true the relation			
\begin{equation}\label{eq:9}
\left( {\hat L_z \Psi ,\hat \varphi \;\Psi } \right) = \left( {\Psi ,\hat L_z \,\hat \varphi \;\Psi } \right) + i\hbar 
\end{equation}
Therefore, for systems described by the wave functions \eqref{eq:4}, the formula 
\eqref{eq:6}   is invalid by its essence. 

     Now note that, even when the condition \eqref{eq:A3}   is not satisfied, according to the QM general formula \eqref{eq:A2}, for the discussed situations it is true the relation 
\begin{equation}\label{eq:10}
\Delta _\Psi  L_z  \cdot \Delta _\Psi  \varphi  \ge \left| {\left( {\delta _\Psi  \hat L_z \Psi ,\,\delta _\Psi  \,\hat \varphi \,\Psi } \right)} \right|
\end{equation}
written in compliance with definitions \eqref{eq:4}  and \eqref{eq:5}. But, attention, in respective situations the last relation \eqref{eq:10} degenerates   into trivial equality $\textit{'0=0'}$. Add here the fact that relation \eqref{eq:10} is completely equivalent with the formula \eqref{eq:C13} deductible within Fourier analysis.

    The above presented details  argue undoubtedly the  view that  in cases  with   $L_z$ -\textit{ndcr}    the $L_z$ - $\varphi$  pair must to satisfy  not the troublesome formula
		\eqref{eq:6} but the QM justified relation \eqref{eq:10}       ( which in fact reduces itself  to banal equality $\textit{'0=0'}$). Such an argued view clarifies all disputes regarding the mentioned cases. Moreover the same view disproves the idea   of some \textit{'entities ... multiplied beyond necessity'} (such are the modified UR \eqref{eq:8}) intended to replace the inoperative relation \eqref{eq:6}.
		
 \subsubsection{$\textit{\textbf{D}}_{6b}$: Case of Quantum Torsion Pendulum (QTP) }
The case of Quantum Torsion Pendulum (QTP) regards a quantum   harmonic oscillator with torsional 
rotations \cite{19,20, 55}. Such an oscillator can be considered as the simplest theoretical model for molecular twisting motion 
(\textit{"`change in the angle between the planes of two groups of atoms "`} \
\cite {57}) . For a QTP oscillating around the z-axis the Hamiltonian operator has the form
\begin{equation}\label{eq:11}
\hat H = \frac{1}{{2I}}\hat L_z ^2  + \frac{1}{2}I\omega _0 ^2 \hat \varphi ^2 
\end{equation}  
Here $\varphi$   denotes the twisting angle with   domain $\varphi \in\left(-\infty, +\infty\right)$  while the operators
 $\hat{L}_z$  and  $\hat{\varphi}$  obey the rules 	\eqref{eq:5}.  The other symbols from 	\eqref{eq:11} are $I$ and $\omega_0$  representing the momentum of inertia respectively the (undamped) resonant frequency ( $ \omega _0  = \sqrt {{\raise0.7ex\hbox{$\kappa $} \!\mathord{\left/{\vphantom {\kappa  I}}\right.\kern-\nulldelimiterspace}
\!\lower0.7ex\hbox{$I$}}} $ , $\kappa$ = torsion elastic modulus). 

By means of Schrodinger equation $E\Psi=\hat{H}\Psi$   one finds that the QTP eigenstates are described by the wave functions
\begin{equation}\label{eq:12}
\Psi _n \left( \varphi  \right) = \Psi _n \left( \xi  \right) \propto \exp \left( { - \frac{{\xi ^2 }}{2}} \right) \cdot \mathcal{H}_n \left( \xi  \right)\;\;,\quad \xi  = \varphi \sqrt {\frac{{I\omega _0 }}{\hbar }} 
\end{equation} 

 These wave functions correspond to the oscillation quantum numbers \\ $n = 0, 1, 2, 3, . . . $ and energy eigenvalues 
$E_n  = \hbar \omega _0 \left( {n + \frac{1}{2}} \right)$  .  In \eqref{eq:12} 
$\mathcal{H}_n\left(\xi\right)$   represent the  Hermite polynomials of  $\xi$ . 

For each of the states 
\eqref{eq:12} for observables $L_z$ and $\varphi$ associated with the operators \eqref{eq:5} one obtains the expressions
\begin{equation}\label{eq:13}
\begin{array}{l}
 \Delta \varphi  = \sqrt {\frac{\hbar }{{I\omega _0 }}\left( {n + \frac{1}{2}} \right)} \;\;\;,\quad \Delta \,L{}_z = \sqrt {\hbar I\omega _0 \left( {n + \frac{1}{2}} \right)} \,\;\;\;,\quad \left| {\left( {\Psi ,\left[ {\hat L_z ,\hat \varphi } \right]} \right)} \right| = \hbar  \\ 
  \\ 
 \quad \quad \quad \quad \quad \quad \quad \quad \quad \Delta \varphi  \cdot \Delta \,L{}_z = \;\hbar  \cdot \left( {n + \frac{1}{2}} \right)\;\; \ge \;\;\frac{\hbar }{2} \\ 
 \end{array}
\end{equation}  

These expressions show the fact that, for each   QTP eigenstate, the \\ $L_z$ - $\varphi$  pair satisfies the relation  \eqref{eq:6}/\eqref{eq:2}. But note that the respective fact is due to the circumstance that in the mentioned case, in relation with the wave functions 
 \eqref{eq:12} , the operators  $\hat{L}_z$ and $\hat{\varphi}$ satisfy a condition of
\eqref{eq:A3} type, i.e.$\left( {\hat L_z \Psi ,\hat \varphi \;\Psi } \right) = \left( {\Psi ,\hat L_z \,\hat \varphi \;\Psi } \right)$.

\subsubsection{$\textit{\textbf{D}}_{6c}$: On degenerate spatial rotations}

Let us now regard the  cases of  $L_z$ –\textit{degenerate-spatial-rotations} ($L_z$ -\textit{dsr}).
    Such kinds of rotations refer \cite{20,21,55} to states of: (i) a particle on a sphere, (ii) a 2D rotator and (iii) an electron in a hydrogen atom.  The respective rotations are $L_z$ - degenerate in sense that the magnetic quantum number $m$ (associated with $L_z$  ) has multiple values   while the other quantum numbers have unique values. A particle on a sphere or a 2D rotator are in  a $L_z$  -\textit{dsr } when  the orbital number $l$ has a unique value greater than zero while $m$ can take all the values $m\in[-l,+l]$. Then the corresponding rotations are described through the global wave function
\begin{equation}\label{eq:14}
\Psi \left( \varphi  \right) = \Psi _l \left( {\vartheta ,\varphi } \right) = \sum\limits_{m = \; - \;l}^{m = \; + \;l} {c{}_m}  \cdot Y_{lm} \left( {\vartheta ,\varphi } \right)
\end{equation}

Here $\vartheta$ and $\varphi$ denote polar respectively azimuthal angles with
$\vartheta\in[0,\pi]$   and $\varphi\in[0,2\pi)$  . In \eqref{eq:14} 
$Y_{lm} \left( {\vartheta ,\varphi }\right)$  denote spherical functions while $c_m$ are coefficients normalized through the condition 
$\sum\nolimits_{m = \; - \;l}^{m = \; + \;l} {\left| {c{}_m} \right|} ^2  = 1$. Also the wave functions $\Psi_l(\vartheta,\varphi)$  from \eqref{eq:14}  have the property $\Psi _l \left( {\vartheta ,0} \right) = \Psi _l \left( {\vartheta ,2\pi  - 0} \right): = \;\mathop {\lim }\limits_{\varphi  \to 2\pi  - 0} \Psi _l \left( {\vartheta ,\varphi } \right)$. In a direct connection with such a property the operators $\hat{L}_z$  and $\hat{\varphi}$  obey the rules \eqref{eq:5}.

Now let us regard what are the peculiarities of the $L_z$- \textit{dsr}  cases in respect with the controversial relation
 \eqref{eq:6}.  Principled, such a regard demands that, by using the formulas \eqref{eq:5} and \eqref{eq:14}, to evaluate the corresponding expressions for the quantities $\Delta _\Psi  L_z $  ,  $\Delta _\Psi  \varphi $  and
				$\left| {\left( {\Psi ,\left[ {\hat L_z ,\,\hat \varphi \;} \right]\Psi } \right)} \right|$. With the respective expressions one finds possibilities that the relation \eqref{eq:6} to be or not to be satisfied. Of course that such possibilities are conditioned by the concrete values of the coefficients $c_m$.  But note that, if the relation \eqref{eq:6} is not satisfied, the fact  appears because essentially   in such a situation the condition 
				\eqref{eq:A3} is not fulfilled.  Add here the important observation that, independently of validity for   relation 
				\eqref{eq:6}, in all cases of $L_z$ -\textit{dsr}  the $L_z$ - $\varphi$ pair obeys the prime  QM relation 
				\eqref{eq:A2} through adequate values for the quantities  $\Delta _\Psi  L_z $  ,  $\Delta _\Psi  \varphi $  and $\left| {\left( {\delta _\Psi  \hat L_z \Psi ,\delta _\Psi  \hat \varphi \;\Psi } \right)} \right|$  .  The previous considerations offer a clear evaluation of the situation for $L_z$- \textit{dsr} cases relatively to the conventional relation 
				\eqref{eq:2}  and precept $BP_4$. 
				
				  \underline{Summing up of deficiencies $\textit{\textbf{D}}_6$ (including $\textit{\textbf{D}}_{6a}$,  $\textit{\textbf{D}}_{6b}$ and
					$\textit{\textbf{D}}_{6c}$}): The above discussion about the three kinds of rotations reveals the deficiencies  of the conventional 
relation \eqref{eq:2} and  of the associated precept $\textit{\textbf{BP}}_4$  in regard with the non-commutable observables $L_z$ and $\varphi$. But such revealing is nothing but a direct and irrefutable incrimination of CIUR doctrine and UR - QMS philosophy. 

\subsection{$\textit{\textbf{D}}_7$: On number and phase observables }

The pair $N$ and $\phi$    (number and phase) is another couple of rebellious non-commutable observables which contradict the corresponding stipulation from the precept  $\textit{\textbf{BP}}_4$  of UR - QMS  philosophy. That contradiction emerged in connection   with the associated operators $\hat{N}$  and $\hat{\phi}$. The respective operators were introduced by means of the   ladder (lowering and raising)    operators $\hat{a}$ and $\hat{a}^+$, destined to convert some QM calculations procedures from an analytical version to an algebraic one. Through the respective connection, by taking as base the relation $\left[ {\hat a,\hat a^ +} \right] = 1$ , it was inferred the commutation formula $\left[ {\hat N,\hat \phi } \right] = i$ . 

      The last noted formula motivated the idea that operators  $\hat{N}$   and $\hat{\phi}$     must satisfy the conventional relation \eqref{eq:2} with both $\Delta_\Psi N$  and $\Delta_\Psi \phi$   as non-null quantities. But afterward it was found the fact that, in the case of a harmonic oscillator eigenstates, one obtains  
$\Delta_\Psi N = 0$ and $\Delta _\Psi  \phi  = \pi  \cdot \left( 3 \right)^{ - \;\frac{1}{2}} $     i.e. a violation of the relation \eqref{eq:2}.  Of course that such a fact leads to a deadlock for harmonization of $N$ - $\phi$    observables with the CIUR doctrine and UR - QMS philosophy. Note that this deadlock is completely analogous with the one regarding to $L_z$ - $\varphi$ observables in the above discussed case of  $L_z$-\textit{ndcr} ($L_z$-  non-degenerate –circular- rotations ).

For avoiding the mentioned $N$ - $\phi$   deadlock in many publications were promoted various
adjustments (see \cite{6,43,48,58,59,60,61} and references therein). But it is easy to observe that the respective adjustments regarded the conventional relation \eqref{eq:2} as an absolute mark and tried to adapt accordingly the pair $N$ - $\phi$    for a description of a harmonic oscillator. So it was suggested to replace the original operators  $\hat{N}$  - $\hat{\phi}$  by some ad hoc \textit{'adjusted'} (\textit{adj}) operators  $\hat{N}_{adj}$  
 and $\hat{\phi}_{adj}$  , able to generate formulas resembling (more or less) with the conventional relation \eqref{eq:2} (examples of such adjusted operators can be found in the literature of recent decades). However it is very doubtfully that the corresponding  '\textit{adjusted observables}' $N_{adj}$ and $\phi_{adj}$   can have natural (or even useful) physical significances. Moreover, until now, it not exist a unanimously agreed conception able to guarantee a true elucidation regarding the status of number-phase observables relatively to terms of CIUR doctrine and UR philosophy.

     Our opinion is that a genuine clarification of the $N$ - $\phi$      problem can be done similarly with the above discussed situation of $L_z$ - $\varphi$ observables in the cases of  
		$L_z$-\textit{ndcr}. More exactly we have to note that the disagreement of $N$ - $\phi$    pair with the conventional relation \eqref{eq:2} results from fact that in such a case the respective relation is mathematically incorrect. The aforesaid incorrectness is due mainly to the circumstance that, in cases of a linear  oscillator eigenstates, the $N$ - $\phi$   pair does not satisfy the essential condition \eqref{eq:A3}. This because in that cases for the   operators  $\hat{N}$  - $\hat{\phi}$    is true the formula  $\left( {\hat N\,\Psi ,\hat \phi \;\Psi } \right) = \left( {\Psi ,\hat N\,\hat \phi \;\Psi } \right) + i$  which evidently infringes the condition \eqref{eq:A3}. But it should be pointed out that, even in the mentioned cases, the   $\hat{N}$  - $\hat{\phi}$  operators satisfy the primary relation \eqref{eq:A2} which degenerates into trivial equality '$\textit{0 = 0}$'.
		
		    We think that the above noted opinion gives a natural and incontestable  solution for the problem regarding the $N$ - $\phi$     pair versus the conventional relation \eqref{eq:2}. Accordingly the fictional operators $\hat{N}_{adj}$ and $\hat{\phi}_{adj}$, of  an ad hoc adjusted essence, proves themselves to be nothing but \textit{'entities ... multiplied beyond necessity'}.  
				
    So it can be said that the situation of observables $N$ and  $\phi$    contradict directly the precept $\textit{\textbf{BP}}_4$ in connection with non-commutable observables. Consequently, the respective situation invalidates completely one of basic elements of CIUR doctrine and UR - QMS philosophy.
			
\subsection{$\textit{\textbf{D}}_8$: Concerning the energy - time pair}
Closely to the conventional views of CIUR doctrine and UR - QMS philosophy the pair of observables    $\textit{E}$ - $\textit{t}$  (energy - time) was subject for a large number of   controversial discussions
 (e.g. in works \cite {5,6,62,63,64}, in their references and,  certainly, in many other publications).  The alluded discussions were generated within following circumstances. On one hand, accordingly to the mentioned views,  $\textit{E}$ and $\textit{t}$  are regarded as conjugated observables, having to be described by the next operators and commutator
\begin{equation}\label{eq:15}
\hat E = i\hbar \frac{\partial }{{\partial t}}\;\;,\quad \;\;\hat t = t \cdot \;\;,\quad \;\;\left[ {\hat E,\hat t} \right] = i\hbar 
\end{equation}
  Then the operators $\hat {E} $ and  $\hat {t} $ should satisfy the conventional relation \eqref{eq:2} in a nontrivial version. On the other hand, because of the fact that, in terms of usual QM, the time $\textit{t}$ is a deterministic but not random variable, for any quantum situation one finds the following expressions $\Delta_{\Psi} E $  = '\textit{a finite quantity}' respectively $\Delta_{\Psi} t \equiv 0$ . But these expressions invalidate the relation \eqref{eq:2} and consequently the $\textit{E}$ - $\textit{t}$   pair shows an anomaly in respect with the alluded conventional ideas, especially with the precept 
	$\textit{\textbf{BP}}_4$. For avoiding the noted anomaly, within the literature about $\textit{E}$ - $\textit{t}$  pair, it was substituted the unsuitable relation \eqref{eq:2} by some adjusted formulas written generically as follows
	\begin{equation}\label{eq:16}
\Xi \;E \cdot \Xi \;t \ge \frac{\hbar }{2}
\end{equation}
 The so introduced  quantities  $\Xi \textit{E}$ and $\Xi \textit{t}$   have various significances such are: (i) line-breadth and  half-life  of  a decaying excited state, (ii) frequency domain and temporal widths  of a wave packet, (iii ) $\Xi \textit{E}$ = $\Delta_{\Psi}\textit{E}$  and $\Xi \;t = \Delta _\Psi  A \cdot \left( {{\raise0.5ex\hbox{$\scriptstyle {d\left\langle A \right\rangle }$}
\kern-0.1em/\kern-0.15em\lower0.25ex\hbox{$\scriptstyle {dt}$}}} \right)^{ - \;1}$,  with $A$ = an arbitrary observable.

       As regards the adjusted formulas \eqref{eq:16}  note firstly the fact that various  of their versions are not  congruent with the original conception of relation \eqref{eq:2}. Also the respective versions are not mutually equivalent from a mathematical (theoretical) viewpoint. So they have no reasonable justification in the true QM framework. Moreover in specific literature none of the formulas    \eqref{eq:16}  is accepted unanimously as a correct (or natural) substitute for conventional relation \eqref{eq:2}.
     
		Now it is the place to present the following clarifying remarks. Even if the $\textit{E}$ - $\textit{t}$ pair is considered to be described by the operators  \eqref{eq:15}, according to the true QM terms, one finds the relation
	\begin{equation}\label{eq:17}
\left( {\hat E\Psi ,\hat t\Psi } \right) = \left( {\Psi ,\hat E\;\hat t\Psi } \right) - i\hbar 
\end{equation}	
	By comparing   this relation   with condition \eqref{eq:A3}  one sees directly that the $\textit{E}$-$\textit{t}$ pair cannot ever satisfy the respective condition.  This is the essential reason because of which for the $\textit{E}$-$\textit{t}$ pair the conventional relation \eqref{eq:2} is not applicable at all. Nevertheless, for the same pair described by the operators \eqref{eq:15}, the  QM relation 
	\eqref{eq:A2} is always true. But because in QM the time $\textit{t}$ is a deterministic (i.e. non-stochastic) variable in all cases the respective true relation degenerates into the trivial equality '$\textit{0 = 0}$ '.
	
	The above noted comments   lead to the next findings:
	
	\begin{itemize}
		\item  In case of the $\textit{E}$-$\textit{t}$ pair the conventional views (of CIUR doctrine and UR - QMS philosophy) are completely nonfunctional.
			\item   Genuinely, within a true QM framework, the time $t$ is in fact a pure deterministic (non-stochastic ) quantity without any standard deviation 
			(or fluctuation).	
	\end{itemize}

		But, taken together, such findings about time - energy pair must be reported as a serious and insurmountable deficiency of CIUR doctrine and UR - QMS philosophy.
		
\subsection{$\textit{\textbf{D}}_9$: Atypical analogues of UR (1) and (2)}
 
     By basic precept $\textbf{BP}_5$ the UR philosophy claims idea that relations \eqref{eq:1} and \eqref{eq:2} possess an essential  typicality represented by their  QM  uniqueness related with the systems of atomic size. 
	 Consequently,  the respective relations should not have analogues in other areas of physics or for systems of radically different sizes.  But the respective idea is definitely denied by some example that we will present below.
\subsubsection{$\textit{\textbf{D}}_{9a}$: Classical  Rayleigh formula}	
As a first example of  an atypical analogue of the UR (1) can be quoted  the formula 
\begin{equation}\label{eq:18}
\sin \alpha  \cong \frac{\lambda }{d}
\end{equation} 	
which expresses \cite{35,39,40,65} the   Rayleigh resolution criterion from classical optics.  In 
\eqref{eq:18} $\alpha$ denotes the '\textit{angular resolution}', $\lambda$  is the wavelength of light, and $d$  represents the diameter of lens aperture. Note that criterion \eqref{eq:18} was introduced in classical optics in 1879, i.e. by long time before the QM appeared.  Later one relation \eqref{eq:1} was introduced by taking in \eqref{eq:18}$  \\ d\thicksim \Delta_{TE} q$  for coordinate uncertainty, respectively 
$\lambda=(\hbar/p)$    for momentum $p$  (through wave-particle duality formula) and
$p\cdot\sin{\alpha}\thicksim  \Delta_{TE} p$   for momentum uncertainty.

\subsubsection{$\textit{\textbf{D}}_{9b}$: Classical  'Gabor's uncertainty relation'}
          An example of an atypical analogue  of \eqref{eq:2} can be found  within  the mathematical harmonic analysis in connection with a pair of random quantities regarded as Fourier conjugated variables (see \cite{66,67} and the Appendix C below). In non-quantum physics such  an   analogue is known \cite{67}  as\textit{ 'Gabor's uncertainty relation' } which can be represented  through the relation
	\begin{equation}\label{eq:19}
\Delta t \cdot \Delta \nu  \ge \frac{1}{{4\pi }}
\end{equation}				
This last relation \eqref{eq:19} shows the fact that for a classical signal,    regarded as a wave packet (of acoustic or electromagnetic nature), the product of the '\textit{uncertainties}' ('\textit{irresolutions}')   
 $\Delta t$ and $\Delta\nu$ in the time and frequency domains cannot be smaller than a specific constant.

\subsubsection{$\textit{\textbf{D}}_{9c}$: A relation regarding  thermodynamic observables }

Another example of an atypical  similar of UR \eqref{eq:2} is given by the following classical formula
	\begin{equation}\label{eq:20}
\Delta _W \mathbb{A} \cdot \Delta _W \mathbb{B} \ge \left| {\left\langle {\delta _W \mathbb{A} \cdot \delta _W \mathbb{B}} \right\rangle _W } \right|
\end{equation}
showed  as relation  \eqref{eq:D3} in Appendix D  of the present article. The elements
 (notations and physical significances)  implied in \eqref{eq:20} are those detailed  in Appendix D. The respective elements are specific  to the phenomenological theory, initiated by Einstein,  about fluctuations  of macroscopic thermodynamic observables 
(see \cite{20,68,69,70,71,72}  and Appendix D  below).  

     Note that, from the perspective of mathematics (more exactly of  probability 
		theory), the macroscopic formula \eqref{eq:20} and UR \eqref{eq:2} are analogue relations, both of them regard the fluctuations of the corresponding observables judged  as random variables.  Moreover they  describe the intrinsic properties of considered systems (of macroscopic-thermodynamic respectively quantum nature) but not aspects of measurements performed on the respective systems. The corresponding measurements can be described through a distinct approaches modeled/depicted as   transmission processes for stochastic data (see below Appendix E and  Section 5   in present article).		
		
    As  regards  the  formula  \eqref{eq:20}, the following notifications should  be  done too. To a some extent the respective formula can be  considered as being   member to a family of so  called  '\textit{thermodynamic UR}' , discussed in a number of    publications from the last century (see \cite{78,79}  and references).  Note that the  alluded membership is true only in respect with the '\textit{regular}'  subset of respective family, derivable from the  Einstein's phenomenological theory. But the mentioned  family  includes moreover a class of '\textit{irregular}' relations.  The most known  such an '\textit{irregular}' relation regards  the conjugate variables energy $U$ and temperature $T$ of a thermodynamic system. It has \cite{78}   the form
\begin{equation}\label{eq:21}
\Delta U \cdot \Delta \left( {\frac{1}{T}} \right) \ge k_B 
\end{equation}
where $k_B$ denote the Boltzmann's constant.

It must be noted now the reality that  fluctuation formula \eqref{eq:20} and 
'\textit{irregular}' relations like is \eqref{eq:21} 
 are completely dissimilar, first of all, due to the important  distinction between reference frames  of their definitions. The respective dissimilarity is pointed out by the following aspects.  On the one hand, the quantities   $\Delta_W \mathbb{A}$  and $\Delta_W \mathbb{B}$   from \eqref{eq:20} are defined by referring to the same state of the considered system. On the other hand the quantities $U$ and  $T$ which appear in \eqref{eq:21} refer to  different states of a system, namely    states characterized by an  energetic  isolation respectively by a thermal contact. Due mainly to the above mentioned dissimilarity :
 "`\textit{a derivation of the uncertainty relation  \eqref{eq:21} analogous to that of the usual Heisenberg relations} (i.e. UR  \eqref{eq:2}) \textit{is impossible}"'\cite{78}.

Add here the fact that, within associate literature, it was reported a number of controversies about the aspects regarding the possible similarities  between the '\textit{thermodynamic UR}' (mainly from the same subset as  \eqref{eq:21} ) and quantum UR \eqref{eq:2} (see \cite{78} and references). Among respective aspects can be quoted : 
\begin{itemize}
	\item compatibility of macroscopic observables, 
	\item commutativity of   thermodynamic variables and
	\item reconstruction of QM from hidden variables theories similarly with the rebuilding of thermodynamics through  subjacent molecular considerations.

\end{itemize}
  
Note that the just  mentioned aspects are not taken  into account (as relevant elements) for  our present survey on deficiencies of prevalent philosophy regarding UR and QMS.

\subsubsection{$\textit{\textbf{D}}_{9d}$: On the so called macroscopic operators}
    In the spirit of conventional precept $\textit{\textbf{BP}}_5$ the uniqueness of UR 
		 \eqref{eq:2} consists in its strict specificity for   micro-particles (of  atomic size), without analogues  in cases of  macroscopic  systems. But, as it is pointed out through relation  \eqref{eq:D12}  from  Appendix D, in case of macroscopic thermodynamic system studied in quantum statistical physics one finds the formula 
	\begin{equation}\label{eq:22}
\Delta _\rho \mathbb{A} \cdot \Delta _\rho \mathbb{ B} \ge \frac{1}{2}\left| {\left\langle {\left[ {\hat {\mathbb{A}}},{\hat {\mathbb{B}}} \right]} \right\rangle _\rho  } \right|
\end{equation}	
This last formula is  similar with the conventional UR \eqref{eq:2}  (more exactly, mathematically, with  its  primary versions  \eqref{eq:A7} and \eqref{eq:B4}).   Due to such a similarity, probably,  some publications (e.g. \cite{74}  and references) have tried  to regard \eqref{eq:22} as a macroscopic UR. But the respective regard was found to be incompatible with the known UR - QMS  philosophy, mainly with the 
precept  $\textit{\textbf{BP}}_4$. 	

    The  alluded incompatibility is pointed out by the following facts. On the one hand, in  spirit of  UR philosophy (precepts $\textit{\textbf{BP}}_1$ - $\textit{\textbf{BP}}_4$),  the  quantities $\Delta_\rho \mathbb{A}$  and  $\Delta_\rho \mathbb{B}$ 
		from \eqref{eq:22} should be  considered  as measuring uncertainties of  macroscopic observables 
			$\mathbb{A}$ and  $\mathbb{B}$.  Additionally when the operators 
					$\hat{\mathbb{A}}$ and  $\hat{\mathbb{B}}$ and  do not commute 
					(i.e. $[	\hat{\mathbb{A}}, \hat{\mathbb{B}} ]\neq 0$ ), according to 
					\eqref{eq:22},  the quantities   $\Delta_\rho \mathbb{A}$  and  $\Delta_\rho \mathbb{B}$     can be never  reduced concomitantly to null values. Consequently, in terms of UR - QMS philosophy, for any situation, the non-commutable macroscopic observables   $\mathbb{A}$ and  $\mathbb{B}$ are allowed to be measurable simultaneously only with non-null and interconnected uncertainties.  But, on the other hand, according to the classical physics any two macroscopic observables can be measured concurrently with unlimited accuracies and without any interrelated uncertainties. 

					For avoiding the above noted incompatibility  some partisans of UR philosophy have suggested  the following expedient. Abrogation of \eqref{eq:22}   by replacement   of genuine  macroscopic operators $\hat{\mathbb{A}}$ and  $\hat{\mathbb{B}}$  	   with another quasi-diagonal operators $\hat{\mathbf{A}}$  and 	$\hat{\mathbf{B}}$ 
					(i.e. with operators whose representations in any base are quasi-diagonal matrices). Such substituting operators should to commute
and so the right hand term in \eqref{eq:22}  to be (quasi) null (i.e. $\left| {\left\langle {\left[ \hat{\mathbf {A}},\hat{\mathbf {B}} \right]} \right\rangle _\rho  } \right|\approx 0 $). Through the mentioned substitution  the inconvenient relation \eqref{eq:22} could be changed  with the more convenient formula						
	\begin{equation}\label{eq:23}
\Delta _\rho \mathbf {A} \cdot \Delta _\rho \mathbf { B} \ge \frac{1}{2}\left| {\left\langle {\left[ \hat{\mathbf {A}},\hat{\mathbf {B}} \right]} \right\rangle _\rho  } \right|\approx 0
\end{equation}					
Then it seems to be possible that the substituted macroscopic uncertainties 
$\Delta_\rho \mathbf{A}$ and   $\Delta_\rho \mathbf{B}$    to be reduced simultaneously to arbitrarily small (even zero) values. Apparently, such a possibility should to harmonize the interpretation of the relation \eqref{eq:23} with the concepts of classical physics. 

      However, in fact, the above mentioned harmonization is not possible and the suggested expedient is useless. This, at least, due to the following reasons:
	
	\begin{itemize}
		\item Firstly, the relations \eqref{eq:22} cannot be abrogated/substituted if the entire mathematical framework of quantum statistical physics is not abrogated/substituted too.
	\item  Secondly, in common practice of studies of quantum statistical systems (e.g.such are the ones investigated in \cite{80,81}) are used  the genuine operators $\hat{\mathbb{A}}$ and  $\hat{\mathbb{B}}$  but not the quasi - diagonal ones 
	$\hat{\mathbf{A}}$ and  $\hat{\mathbf{B}}$.
		\item As a third reason, the following fact can be also noted. Even in certain situations when the original operators $\hat{\mathbb{A}}$ and  $\hat{\mathbb{B}}$    are quasi-diagonal in the sense of the mentioned expedient, the relation \eqref{eq:23} does not turn into a form having a null term in the right hand side. Such a situation can be found \cite{20}   in case regarding a macroscopic paramagnetic system made of a huge number  of independent $1/2$-spins. In such a case as macroscopic operators appear the Cartesian components $\hat{\mathbb{M}}_\alpha$   
($\alpha = x,y,z$ ) of the system magnetization. Note that the operators 
$\hat{\mathbb{M}}_\alpha$  are quasi-diagonal in the sense required by the aforesaid expedient/substitution. But, for all that, the respective operators do not commute because
 $[\hat{\mathbb{M}}_\alpha,\hat{\mathbb{M}}_\beta] = i\hbar\gamma\cdot\epsilon_{\alpha\beta\mu}\cdot\hat{\mathbb{M}}_\mu$ 
($\gamma$ = magneto-mechanical factor and    
$\epsilon_{\alpha\beta\mu} $ denotes the Levi-Civita tensor).
\end{itemize}
			
By taking into account  the above pointed out  deficiencies $\textit{\textbf{D}}_9$  (including $\textit{\textbf{D}}_{9a}$, $\textit{\textbf{D}}_{9b}$, $\textit{\textbf{D}}_{9c}$  and
 $\textit{\textbf{D}}_{9d}$)  one  may record the following conclusion. The relations
\eqref{eq:D12}/\eqref{eq:22} are relations regarding macroscopic areas of physics but not pieces which  should be adapted to the requirements  of  prevalent philosophy about UR and QMS. 
	\subsection{$\textit{\textbf{D}}_{10}$: On the uniqueness  of quantum measurements}	
		     Let us refer now to the uniqueness character of conventional relations 
				\eqref{eq:1} and \eqref{eq:2} with regard to the measurements peculiarities at quantum level. The aforesaid character was largely debated in literature and it has generated the still open questions about   the main characteristics (conceptual relevance and description procedures) of Quantum Measurements (QMS). By promoting all the assertions from percepts $\textit{\textbf{BP}}_1$ - $\textit{\textbf{BP}}_4$  the UR - QMS philosophy tried to enforce the opinion that relations \eqref{eq:1} and \eqref{eq:2}  are closely linked with the   measuring particularities that are unique in quantum context, without any correspondence (analogy) in non-quantum domains of physics. The mentioned opinion, often promoted as a true dogma, dominates the mainstream of existing publications.

     On the other hand, as we have argued above through the deficiencies \\
		$\textbf{\textit{D}}_1$ - $\textbf{\textit{D}}_9$, the alluded opinion is completely unfounded because, genuinely, the respective relations are :
		
		\begin{itemize}
			\item  either an old-fashioned (and removable) empirical convention  
			(in case of \eqref{eq:1}), 
			\item or simple	(non-magistral) theoretical   formula (in case of \eqref{eq:2}).
		\end{itemize}
	 
	Within UR - QMS prevalent philosophy, as a widespread belief, the uniqueness peculiarities of QMS  are motivated  through the so called 'observer effect'. The respective effect is presented as a perturbing influence of observer ( by  experimental devices) on investigated systems and  on measuring results. It is presumed to differentiate radically the QMS from classical  measurements (of macroscopic physics). Such effects  are    absolutely  unavoidable and affected by notable uncertainties  in quantum  contexts but  entirely preventable and with negligible inaccuracies  in classical situations.

The above mentioned belief  is categorically disproved by the following observations. The 'observer effect' appear not only in QMS but also in  
some classical  measurements  (e.g. \cite{82} in electronics or in thermodynamics). Of course that in classical cases the measuring inaccuracies can be made negligible (by adequate improvements of experimental devices and/or procedures). It should be noted, that, in principle, quantum uncertainties can be also diminished ( for example, with the super-resolution  techniques discussed above in $\textit{\textbf{D}}_1$ ).
	
	Then the idea of uniqueness quantum measuring character for conventional relations 
		\eqref{eq:1}  and \eqref{eq:2},  promoted by UR philosophy through 
		$\textit{\textbf{BP}}_5$,  proves oneself  as being a groundless fiction which should be disregarded. But such a disregard come to fortify the J. Bell's thinking \cite{83,84} that: "`\textit{the word 'measurement' should be avoided (} or even ... \textsl{banned)  altogether in quantum mechanics}"'. Some annotations about the respective thinking are given below in Section 5 where we will present briefly a non-conventional approach of QMS problems.

\subsection{$\textit{\textbf{D}}_{11}$: On the  uniqueness of Planck's constant}	
	  Another aspect of quantum uniqueness invoked in precept 
		$\textit{\textbf{BP}}_5$   regards the presence of Planck's constant $\hbar$ as a specific symbol in conventional quantum relations \eqref{eq:1}  and \eqref{eq:2}, comparatively with a total absence of some similar symbols in all classical (non-quantum) formulas. We shall examine the alluded aspect in regard with the relation \eqref{eq:2}.  Then of prime importance is to notify the fact that, mathematically, quantum observables from the relation \eqref{eq:2} have a stochastic (non-deterministic) character. But a completely similar character one finds in cases of macroscopic observables implied in formula \eqref{eq:20} regarding fluctuations specific to macroscopic thermodynamic systems.
				
				     Both kinds of mentioned stochastic observables describe fluctuations (at quantum respectively macroscopic scale).  The mentioned   fluctuations are characterized quantitatively by the corresponding standard deviations such are $\Delta_\Psi A$    or  $\Delta_W \mathbb{A}$. But, mathematically, the standard deviation indicates quantitatively the stochasticity (randomness) degree of an observable. This in the sense that the alluded deviation has a positive or null value as the corresponding observable is a random or, alternatively, a deterministic (non-stochastic) variable. Consequently the quantities  $\Delta_\Psi A$   and  $\Delta_W \mathbb{A}$ can be regarded as similar indicators of stochasticity for  quantum respectively macroscopic observables.
	
						      In principle for macroscopic thermal fluctuations the standard deviations like is  $\Delta_W \mathbb{A}$ can have various expressions (depending on system, state and observable). Apparently, it would seem that the respective expressions do not contain any common element. Nevertheless such an element can be found as being materialized by the Boltzmann's constant $k_B$   (see  relation \eqref{eq:D4} in Appendix D below and articles \cite{71,73} ). So, for any macroscopic fluctuating observable $A$, the quantity $(\Delta_W \mathbb{A})^2$  
									(i.e. dispersion = square of the standard deviation) appears as a product of Boltzmann's constant $k_B$ with factors which are independent of  $k_B$ .
									
									This means that the quantity $(\Delta_W \mathbb{A})^2$, in its quality of quantitative indicator of thermal fluctuations, is directly proportional with $k_B$. Consequently $(\Delta_W \mathbb{A})^2$ has a non-null respectively null value as $k_B\neq 0$ or  $k_B\rightarrow 0$ (Note that because 
									$k_B$  is a physical constant the limit $k_B\rightarrow 0$ means that the quantities directly proportional with $k_B$   are negligible comparatively with other quantities of same dimensionality but independent of
									$k_B$ ). On the other hand,  the standard deviation 
										$\Delta_W \mathbb{A}$ is a particular indicator for macroscopic stochasticity revealed through thermal fluctuations. 
										
										Bringing together the above noted aspects it can be said that $k_B$  has the qualities of an authentic generic indicator for thermal stochasticity which is specific for classical macroscopic fluctuating systems.
										
	Now let us discuss about the quantum stochasticity whose particular indicators are the standard deviations $\Delta_\Psi A$. Based on the relations \eqref{eq:13} one can say that in many situations the expressions for dispersions 
$ (\Delta_\Psi A)^2$ consist in products of Planck constant $\hbar$ with factors which are independent of $\hbar$. Then, by analogy with the above discussed macroscopic situations, $\hbar$ places itself in the posture of generic indicator for quantum stochasticity.
	
	   The mentioned roles as generic indicators for $k_B$ and $\hbar$ (in direct connections with the quantities $\Delta_W \mathbb{A}$ and $\Delta_\Psi A$ ) regard the one-fold (simple) stochasticity, of thermal and quantum nature respectively. But in physics is also known a twofold (double) stochasticity, of a combined thermal and quantum nature. Such a kind of stochasticity one finds in cases of macroscopic thermodynamic systems composed of statistical assemblies of quantum micro-particles. The alluded twofold stochasticity can be evaluated in a way through the dispersions $(\Delta_\rho \mathbb{A}_j)^2$   which estimate the level of fluctuations in the mentioned systems
		(see \cite{20,73,76} and Appendix D below). As it is noted in relation 
		\eqref{eq:D13} the dispersions $(\Delta_\rho \mathbb{A}_j)^2$   can be given through of products containing the function
		$\mathfrak{f}(k_B,\hbar) = {\hbar\cdot coth(\frac{\hbar\omega}{2k_B T})}$    and factors which are independent of  both $k_B$ and $\hbar$.
		
		Then it results that  $k_B$ and $\hbar$  considered together turn out to be a couple of generic indicators for the twofold (double) stochaticity of thermal and quantum nature. Such a kind of stochaticity is significant or negligible in situations when $k_B\neq 0$ and $\hbar\neq 0$  respectively if
				$k_B\rightarrow 0$ and $\hbar\rightarrow 0$.
				
      Now we can note   the indubitable remark that Planck's constant $\hbar$    has an authentic classical analog represented by the Boltzmann's constant
				$k_B$,  both $\hbar$ and $k_B$  having relevant significances as generic indicators of stochaticity. But such an analogy contradicts directly the basic precept $\textit{\textbf{BP}}_5$  of  CIUR doctrine and UR - QMS philosophy.
				
				\subsection{$\textit{\textbf{D}}_{12}$: On the excessive ranking of UR}
			
			 The ranking of UR  to a position of  principle,  is widespread in the dominant literature, mainly through the authoritative and normative writings of many leading scientists. Surprisingly the respective  ranking is  argued  merely in few occasions (e.g. in \cite{10}) but only partially and not convincingly.
							
			       However, in  \cite{10},   it was signaled the fact that "`\textit{over the years, some authors and foremost K. Popper, have contested this view }, of such a \\ …'ranking' "'. The mentioned contestation seems to have been motivated by the assertion: "`\textit{ uncertainty relations cannot be granted the status of a principle on the grounds that they are derivable from the theory }('QM'), \textit{whereas one cannot obtain the theory from the uncertainty relations}"'. The  aforesaid motivation was minimized and repudiated \cite{10}  through  of  the  conventional (and prevalent)  opinion that : "`\textit{there are many statements in physical theories which are called principles even though they are in fact derivable from other statements in the theory in question}"'.  Note that in spite of the mentioned repudiation,    it was added in \cite{10}  the noteworthy observation that "`\textit{ Serious attempts to build up quantum theory as a full-fledged Theory of Principle on the basis of the uncertainty principle have never been carried out }"`.
						
					As regards the above presented controversy our belief can be expressed as follows. The  Popper's contestation of  UR ranking (i.e., in fact, of  the  precept $\textit{\textbf{BP}}_6$ ) has a genuine character  while the opposing conventional opinion is nothing but a questionable (and unfounded) attempt to preserve a predominant traditionalist doctrine (dogma).

     Now, from another   perspective, we wish to point out a new important aspect. On the one hand a true scientific conception   attests indubitably the idea that: "`\textit{A principle is statement which is taken to be true at all times and all places where it is applicable}"' \cite{85}. On the other hand   all previously proved deficiencies $\textit{\textbf{D}}_1$ - $\textit{\textbf{D}}_{11}$   show that  usual philosophy of UR is not valid in a wide class of situations where they should to be applied. Therefore such a philosophy cannot provide (generate) a principle (fundamental concept) applicable in an unquestionable manner for a large area of situations. That is why it turns out to be totally unacceptable (and useless) the idea to raise the entire UR philosophy to a rank of fundamental principle for QM.  

Consequently, the precept   $\textit{\textbf{BP}}_6$    shows oneself as being nothing but an unjustified thesis. At the same time, from a true scientific perspective, it is outside of acceptable usages to put in practice an idea such is  
\cite{10}  : \textit{"`we use the name 'uncertainty principle' simply because it is the most common one in the literature"'}.

\section{Which is really the true significance of UR?}

Summing  all the  discussions incorporated within deficiencies 
$\textit{\textbf{D}}_1$ - $\textit{\textbf{D}}_{12}$ one can notify the following evident   remarks:
	
\begin{itemize}
	\item 	There are profound deficiencies regarding   all the basic elements and precepts of the conventional conceptions (CIUR doctrine and UR - QMS philosophy). 
\item	In their essence the respective deficiencies are unavoidable and insurmountable within own   framework of respective conceptions.
\item	Consequently the mentioned conceptions prove themselves as being undoubtedly in a failure situation which impose their abandonment.
\end{itemize}

      The above argued abandonment of   conventional conceptions points out   very clearly the indubitable ending of the existing prevalent philosophy  about UR. But a fair evaluation of such an ending requires an adequate epilogue regarding the future scientific status of the respective philosophy and of its constitutive and associate concepts.
			
     The alluded epilogue demands firstly, detailed  re-evaluations of the generative relations \eqref{eq:1} and \eqref{eq:2} from which have been expanded themselves the mentioned philosophy and concepts.  The respective re-evaluations have to be done and argued by taking into account all the aspects noted previously within the texts of deficiencies 	$\textit{\textbf{D}}_1$ - $\textit{\textbf{D}}_{12}$. Doing so one arrives to the following observations:
	
\begin{itemize}
	\item 	Relation \eqref{eq:1} is nothing but   an old-fashioned (and removable) empirical convention. It persists as a  piece of historical reminiscence,  destitute of any wonderful  status/significance for actual and future physics. 
\item 	Relation \eqref{eq:2}  proves to be   only an ordinary QM formula, of well-defined (but not universal) validity. In such a posture it describes in a simple manner the connections between fluctuation characteristics of two quantum observables.
\item In fact the relations \eqref{eq:1} and \eqref{eq:2}  have not any crucial significance, for  QM concretely and less so for physics in general.
\item  Relations \eqref{eq:1} and \eqref{eq:2} or  their 'adjustments' have not any connection with  genuine descriptions of QMS. 
\item   Particularly the respective relations do not depict in any way the so called 'observer effect' (i.e. perturbing influence of  'experimenter' on the investigated system ).
 \end{itemize}

\section{ Considerations on quantum measurements}
       Besides the main discussions about the   meaning of early  relations \eqref{eq:1} and \eqref{eq:2}, the conventional UR philosophy   generated also many collateral debates on Quantum Measurements (QMS) 
			( see \cite{1,2,3,4,5,6,7,8,9,10,11,12,86,87,88} and references). The respective debates, still active in writings of  many scientists, promoted an appreciable diversity of viewpoints about conceptual significance  and practical importance of QMS. But in the same  context, were recorded  observations like is the following one
\begin{itemize}
		 \item \textit{"`Despite long efforts, no progress has been made. . . for . . . the understanding of quantum mechanics, in particular its measurement process and interpretation"'}\cite{89}.
	 \end{itemize}
	Nevertheless, beyond the mentioned debates, the respective subject of QMS involves also  a matter of real interest for physics. That matter regards     the natural interest in developing adequate theoretical description(s) for QMS, which should to be  proved through viable  arguments and which have to become  of  suitable  utility for scientific and technical activities.
	
	The above signaled situation  have motivated  interest  for both conventional and non-conventional  approaches of QMS  problem. A modest non-conventional  approach was put in work progressively in our investigations over many years (see \cite{17,18,19,20,47,55,90,91,92,93,94} ). Here, as well as in all sections of present article, we try to gather, extend, systematize   and improve the results of mentioned investigations in order to present  argued viewpoints about  the main aspects  of QMS matter .	
	
	\subsection{$\textit{\textbf{D}}_{13}$ : The  incorrect association of  QMS with UR}
	
	As a first main aspect of the so much debated QMS problem   is  fact that it has a theoretical essence. Namely, it  is focused  around the idea of developing  a general theoretical model for describing measurements on quantum systems. The respective model should have 
	some  similarity (a bit of reference) with the one centered on Schrodinger equation within QM.
	
		From the  perspective of the such supposed similarity  most  of  publications  promoted or accepted the opinion that QMS  have a basic essentiality  for QM in itself.
  During the years were recorded even assertions like the following one : 
	\begin{itemize}
	\item '\textit{the description of QMS is "`probably the most important part of the theory (QM)"'} ' \cite{5}. 
	\end{itemize}
But note that both the mentioned opinion and assertion are grounded on the belief that, mainly, the claimed essentiality/importance of QMS for QM is given by relations  \eqref{eq:1} and \eqref{eq:2} in terms of precepts $\textit{\textbf{BP}}_1$ - $\textit{\textbf{BP}}_6$. 

On the other hand, it is easy to see that the respective belief is invalidated by the  arguments from  the entire collection of deficiencies $\textit{\textbf{D}}_1$ - $\textit{\textbf{D}}_{12}$ notified  by us above in Section 3. 
	
	Now, besides the aforesaid notifications, for starting our  non-conventional  approach of QMS subject, we take into account the following remarks of J.S.Bell:	
\begin{itemize}
	\item "'\textit{I agree with what you say about the uncertainty principle : it has to do with the uncertainty in predictions rather the accuracy of 'measurement'. \textbf{I think in fact that the word 'measurement' has been so abused in quantum mechanics that it would good to avoid it altogether}}"'(see\cite{83} and Appendix I below).
\item  "`\textit{...The word ('measurement') has had such a damaging effect on the discussions that. . . it should be banned altogether in quantum mechanics}"'\cite{84}.
\end{itemize}
A similar account we give also to  the next remark:		
\begin{itemize}
	\item  \textit{'the procedures of measurement (comparison with standards) has a part which cannot be described inside the branch of physics where it is used'}.\cite{95}
\end{itemize}
The just noted remarks consolidate for us the following  key view

\begin{itemize}
	\item The significance of UR is an intrinsic question of QM
while the description of QMS constitutes an adjacent
but distinct subject comparatively with QM in itself.
\end{itemize}

As another  reference element for starting our  approach we agree the following observation:
 
	\begin{itemize}
		\item "`\textit{it seems essential to the notion of measurement that it answers a question about the given situation existing before the measurement. Whether the measurement leaves the measured system unchanged or brings about a new and different state of that system is a second and independent question}"'\cite{96}.
\end{itemize}
In sense of above observation for a measured physical system the 'situation existing before the measurement' regards the intrinsic properties of that system. The characteristics of the respective properties   play a role of input data (information) for measuring actions. On the other hand  for the same system, the 'answer (i.e. result) of measurement'  is accumulated in  'output data (information)' that are provided  by  measuring process. Correspondingly  the whole  measurement can be considered as a transmission process for information (stochastic data), while the measuring device appears as  a communication channel (viewed as in \cite{97}).
  
So the  whole image of a measurement can be depicted through the scheme
\begin{equation}\label{eq:24}
\left| \begin{array}{l}
 input \\ 
 data \\ 
 \end{array} \right\rangle  \Rightarrow \left[ \begin{array}{l}
communication \\ 
 \;\;\;\;channel \\ 
 \end{array} \right] \Rightarrow \left[ \begin{array}{l}
 output \\ 
 \;data \\ 
 \end{array} \right]
\end{equation}
 
For giving  concrete  descriptions of the above scheme in cases of QMS 
(measurements on quantum systems) it should also  to take into view the next remark
\begin{itemize}
	\item \textit{"`To our best current knowledge the measurement process in quantum mechanics is non-deterministic"'}\cite{89}.	
\end{itemize}
In such a view the  mentioned input and output  data as well the description of a QMS have to be presented by means of some non-deterministic  (stochastic or random) entities.  For a measured quantum system the totality of   input data  can be considered as being   comprised in  its specific (intrinsic)  wave function $\Psi_{in}$,   with   known stochastic/probabilistic own significance. As regards  the same system the output data  should be represented by some quantities having also stochastic features. Formally, such quantities  can be considered as being incorporated  in an  output wave function $\Psi_{out}$.  Then  the measuring process appear as communication channel which 	transposes the wave function    from a $\Psi_{in}$ reading into  a $\Psi_{out}$ image. So it can be suggested  that, in case of a QMS,   the scheme \eqref{eq:24}  can be represented through the following generic  pattern:
\begin{equation}\label{eq:25}
\left| \begin{array}{l}
 \;{\rm probabilistic} \\ 
 {\rm content}\;{\rm of }\;\Psi _{{\rm in}}  \\ 
 \end{array} \right\rangle  \Rightarrow \left[ \widehat{\textbf{\textit{SCC}}} \right] \Rightarrow \left[ \begin{array}{l}
 \;{\rm probabilistic} \\ 
 {\rm content}\;{\rm of }\;\Psi _{{\rm out}}  \\ 
 \end{array} \right]
\end{equation}
where   $\widehat{\textbf{\textit{SCC}}}$,   depicts the \textit{'\textbf{\textit{S}}tochastic \textbf{\textit{C}}ommunication \textbf{\textit{C}}hannel'} regarded as an \textit{'operator'} which describe the  measuring process. 

The above suggested pattern regarding QMS can be particularized for various concrete situations by using QM terminology. Two such particularization will be detailed below in the Subsections 5.2 and 5.4.

\subsection{On   an observable with  discrete spectrum}

 Let us refer to the case of a  QMS for  a single quantum observable $A$ 
endowed with a non-degenerate discrete  spectrum of eigenvalues $\left\{ {a_j } \right\}_{j = 1}^ n $. The respective observable is described by the operator $\hat{A}$ 
which satisfy the equations $\hat{A}\varphi_j = a_j\cdot\varphi_j$, where 
$\left\{ {\varphi_j } \right\}_{j = 1}^ n $ signify the corresponding eigenfunctions.

If the set of eigenfunctions $\left\{ {\varphi_j } \right\}_{j = 1}^ n $ is regarded   as an  orthonormal basis the wave functions $\Psi_{in}$  and   $\Psi_{out}$ can be represented as follows
\begin{equation}\label{eq:26}
\begin{array}{l}
 \Psi _{in}  = \sum\limits_{j = 1}^n {\alpha _j } \varphi _j \quad ,\quad \sum\limits_{j = 1}^n {\left| {\alpha _j } \right|^2 }  = 1 \\ 
  \\ 
 \Psi _{out}  = \sum\limits_{k = 1}^n {\beta _k } \varphi _k \quad ,\quad \sum\limits_{k = 1}^n {\left| {\beta _k } \right|^2 }  = 1 \\ 
 \end{array}
\end{equation}
 Then the the pattern \eqref{eq:25} appears as a transformation of the  corresponding probabilities from  \textit{in}-readings
 $\left\{ {\left| {\alpha _j } \right|^2 } \right\}_{j = 1}^n $ 
 into  \textit{out}-images $
\left\{ {\left| {\beta _k } \right|^2 } \right\}_{k = 1}^n $ . According to mathematics (probability and information theories) the mentioned transformation (i.e.the operator
$\widehat{\textbf{\textit{SCC}}}$)  can be depicted by means of a doubly stochastic matrix
$M_{kj}$ ($k,j=1,2,...,n$), interpreted as in \cite{98}. Such a depiction has the form
\begin{equation}\label{eq:27}
\left| {\beta _k } \right|^2  = \sum\limits_{j = 1}^n {M_{kj} }  \cdot \left| {\alpha _j } \right|^2 
\end{equation}
where the matrix $M_{jk}$ satisfies the conditions
$\sum\limits_{k = 1}^n {M_{kj} }  = 
\sum\limits_{j = 1}^n {M_{kj} }  = 1$.

As above described  a QMS appear as being ideal respectively non-ideal, accordingly as 
$M_{kj}=\delta_{kj}$ or $M_{kj}\neq\delta_{kj}$, where $\delta _{kj}$ denotes   a Kronecker delta.

By using \eqref{eq:26} and \eqref{eq:27} for the $\eta$-expected  values 
$\left\langle A\right\rangle_\eta = \left(\Psi_\eta,\hat{A}\Psi_\eta\right)$, ($\eta= in, out$),  of observable $A$ one obtains
\begin{equation}\label{eq:28}
\begin{array}{l}
 \quad \quad \quad \quad \left\langle A \right\rangle _{in}  = \sum\limits_{j = 1}^n {a_j  \cdot \left| {\alpha _j } \right|^2 }  \\ 
  \\ 
 \left\langle A \right\rangle _{out}  = \sum\limits_{k = 1}^n {a_k  \cdot \left| {\beta _k } \right|^2 }  = \sum\limits_{k = 1}^n {\sum\limits_{j = 1}^n {a_k  \cdot M_{kj} } }  \cdot \left| {\alpha _j } \right|^2  \\ 
 \end{array}
\end{equation}
In terms of above notations the error for the expected value of $A$ is:
\begin{equation}\label{eq:29}
\mathcal{E}\left\{ {\left\langle A \right\rangle } \right\} = \left\langle A \right\rangle _{out}  - \left\langle A \right\rangle _{in}  = \sum\limits_{k = 1}^n {\sum\limits_{j = 1}^n {a_k  \cdot \left( {M_{kj}  - \delta _{kj} } \right)} }  \cdot \left| {\alpha _j } \right|^2 
\end{equation}
where $\delta_{jk}$ signifies  a Kronecker delta.

Because, mathematically,  the observable $A$ is a random variable it is characterized also by the  standard deviations $ \Delta_\eta A $ ($\eta= in, out$), defined as follows
\begin{equation}\label{eq:30}
\begin{array}{l}
 \left( {\Delta _{in} A} \right)^2  = \left\langle {\left( {A - \left\langle A \right\rangle _{in} } \right)^2 } \right\rangle _{in}  = \sum\limits_{j = 1}^n {a_j^2 }  \cdot \left| {\alpha _j } \right|^2  - \left( {\sum\limits_{j = 1}^n {a_j  \cdot \left| {\alpha _j } \right|^2 } } \right)^2  \\ 
  \\ 
 \quad \quad \quad \quad \left( {\Delta _{out} A} \right)^2  = \left\langle {\left( {A - \left\langle A \right\rangle _{out} } \right)^2 } \right\rangle _{out}  =  \\ 
  \\ 
 \quad \quad  = \sum\limits_{k = 1}^n {\sum\limits_{j = 1}^n {a_k^2  \cdot M_{kj}  \cdot \left| {\alpha _j } \right|^2 } }  - \left( {\sum\limits_{k = 1}^n {\sum\limits_{j = 1}^n {a_k  \cdot M_{kj} \left| {\alpha _j } \right|^2 } } } \right)^2  \\ 
 \end{array}
\end{equation}
So for   error $\mathcal{E}\left\{\Delta A\right\}$ of standard deviation regarding the observable $A$ one finds

\begin{equation}\label{eq:31}
\begin{array}{l}
 \quad \quad \quad \quad \mathcal{E}\left\{ {\Delta \;A} \right\} = \Delta _{out} A - \Delta _{in} A =  \\ 
  \\ 
  = \sqrt {\sum\limits_{k = 1}^n {\sum\limits_{j = 1}^n {a_k^2  \cdot M_{kj}  \cdot \left| {\alpha _j } \right|^2 } }  - \left( {\sum\limits_{k = 1}^n {\sum\limits_{j = 1}^n {a_k  \cdot M_{kj} \left| {\alpha _j } \right|^2 } } } \right)^2 }  -  \\ 
  \\ 
 \quad \quad \quad \quad \quad \sqrt {\sum\limits_{j = 1}^n {a_j^2 }  \cdot \left| {\alpha _j } \right|^2  - \left( {\sum\limits_{j = 1}^n {a_j  \cdot \left| {\alpha _j } \right|^2 } } \right)^2 }  \\ 
 \end{array}
\end{equation}

Now note the fact that, to some extent, the above presented  model of a QMS description has general features. This  because, excepting the conditions of being doubly stochastic, the measuring matrix $M_{kj}$  can consists of arbitrary components.  The mentioned generality/arbitrariness should be reduced when one refers to the relatively accurate measurements. Such a reduction can be modeled if the measuring matrix elements  $M_{kj}$ are taken of the forms 
\begin{equation}\label{eq:32}
\begin{array}{l}
 \quad \quad \quad M_{kj}  = \delta _{kj}  + \tau _{kj}  \\ 
  \\ 
 \left| {\tau _{kj} } \right| <  < 1\quad ,\quad \quad \sum\limits_{k = 1}^n {\tau _{kj}  = } \sum\limits_{j = 1}^n {\tau _{kj}  = } 0 \\ 
 \end{array}
\end{equation}
where $\delta_{kj}$ signifies  the a Kronecker delta and $\tau _{kj}$ are real and dimensionless   quantities of (very) small values.

When the matrix elements $M_{jk}$ are approximated as in \eqref{eq:32}  the errors 
$\mathcal{E}\left\{\left\langle A\right\rangle \right\}$ and
 $\mathcal{E}\left\{ {\Delta \;A} \right\}$ from \eqref{eq:29} and \eqref{eq:31} can be estimated through  a direct calculation, respectively by means of the first order term in  Taylor series. Then  one finds
\begin{equation}\label{eq:33}
\begin{array}{l}
 \mathcal{E}\left\{ {\left\langle A \right\rangle } \right\} = \sum\limits_{k = 1}^n {\sum\limits_{j = 1}^n {a_k } }  \cdot \tau _{kj}  \cdot \left| {\alpha _j } \right|^2  \\ 
  \\ 
 \mathcal{E}\left\{ {\Delta A} \right\} \approx \sum\limits_{k = 1}^n {\sum\limits_{j = 1}^n {\left[ {\frac{{\partial \mathcal {E}\left( {\tau_{kj}} \right)}}{{\partial \tau _{kj} }}} \right]} } _{\tau _{kj}  = 0}  \cdot \tau _{kj}  \\ 
 \end{array}
\end{equation}
where $\mathcal {E}\left( {\tau_{kj}} \right) $ signifies  the 
standard-deviation error   $\mathcal{E}\left\{ {\Delta \;A} \right\}$ from \eqref{eq:31} in which one uses the approximations   \eqref{eq:32}.

Relations \eqref{eq:33} show that within  mentioned  approximations  the parameters $\tau_{jk}$ appear as significant indexes regarding the  measuring accuracies. So the discussed measurement can be regarded as ideal when $\tau_{kj} =0$ for all $k$ and $j$, respectively as non-ideal when $\tau_{kj}\neq 0$ at least for some \\ values of $k$ or $j$.

\subsection{$\textit{\textbf{D}}_{14}$ : On the measuring  scenarios with a unique sampling } 
As it was  pointed out   in Subsection 5.1, a QMS is essentially a non-deterministic process. Due to the mentioned essentiality,  the 'result' of such a process must be represented in terms of some stochastic (probabilistic) output data.  But, surprisingly, in conventional publications  \cite{99,100,101,102,103,104,105,106} a QMS is regarded as a scenario (i.e. an imagined sequence of possible events) conceived as  a single sampling (i.e. as  a unique-deterministic selection from a set of random data ). So regarded, a QMS gives as its result (outcome) a single value in which falls (collapses) the whole physical content of the measured observable. The referred falling scenarios are illustrated  by two widely  debated themes  regarding    the Wave Function Collapse ( WFC) \cite{99,100,101,102,103} respectively  the  Schrodinger's Cat Thought Experiment (SCTE) \cite{104,105,106}.   Historically, both the respective themes have  occurred in a direct connection with the establishing of basic precepts  
$\textit{\textbf{BP}}_1$ - $\textit{\textbf{BP}}_6$ of CIUR doctrine and UR - QMS philosophy. Therefore, by taking into account the deficiencies of precepts $\textit{\textbf{BP}}_1$ - $\textit{\textbf{BP}}_6$ , revealed above in Section 3, it is here the place  to investigate also  the possible deficiencies of the aforesaid scenarios.  

 Let us begin the announced investigation by   referring to the WFC-measuring-scenario .  The respective scenarios  germinated  from the hypothesis  that, due to unavoidable measuring perturbations, all QMS cause specific collapses (falls, jumps) in states of the measured quantum systems. It can be presented succinctly in usual terms of QM as follows. 

Consider a measuring investigation focused on the system and observable $A$ discussed in the previous Subsection 5.2 . For the respective system in WFC-scenario the 'situation existing before measurement'  is inscribed in its intrinsic wave function $\Psi_{in}$. The  probabilistic content of $\Psi_{in}$ play the role of input data (information) for investigation  actions.  But, attention, within  the WFC-scenario, the measuring actions are imagined as providing as result  an \textbf{\textit{u}}nique \textbf{\textit{d}}etertministic \textbf{\textit{o}}utcome (\textbf{\textit{udo}}) namely  $a_k$.

 Note that $a_k$  is one of the eigenvalues  $\left\{a_j\right\}_{j=1}^n$   from the spectrum  of  $A$. The eigenvalues $\left\{a_j\right\}_{j=1}^n$  are defined through the relations 
$\hat A \varphi_j = a_j\cdot\varphi_j$ $\left(j=1,2,...,n\right)$, where 
$\left\{\varphi_j\right\}_{j=1}^n$ denote the eigenfunctions of operator $\hat A$ associated to the observable $A$. Then, in terms detailed previously in Subsection 5.2, the whole WFC-scenario can be illustrated through the following two schemes
\begin{equation}\label{eq:34}
\left| {\left\{ {a_j } \right\}_{j = 1}^n  \cup \left\{ {\left| {\alpha _j } \right|^2 } \right\}_{j = 1}^n } \right\rangle  \Rightarrow \left[ \widehat{\textbf{\textit{udo}}}  \right] \Rightarrow a_k 
\end{equation}
\begin{equation}\label{eq:35}
\left| {\Psi _{in}  = \sum\limits_{j = 1}^n {\alpha _j  \cdot \varphi _j } } \right\rangle  \Rightarrow \left[ \widehat{\mathfrak{{udo}}} \right] \Rightarrow \varphi _k 
\end{equation}
where $\widehat{\textbf{\textit{udo}}}$ symbolize an operator which describe the mesuring actions in WFC-scenario.
 
 On the one hand, firstly, the  schema \eqref{eq:34} regards the measurement of observable $A$.  It show a falling of the  respective observable from a whole spectrum of values $\left\{a_j\right\}_{j=1}^n$, having probabilities $\left\{\left|\alpha_j\right|^2\right\}_{j=1}^n$ in measured state, to a unique value $a_k$ as result of the scenario. Secondly, on the other hand, the schema \eqref{eq:35} refers to the evolution of the considered system from a state 'existing before the measurement' ( at the beginning of scenario) in an 'after measurement' state (in the end of scenario). 

Specify here the fact that  conventional publications (see \cite {99,100,101,102,103} and references) regard relation \eqref{eq:35} as being the essential symbol of WFC. That is why the mentioned publications tried to done  analytical representations of the respective relation  considered as image of a dynamical physical process. For such representations were promoted various inventions, e.g.   nonlinear extensions of Schrodinger equation or even appeals to new kinds of fundamental physical constants.

The above mentioned WFC-scenario regarding  QMS can be admonished through the following  remarks.

Firstly note  that quantum observables are stochastic variables. Consequently a true measurement of such an observable should be regarded as being provided not by an  \textbf{\textit{udo}} ( \textbf{\textit{u}}nique \textit{\textbf{d}}eterministic 
\textit{\textbf{o}}utcome) but by  an adequate probabilistic set of such outcomes. The data given  by the respective set are expected to provide 
relevant (and as complete as possible) information about the considered observables.

	Secondly, the idea of describing QMS through an analytical representation of the WFC schema  \eqref{eq:35} proves oneself as being an extravagance without solid arguments or credible hypotheses. Some main aspects of the respective extravagance can be revealed by taking into account the stochastic similitude between quantum and thermal (macroscopic) random observables. Such a reveal we point out here as follows.
 
     Let us refer to a macroscopic thermodynamic  system described in terms of phenomenological theory of fluctuations (see below the Appendix D). For simplicity the system will be considered to be characterized by a single macroscopic thermodynamic observable $\mathbb{A}$  .  Mathematically the macroscopic fluctuations of  $\mathbb{A}$  are accounted   by a real random variable $\mathcal{A}$   and described by the probability density $W = W (\mathcal{A})$.  Through the before specified terms can be pointed out the analogy between  measuring acts regarding the stochastic  observables of   quantum and macroscopic nature. 
		An \textit{\textbf{udo}}, specific to WFC-scenario, for a quantum observable was discussed succinctly above in connection with the relations  \eqref{eq:34} and\eqref{eq:35}.  A completely similar \textbf{\textit{udo}} regarding a macroscopic observable $\mathbb{A}$ can be depicted as follows. By means of an \textbf{\textit{udo}} for the variable $\mathcal{A}$ one obtains a unique value say $\mathcal{A}_0$.  Then for $\mathbb{A}$  the respective 
		\textit{\textbf{udo}} can be illustrated through the following relations 
\begin{equation}\label{eq:36}
\left| {\mathcal{A} \in \left( { - \infty , + \infty } \right)} \right\rangle  \Rightarrow \left[ {\widehat{\textit{\textbf{udo}}}} \right] \Rightarrow \mathcal{A}_0 
\end{equation}
\begin{equation}\label{eq:37}
\left| {W\left( \mathcal{A} \right)} \right\rangle  \Rightarrow \left[ {\widehat{\textit{\textbf{udo}}}} \right] \Rightarrow \delta \left( {\mathcal{A} - \mathcal{A}_0 } \right)
\end{equation}
where $\delta(X)$   denotes the Dirac's $\delta$ - function  of $X$.

  In principle, the aspects of quantum and macroscopic observables, depicted by 
	\eqref{eq:34}  and  \eqref{eq:35} respectively 	\eqref{eq:36}  and  \eqref{eq:37}  are completely similar. Therefore the discussions regarding the two kinds of 
	\textit{\textbf{udo}}  should be similarly too. But in the macroscopic case the relation \eqref{eq:37}  is not considered at all as illustrating  a dynamic process.
Moreover within the corresponding macroscopic studies there is no interest  for giving an analytical representation (through some evolution equations) regarding a scenario   of type  \eqref{eq:37}. This even if for the investigation  of macroscopic observables one can use  in principle  a subjacent description given by  classical statistical mechanics. Then, by virtue of above noted similarity, it can be said that  the quantum scenario  \eqref{eq:35} should be not considered as a dynamic process. Consequently the QM studies have to be not concerned about the analytical representation  (by some evolution equations) of an \textit{\textbf{udo}} as the one illustrated by \eqref{eq:35}. Such  regards about the scenario \eqref{eq:35} are required, with all the more, as QM is not complemented (until today) by any subjacent theory of sub-quantum essence. Furthermore, for a true physical approach, the result of the respective  \textit{\textbf{udo}} must be gathered together with the answers of a significant statistical group of many other akin 
\textit{\textbf{udo}}.  The respective answers should  allow to find adequate probabilistic  estimators of the investigated quantum observable.

 Regarding the problem of QMS description, in the category of falling  scenarios, along with the WFC idea one finds  also the famous problem of SCTE (Schrodinger's Cat Thought Experiment).  The respective problem, known also as Schrodinger's cat paradox,  
has retained the attention of many debates over the decades (see \cite{104,105,106} and references). The essential element in SCTE is represented by a single decay of an individual  radioactive atom (which, through some macroscopic machinery, kills an initially living  cat).  But the individual lifetime of a single decaying atom  is a stochastic (random) variable. That is why the mentioned killing decay  is in fact a twin analogue of the  above  mentioned
 \textit{\textbf{udo}} taken into account by the WFC-scenario. So, the above considerations reveal the notifiable fact that, for a true evaluation of a stochastic observable (such is the mentioned decay lifetime), is worthlessly to operate with an 
\textit{\textbf{udo }}  which gives an unique  result of measurement. Accordingly, the SCTE problem  appears as a twin analogue of the IWFC-scenario,  i.e. as a fiction (figment) without any real scientific value.

	     The aforesaid fictional character of the SCTE can be pointed out once more by observation \cite{93,94}  that it is possible to imagine a macroscopic thought-experiment completely analogous with the SCTE. Within the respective macroscopic analogue, a cousin of Schrodinger's cat can be killed through launching a single macroscopic ballistic projectile. More specifically, the killing machinery is activated by an uncontrollable (unobservable) sensor located within the 'circular error probable' (CEP) \cite{109} of a ballistic projectile trajectory. The hitting point of the projectile is expected to arrive within CEP with the probability
			 $50 \%$ .  That is why the murderous action of a single launched projectile is just as much unpredictable as that of the unique radioactive atom within original SCTE. Therefore, the mentioned macroscopic analogy makes clear once more the fictional character of the SCTE.
			
   According to the above-noted remarks, it should be regarded as worthless statements some assertions such as: \textit{"`The Schrdinger's cat thought experiment remains a defining touchstone for modern interpretations of quantum mechanics"'} \cite{106}. Note that such or similar assertions can be found in many popular publications or in the texts disseminated via the Internet (e.g.\cite{110}). 
	
	Therefore SCTE problem as well as its similar  WFC idea, discussed previously, prove themselves to be not real scientific topics but rather useless exercises (fictive  scenarios), without any  conceptual or practical significance.
	
	\subsection {About  observables with continuous spectra}

	As it was noted in the beginning of this Section 5, for physics, development of  suitable models for QMS description present a  natural necessity. Above, in Subsection 5.2 of this article, it is detailed such a model regarding the measurement of an observable endowed with a discrete  non-degenerate spectra. 
Here below we try to propose a   measuring model with similar purpose (QMS description)  but regarding observables having continuous spectra of values.	
	
		As in case with discrete spectrum for here regarded measuring situation we adopt the same generic pattern depicted in \eqref{eq:25}.  The probabilistic content of wave functions $\Psi_{in}$ and $\Psi_{out}$ incorporate information  (data) about the intrinsic state of the measured system respectively concerning the results provided by measurement.
		We will  restrict our considerations to the measurements of orbital characteristics for  a quantum spin-less micro-particle, supposed  in a unidirectional motion along the x-axis. Note that the announced considerations can be easily extended for measurements regarding systems with spatial  orbital motions. Then the wave functions $\Psi_\eta$ 
		($\eta = in, out$) will be taken of the form  $\Psi_\eta = \Psi_\eta (x)$
  (note that here we omit to specify the time t  as visible variable because the  considered state of system  refers to a given ante-measurement instant). 
	
     Note now the fact that according QM rules the wave functions $\Psi_\eta$   have only significance of probability amplitudes but not a direct probability meaning. Therefore, in the case of interest here,  the  picture \eqref{eq:25} of QMS   should be detailed not in terms of  wave functions $\Psi_\eta$, but by means of some  entities  with  direct probabilistic meanings. This  especially because the real measuring devices report the  occurrence of  some  random values for investigated  observables. In usual  terms of QM  entities  with direct probabilistic significance
 are  carriers of stochasticity: probability densities $\rho_\eta$  and probability currents $j_\eta$  ($ \eta = in , out$). Let us write  the wave functions  $\Psi_\eta$ as
 $\Psi _\eta  \left( x \right) = \left| {\Psi _\eta  \left( x \right)} \right| \cdot \exp \left\{ {i\,\Phi_\eta \left( x \right)} \right\}$. Then, for a micro-particle with mass $m$ considered as measured system,   the alluded $\rho_\eta$  and $j_\eta$ are given by relations:
\begin{equation}\label{eq:38}
\rho _\eta   = \rho _\eta  \left( x \right) = \left| {\Psi _\eta  \left( x \right)} \right|^2 \quad ,\quad j_\eta   = j_\eta  \left( x \right) = \frac{\hbar }{m}\left| {\Psi _\eta  \left( x \right)} \right|^2  \cdot \nabla _x \Phi_\eta \left( x \right)
\end{equation}
where $\nabla_x = \frac{\partial}{\partial x}$

      Now it must to specify that  $\rho_\eta$  and $j_\eta$ refer to the positional and the motional kinds of probabilities respectively. Experimentally the two kinds can be regarded as measurable by distinct devices and procedures. The situation is similar with that of    electricity studies where the aspects regarding  position and mobility  of electrical charges are evaluated through completely different devices  and procedures. Due to the aforesaid specifications it results that in fact the generic pattern depicted in \eqref{eq:25} has to be amended  as follows 
\begin{equation}\label{eq:39}
\left| {\rho _{in} \left( x \right) \cup j_{in} \left( x \right)} \right\rangle  \Rightarrow \left[ {\widehat{\textbf{\textit{SCC}}}} \right] \Rightarrow \left[ {\rho _{out} \left( x \right) \cup j_{out} \left( x \right)} \right]
\end{equation}

	Mathematical considerations about the   relations \eqref{eq:25} and \eqref{eq:E1},   (early referred also in \cite{107})  can be applied by similarity for the  pattern \eqref{eq:39}. So the respective pattern (i.e. the operator ${\widehat{\textbf{\textit{SCC}}}}$)can be represented through the 
next two transformations:		
\begin{equation}\label{eq:40}
\begin{array}{l}
 \rho _{out} \left( x \right) = \int\limits_{ - \infty }^{ + \infty }
 {\Gamma \left( {x,x'} \right)}  \cdot \rho _{in} \left( {x'} \right) \cdot dx' \\ 
  \\ 
 j_{out} \left( x \right) = \int\limits_{ - \infty }^{ + \infty }
 {\Lambda \left( {x,x'} \right)}  \cdot j_{in} \left( {x'} \right) \cdot dx' \\ 
 \end{array}
\end{equation}
Here $\Gamma \left( {x,x'} \right)$  and $\Lambda \left( {x,x'} \right)$    represent the corresponding  doubly stochastic kernels ( in sense defined in \cite{108}). This means that the kernels 
$\Re  = \left\{ {\Gamma \,,\;\Lambda } \right\}\ $  satisfy the following relations $ \int\limits_{ - \infty }^{ + \infty } {\Re \left( {x,x'} \right)} \,dx = \int\limits_{ - \infty }^{ + \infty } {\Re \left( {x,x'} \right)} \,dx' = 1 $ .
The mentioned kernels incorporate some extra - QM elements regarding the characteristics of measuring devices and procedures. Such elements do not belong to the usual QM framework which refers to the intrinsic (own) characteristics of the measured micro-particle (system).

Through the  above considerations can be  evaluated  the effects induced by QMS. The respective effects regards  the probabilistic estimators for   orbital observables 
$A_j$  of considered quantum system.  Such observables are described by the operators  $\hat{A}_j$  ($j=1,2,...,n$). As in case of  classical measuring model  (see the Appendix E), without any loss of generality, here  one can  suppose that the quantum observables have identical spectra of values in both \textit{in}-and
 \textit{out}-situations.  In terms of QM the mentioned supposition means that the operators  $\hat{A}_j$  have the same mathematical expressions in both \textit{in}- and \textit{out}-readings, i.e. that the respective expressions remain invariant under the transformations which describe QMS.   In the here discussed case of a system with rectilinear orbital motion  the mentioned expressions depend on $x$  and  $\nabla_x $.

So one can say that in the situations associated with the wave functions 
$\Psi_\eta = \Psi_\eta \left(x\right) $ ( $\eta = in , out$) the mentioned quantum observables $A_j$,  can  characterized  by the following lower order estimators (or numerical parameters) :  mean values $\left\langle A_j\right\rangle_\eta$ ,  correlations $C_\eta\left(A_j,A_k\right)$    and standard deviations 
$\Delta_\eta A_j$ . We use the common notation  $\left(f,g\right)$ for  scalar product of functions $f$ and $g$, i.e. $\left( {f,g} \right) = \int\limits_{ - \infty }^{ + \infty } {f^ *  } \left( x \right) \cdot g\left( x \right) \cdot dx $ . Then  the mentioned estimators are defined by the relations
\begin{equation}\label{eq:41}
\begin{array}{l}
 \quad \quad \left\langle {A_j } \right\rangle _\eta   = \left( {\Psi _\eta  ,\hat A_j \Psi _\eta  } \right)\quad ,\quad \quad \delta _\eta  \hat A_j  = \hat A_j  - \left\langle {A_j } \right\rangle _\eta   \\ 
  \\ 
 C_\eta  \left( {A_j ,A_k } \right) = \left( {\delta _\eta  \hat A_j \Psi _\eta  ,\delta _\eta  \hat A_k \Psi _\eta  } \right)\quad ,\quad \Delta _\eta  A_j  = \sqrt {C_\eta  \left( {A_j ,A_j } \right)}  \\ 
 \end{array}
\end{equation}

     Note here the fact that, on the one hand, the \textit{in}-version of discussions the estimators \eqref{eq:41}   are calculated by means of the wave function
		$\Psi_{in}$ . The respective function is supposed as being  known from the considerations about the intrinsic properties of the investigated system (e.g. by solving the corresponding Schrodinger equation).
		
		       On the other hand, apparently, the evaluation  of estimators \eqref{eq:41}
					in $\eta$= \textit{out}-version requires 
to operate with the wave  function $\Psi_{out}$ . But the respective appearance can be surpassed \cite{20} through operations which use the   probability density 
$\rho_{out}$   and current  $j_{out}$.   So if  an operator $\hat{A}_j$  does not depend on $\nabla _x  = \frac{\partial }{{\partial x}}$  (i.e. $\hat{A}_j = \hat{A}_j\left(x\right)$  ) in evaluating the scalar products from \eqref{eq:41} can be
used the evident equality $\Psi_{out}^\ast \hat{A}_j\Psi_{out} = \hat{A}_j\cdot\rho_{out}$. Additionally, when $\hat{A}_j$  depends on 
$\nabla _x  = \frac{\partial }{{\partial x}}$
(i.e.$\hat{A}_j=\hat{A}_j(\nabla_x)$  ), in the same products the expressions of the type $\hat{A}_j(\nabla_x)\Psi_{out}(x)$   can be converted  in terms of $\rho_{out}(x)$  and $j_{out}(x)$. Namely  from \eqref{eq:38} one finds directly:
\begin{equation}\label{eq:42}
\nabla _x \left| {\Psi _{out} \left( x \right)} \right| = \nabla _x \:\sqrt[{}]{{\rho _{out} \left( x \right)}}\quad ,\quad \quad \nabla _x \Phi_{out} \left( x \right) = \frac{m}{\hbar }\frac{{j_{out} \left( x \right)}}{{\rho _{out} \left( x \right)}}
\end{equation}
  By a single or repeated application of these formulas, any expression of type
		$\hat{A}_j (\nabla_x)\Psi_{out}(x)$ can be transcribed in terms  of $\rho_{out}$   and $j_{out}$.
		
		   The aforesaid  discussion should be supplemented by specifying some indicators able to characterize  the errors (uncertainties) of  considered QMS. For  the  above quoted observables $A_j$ such indicators are the following ones:
\begin{equation}\label{eq:43}
 \left. \begin{array}{l}
 \mathcal{E}\left\{ {\left\langle {A_j } \right\rangle } \right\} = \left\langle {A_j } \right\rangle _{out}  - \left\langle {A_j } \right\rangle _{in}  \\ 
  \\ 
 \mathcal{E}\left\{ {C\left( {A_j ,A_k } \right)} \right\} = C_{out} \left( {A_j ,A_k } \right) - C_{in} \left( {A_j ,A_k } \right) \\ 
  \\ 
 \mathcal{E}\left\{ {\Delta \,A_j } \right\} = \Delta _{out} A_j  - \Delta _{in} A_j  \\ 
 \end{array} \right\}
\end{equation}
    The above presented model regarding the description of  QMS for observables with continuous spectra is illustrated on a simple example  in the Appendix G below. 
	
	\section{Some concluding remarks  }
	
	The present paper was motivated by the existence of many unclearnesses  (unfinished controversies and unelucidated questions) about  of UR and QMS. It was built   as a survey on deficiencies of actual prevalent philosophy in matter. So were re-evaluated the main ideas claimed within the mentioned  philosophy. The basic results of  the respective re-evaluations can be summarized  through the following \textit{\textbf{C}}oncluding 
	\textit{\textbf{R}}emarks ($\textit{\textbf{CR}}$) :
	
	$\bullet$ $\textit{\textbf{CR}}_1$ :  Firstly, through multiple arguments, we have proved the observation that the UR \eqref{eq:1} and \eqref{eq:2} have   not  any essential  significance for physics. Namely the respective UR  are revealed as being  either old-fashioned, short-lived (and removable) conventions (in empirical, thought-experimental justification)  or simple   (and limited ) mathematical formulas (in theoretical vision). But  such an observation  comes to advocate and consolidate the Dirac's intuitive 
	prediction  \cite{23} :\textit{ "`I think one can make a safe guess that uncertainty relations in their present form will not survive in the physics of 
	future"'}. Note that the respective  prediction was founded not on some considerations about the UR essence but on an intuition about the future role in physics of Planck’s constant $\hbar$.  Dirac   predicted  that $\hbar$ will become a derived (secondary) quantity while $c$ and $e$ will remain as fundamental constants ($c$ = speed of light and $e$ = elementary electric charge). $\blacksquare$

$\bullet$ $\textit{\textbf{CR}}_2$ :  A significant idea that emerges from previous discussions is the one that  neither UR \eqref{eq:1} and \eqref{eq:2} nor various 'generalizations'  of them, have  not  any connection with genuine  descriptions of quantum measurements (QMS). All the respective  descriptions should be considered  as a distinct (and additional)  subject which must be investigated separately  but somewhat in  association  with   QM. Examples  of  such   description are presented  briefly, in Subsection, 5.2 and 5.4, for observables having discrete respectively continuous spectra.  $\blacksquare$

$\bullet$ $\textit{\textbf{CR}}_3$ :  Note that, in all of their aspects, the discussions from Subsection 5.2 and 5.4 have a theoretical essence. This means that, the entities like wave function  $\Psi_{in}$ as well as the  measuring indicators $M_{jk}$, 
 $\Gamma \left( {x,x'} \right)$ and $\Lambda \left( {x,x'} \right)$,   are nothing but abstract concepts   which enable elaboration of   theoretical models   regarding the descriptions of QMS . On the one hand   $\Psi_{in}$ refers to the intrinsic data about the  studied system. It is evaluated by means of some known theoretic  procedures (e.g. by means of the corresponding Schrodinger equation).  On the other hand the  indicators $M_{jk}$ , $\Gamma \left( {x,x'} \right)$  and $\Lambda \left( {x,x'} \right)$ are introduced as theoretical entities for modeling the characteristics of the considered measuring devices/processes. $\blacksquare$

$\bullet$  $\textit{\textbf{CR}}_4 $ :  Correlated with the previous $\textbf{\textit{CR}}_2$ and
 $\textit{\textbf{CR}}_3$ it must be specified  that, in relation with QMS, the  inventions of  Wave Function Collapse (WFC) and  Schrodinger's Cat Thought Experiment (SCTE) are nothing but unnatural falling scenarios.  Consequently, as we have argued above in Subsections 5.3,  both idea of WFC  and SCTE problem prove themselves as being  not real scientific subjects but  rather unnecessary figments .$\blacksquare$

$\bullet$ $\textit{\textbf{CR}}_5$ : It is interesting to note here the fact that the history of quantum mechanics was abounded by an impressive number of publications related to UR - QMS philosophy. So, for  the years between 1935 and 1978, as regards    EPR (Einstein-Podolsky-Rosen) paradox, associated \cite{112} with the situation of non-commuting observables, some authors \cite{113} noted  that \textit{"` $\geq 10^6$   papers have been written  "`} - i.e.  $ \geq 63$ papers per day (! ?). Also the same  publishing abundance about UR - QMS  matters motivates remarks such is the following one :"`\textit{  While the stunningly elegant formalism of quantum mechanics can be written down on a napkin, attempts to make sense of it fill entire libraries} "' \cite{114}. Probably that, in some future, the alluded  abundance   will be investigated from historic and sociological perspectives. 

	$\bullet$ $\textit{\textbf{CR}}_6$ : Over the  years original  UR \eqref{eq:1} and \eqref{eq:2}  were supplemented with  many kinds of  'generalizations' 
	(see \cite{115,116,117,118,119,120} and references). Until today, the respective 'generalizations' appear    as being  de facto  only  extrapolation  mathematical 'constructs' (often of  impressive inventiveness).  As a rule, they  are not pointed out  as having  significance    for  some concrete physical questions (of conceptual or    experimental relevance). But the existence of such  significance  is absolutely necessary in order to associate the mentioned 'generalizations' with matters of certain importance for physics. In the light of the discussions from the present paper one can say that the sole physical significance of some from the referred  'generalizations' seems to be their meaning as quantitative indicators of   fluctuations (i.e. of stochasticity). But from a practical perspective among   the respective indicators   of  practical usage  are only the ones of relative lower order. Therefore, for tangible interests of physics, all the discussed 'generalizations'  seem to be  rather  excessive pieces. They  remain only as interesting mathematical 'constructs', which ignore the desideratum:
\textit{"`Entities are not to be multiplied beyond necessity"'}.$\blacksquare$ 

$\bullet$ $\textit{\textbf{CR}}_7$ : In discussions and revaluations proposed in this article, we have referred only to the prevalent philosophy of UR and QMS regarding the primarily the foundations and interpretation of QM.
But, as it is known,  the mentioned  philosophy has been extrapolated  in other 'extra muros' domains,  outside of QM. As aforesaid domains can be quoted the following ones: mathematical computations, biology and medical  sciences, economy and finance, human behavior, social sciences and even politics. A relevant bibliography regarding the mentioned extramural extrapolations can be accessed easy via internet from Google sites.  Note that our above  reevaluations of UR - QMS philosophy do not contain analyzes referring to the mentioned extrapolations. Such analyzes remain as task for scientists  working in the respective domains. Here we want to point out only one noticeable aspect that differentiates the extramural UR from the primary ones. On the one hand, according to their origin, the primary UR from QM are strongly associated with a cardinal marker represented by the Planck constant $\hbar$. On the other hand, as far as we know, for  extramural projections of UR, the  existence of similar markers, represented by cardinal indicators of the corresponding scientific domains, were not reported. $\blacksquare$

$\bullet$ $\textit{\textbf{CR}}_8$ : In their essence, the above argued revaluations of  UR and QMS,  do not disturb in any way the basic  framework of usual QM.  This means that QM keeps its known specific elements:  concepts  (wave functions, operators) with their significances (of stochastic essence),  principles  and theoretical models (Schrodinger equation),  computing rules (exact or approximate) and investigate  systems (atoms, molecules, mesoscopic structures). We recall here that the basic framework of QM can be deduced \cite{121} from direct physical considerations, without appeals to ambiguous discussions about UR or QMS.The alluded considerations start from real physical facts (particle-wave duality of atomic size systems). Subsequently   they use the continuity equations for genuine probability density and current. After that one obtains the whole framework of QM (i.e. the Schrodinger equation, expressions of operators as descriptors of quantum observables and all the practical rules of  QM regarded as a theoretical model for the corresponding  investigated systems).
	
	In the mentioned perspective, we dare to believe that, to some extent, the revaluations of UR and QMS promoted by us can give modest support for genuine reconsiderations regarding the interpretation and foundations of QM.$\blacksquare$ 
	
	\appendix
\appendixpage
\addappheadtotoc

\begin{subappendices}

\renewcommand{\theequation}{A\arabic{equation}}
\setcounter{equation}{0}
\subsection{A brief compendium of  
some  QM  elements}

Here  we remind briefly some significant elements, selected from the usual theoretical framework \cite{5,29,30} of  Quantum Mechanics (QM).  In this  appendix we use Traditional Notations (TN), taken over from  mathematical algebra developed long before QM appeared.  Few specifications  about  the more recent Dirac's  braket formalism  are given in  Appendix B . 
  
       So, in terms of TN,  we consider a QM micro-particle whose state (of orbital nature) is described  by the wave 
			function $\Psi$. Two observables  Aj (j = 1, 2)  of the respective particle will be described by the operators 
			$\hat A_j$. The \\ notation $(f,g)$ will be used for the scalar (inner) product  of  the functions $f$ and  $g$. Correspondingly, the quantities 
			$\left\langle {\hat A_j } \right\rangle _\Psi   = \left( {\Psi ,\hat A_j \Psi } \right)$  and  
			$\delta _\Psi  \hat A_j  = \hat A - \left\langle {\hat A_j } \right\rangle _\Psi  $  will depict the  mean (expected) value respectively the deviation-operator of
the observable  Aj  regarded as a random variable. Then, by denoting  two observables with 
$A_1 = A $  and $A_2 = B $, one  can be written the following formula  : 
\begin{equation}\label{eq:A1}
\begin{array}{l}
\left( {\delta _\Psi  \hat A\Psi ,\delta _\Psi  \hat A\Psi } \right) \cdot \left( {\delta _\Psi  \hat B\Psi ,\delta _\Psi  \hat B\Psi } \right) \ge  \\ 
 \quad \quad \quad \quad \quad \quad  \ge \;\;\;\left| {\left( {\delta _\Psi  \hat A\Psi ,\delta _\Psi  \hat B\Psi } \right)} \right|^2  \\
\end{array} 
\end{equation}
which is nothing  but a relation of  Cauchy-Schwarz  type from mathematics.  

     For an observable  Aj considered as a random variable, in a mathematical sense,  the  quantity $\Delta _\Psi  A_j  = \left( {\delta _\Psi  \hat A_j \Psi ,\delta _\Psi  \hat A_j \Psi } \right)^{\frac{1}{2}} $  signifies its standard  deviation.  From \eqref{eq:A1} it results directly that the standard deviations $\Delta_\Psi A$     and  $\Delta_\Psi B$  of the mentioned observables satisfy the formula
\begin{equation}\label{eq:A2}
\Delta _\Psi  A \cdot \Delta _\Psi  B \ge \;\left| {\left( {\delta _\Psi  \hat A\Psi ,\delta _\Psi  \hat B\Psi } \right)} \right|
\end{equation}
This last  formula,  with  quantities $\Delta_\Psi A$  and   $\Delta_\Psi B$  regarded together, play an influential  role in QM debates within UR and QMS philosophy. That is why the relation   \eqref{eq:A2} can be called \textit{Cauchy-Schwarz  Quantum  Formula }(CSQF).  Note that formulas \eqref{eq:A1}    and \eqref{eq:A2}  are  always valid, i.e.  for all observables, particles and states. Therefore they must be considered as primary QM formulas.

      For the discussions regarding the  UR - QMS philosophy it is helpful to present  the particular versions of formula \eqref{eq:A1} in the cases when  the  operators $\hat A = \hat {A}_1$  and  $\hat B = \hat {A}_2$   satisfy the  conditions 
\begin{equation}\label{eq:A3}
\;iff\,:\quad \left( {\hat A_j \Psi ,\hat A_k \Psi } \right) = \left( {\Psi ,\hat A_j \hat A_k \Psi } \right)\quad ,\quad \left( {j,k = 1,2} \right)
\end{equation}
(where iff $\equiv$ if and only if).
In the alluded cases  it is true the next formula
\begin{equation}\label{eq:A4}
\left( {\delta _\Psi  \hat A\,\Psi ,\delta _\Psi  \hat B\,\Psi } \right) = \frac{1}{2}\left( {\Psi ,\left\{ {\delta _\Psi  \hat A\,,\delta _\Psi  \hat B} \right\}\Psi } \right) - \frac{i}{2}\left( {\Psi ,i\left[ {\hat A,\hat B} \right]\Psi } \right)
\end{equation}
Here $\left\{ {\hat A,\hat B} \right\} = \hat A\hat B + \hat B\hat A$  and 
$\left[ {\hat A,\hat B} \right] = \hat A\hat B - \hat B\hat A$ signify the anti-commutator  respectively commutator of the operators $\hat A$ and $\hat B$ . Now note the fact that the two terms from the right hand side of \eqref{eq:A4} are  purely real and strictly imaginary quantities respectively.  Therefore in the mentioned cases from \eqref{eq:A2} follows directly the  enlarged inequality
\begin{equation}\label{eq:A5}
\left( {\Delta _\Psi  A} \right)^2  \cdot \left( {\Delta _\Psi  B} \right)^2  \ge \frac{1}{4}\left| {\left\langle {\left\{ {\delta _\Psi  \hat A,\delta _\Psi  \hat B} \right\}} \right\rangle _\Psi  } \right|^2  + \frac{1}{4}\left| {\left\langle {\left[ {\hat A,\hat B} \right]} \right\rangle _\Psi  } \right|^2 
\end{equation}
Sometimes this relation is referred to as the Schrodinger inequality. It imply subsequently the next   two  truncated inequalities
\begin{equation}\label{eq:A6}
\Delta _\Psi  A \cdot \Delta _\Psi  B \ge \frac{1}{2}\left| {\left\langle {\left\{ {\delta _\Psi  \hat A,\delta _\Psi  \hat B} \right\}} \right\rangle _\Psi  } \right|
\end{equation}
\\
\begin{equation}\label{eq:A7}
\Delta _\Psi  A \cdot \Delta _\Psi  B \ge \frac{1}{2}\left| {\left\langle {\left[ {\hat A,\hat B} \right]} \right\rangle _\Psi  } \right|
\end{equation}
One observes that  \eqref{eq:A7}  is nothing more than the conventional Robertson-Schrodinger   relation  \eqref{eq:2}, commonly quoted in the literature of CIUR doctrine and UR - QMS  
philosophy. Note that in the respective literature  besides the relation 
\eqref{eq:2}/\eqref{eq:A7}  sometimes the formula  \eqref{eq:A5} is also mentioned. But, as a fact, the respective  mention is not accompanied with the important specification that   formula \eqref{eq:A5} is valid iff (if and only if) the condition \eqref{eq:A3} is fulfilled.

    In the end of this appendix  we note the  cases of more than two observables, i.e. for a set Aj  (j = 1, 2,..., n ;  n $\geq$ 3),  when the quantities 
$\alpha _{jk}  = \left( {\delta _\Psi  \hat A_j \Psi ,\delta _\Psi  \hat A_k \Psi } \right)$  constitute the components of a positive semi definite matrix. In such cases,  similarly with \eqref{eq:A1},  are  true the formulas
\begin{equation}\label{eq:A8}
\det \left[ {\left( {\delta _\Psi  \hat A_j \Psi ,\delta _\Psi  \hat A_k \Psi } \right)} \right] \ge 0\quad ;\quad \left( {j,k = 1,2,...,n} \right)
\end{equation}
where $\det \left[ {\alpha _{jk} } \right]$   is the determinant whose components are the quantities $\alpha _{jk}$ .

       Note that within  dominant publications promoted by the UR - QMS philosophy  the interpretation of many-observable relations \eqref{eq:A8} is frequently omitted. The omission is due most probably to the fact that the idea of referring to simultaneous measurements for more than two observables is not supported convincingly by the current practice of experimental physics.

		\begin{flushleft}
		\textbf{	Addendum}:\\
		\end{flushleft}
		Sometimes, in QM practice, a wave function $\Psi$ is  represented as a superposition of the form
	\begin{equation}\label{eq:A9}
\Psi  = \sum\limits_n {\alpha _n }  \cdot \varphi _n \quad ,\quad \quad \sum\limits_n {\left| {\alpha _n } \right|^2 }  = 1
\end{equation}	
were  $\left\{\varphi _n\right\}$ denote  a complete set of orthonormal basic functions for which  $(\varphi _n, \varphi _m)=\delta_{nm} = $ a    Kronecker delta.

Then, in a state described by  $\Psi$, the mean value of  an observable $A$  is written as 
\begin{equation}\label{eq:A10}
\left\langle A \right\rangle _\Psi   = \sum\limits_{n,m} {\alpha _n^ *   \cdot A_{nm}  \cdot \;} \alpha _m \quad ,\quad \quad A_{nm}  = \left( {\varphi _n ,
\hat A\,\varphi _m } \right)
\end{equation}
with $A_{nm}$ indicating the matrix elements of operator $\hat A$ in representation given by $\left\{\varphi _n \right\}$.

When $\left\{\varphi _n\right\} $ are eigenfunctions of $\hat A$ the following formulas can be written
\begin{equation}\label{eq:A11}
\hat A\;\varphi _n  = a_n  \cdot \varphi _n \quad ,\quad \quad \left\langle A \right\rangle _\Psi   = \sum\limits_n {\left| {\alpha _n } \right|} ^2  \cdot a_n 
\end{equation}
 where $a_n$ signify the  eigenvalue of $\hat A$ in respect with the eigenfunction $\varphi _n $.

Note that the notations and formulas reminded in this 'Addendum' can be used in connection with all quantities discussed above in   present Appendix.
						
\renewcommand{\theequation}{B\arabic{equation}}
\setcounter{equation}{0}
\subsection{ On  the omission of conditions \eqref{eq:A3} within current literature}

The mentioned omission is encountered in many generally agreed publications on QM (especially in textbooks , e.g. \cite{29} ). It appears when  the  conventional Robertson-Schrodinger  relation \eqref{eq:A7}  is established by starting from the correct formula
\begin{equation}\label{eq:B1}
\left\| {\left( {\left( {\delta _\Psi  \hat A + i\lambda \delta _\Psi  \hat B} \right)\Psi } \right)} \right\| \ge 0
\end{equation}
for the norm $||f||$ of function $ f = \left( {\delta _\Psi  \hat A + i\lambda \delta _\Psi  \hat B} \right)\Psi $. In \eqref{eq:B1} are used the notations presented in the previous Appendix A and $\lambda$  denote a real and arbitrary parameter. In order to go on from this last formula to the relation \eqref{eq:A5}, it is presumed the equality
\begin{equation}\label{eq:B2}
\begin{array}{l}
 \left( {\left( {\delta _\Psi  \hat A + i\lambda \delta _\Psi  \hat B} \right)\Psi ,\left( {\delta _\Psi  \hat A + i\lambda \delta _\Psi  \hat B} \right)\Psi } \right) =  \\ 
  \\ 
 \left( {\Psi ,\left( {\delta _\Psi  \hat A} \right)^2 \Psi } \right) + \lambda ^2 \left( {\Psi ,\left( {\delta _\Psi  \hat B} \right)^2 \Psi } \right) - i\lambda \left( {\Psi ,\left[ {\hat A,\hat B} \right]\Psi } \right) \\ 
 \end{array}
\end{equation}
Then, due to the fact that $\lambda$  is a real and arbitrary quantity, from  \eqref{eq:B1}  it results the relation
\begin{equation}\label{eq:B3}
\left\langle {\left( {\delta _\Psi  \hat A} \right)^2 } \right\rangle _\Psi   \cdot \left\langle {\left( {\delta _\Psi  \hat B} \right)^2 } \right\rangle _\Psi   \ge \frac{1}{4}\left| {\left\langle {\left[ {\hat A,\hat B} \right]} \right\rangle _\Psi  } \right|^2 
\end{equation}
	In terms of notations from Appendix A this last relation gives directly the formula

\begin{equation}\label{eq:B4}
\Delta _\Psi  A \cdot \Delta _\Psi  B \ge \frac{1}{2}\left| {\left\langle {\left[ {\hat A,\hat B} \right]} \right\rangle _\Psi  } \right|
\end{equation}
which is nothing but the relation \eqref{eq:A7} from the previous Appendix.
\\
   $ \underline{\textit{Observation}} $: Note here the next two 
	aspects: (i) Introduction of \eqref{eq:B4} demands with necessity the existence of equality \eqref{eq:B2},   (ii)  The respective equality is true only when  the
	operators $\hat A$  and $\hat B$  satisfy the conditions \eqref{eq:A3}. The noted aspects must be signalized as  omissions of the current literature.

         Another context in which appears  the omission of conditions \eqref{eq:A3}  is connected with the \textit{'braket notation'}  frequently used in QM literature. Within the respective notation, known also as \textit{ Dirac's Notation }(DN) , the  scalar (inner) product of two functions \textit{f }and \textit{g} is depicted as 
$ \left\langle {f}\mathrel{\left | {\vphantom {f g}}
 \right. \kern-\nulldelimiterspace}{g} \right\rangle $
					(see  \cite{29,30,31}).   Of course DN  was used in many texts regarding UR - QMS philosophy. But it must be pointed out the fact  that in those texts the condition
				\eqref{eq:A3}, justified in the previous Appendix, is totally omitted and its implications are not analyzed at all.  It is easy to notice that such an omission is due to the fact that, within the DN, both terms (from left-hand and right-hand sides) of the condition \eqref{eq:A3}  have the same transcription, namely :
\begin{equation}\label{eq:B5}
\left( {\hat A_j \Psi ,\hat A_k \Psi } \right) = \left\langle {\Psi \left| {\hat A_j \hat A_k } \right|\Psi } \right\rangle \quad and\quad \left( {\Psi ,\hat A_j \hat A_k \Psi } \right) = \left\langle {\Psi \left| {\hat A_j \hat A_k } \right|\Psi } \right\rangle 
\end{equation}       
Obviously, such transcriptions create confusion and obstruct the just consideration of the condition \eqref{eq:A3}  for cases where it is absolutely necessary in debates about UR - QMS  philosophy.
      In order to avoid the above mentioned confusion in \cite{32} we suggested that DN to be replaced by an Improved Dirac Notation (IDN). For such an IDN we proposed, that within scalar product of two functions \textit{f} and\textit{ g }, to insert additionally the symbol
			"`$\bullet$"' so that the respective  product to be depicted as 
			$<\textit{f} \: |\bullet |\textit{g} >$ . In such a way it becomes directly visible  the separation of the entities implied in that product.
       Then, inside of IDN,  the two terms from  \eqref{eq:A3}  are transcribed as 
\begin{equation}\label{eq:B6}
\left( {\hat A_j \Psi ,\hat A_k \Psi } \right) = \left\langle {\Psi \left| {\hat A_j  \bullet \hat A_k } \right|\Psi } \right\rangle \;and\;\left( {\Psi ,\hat A_j \hat A_k \Psi } \right) = \left\langle {\Psi \left| { \bullet \hat A_j \hat A_k } \right|\Psi } \right\rangle 
\end{equation}
Now one observes that in terms of IDN  the condition \eqref{eq:A3}   appears  in  the form
\begin{equation}\label{eq:B7}
{\rm iff}\quad \quad \left\langle \Psi  \right|\hat A_j  \bullet \hat A{}_k\left| \Psi  \right\rangle  = \left\langle \Psi  \right| \bullet \hat A_j \hat A{}_k\left| \Psi  \right\rangle 
\end{equation}
which no longer generates confusions in discussions about UR - QMS philosophy.

			\renewcommand{\theequation}{C\arabic{equation}}
\setcounter{equation}{0}
\subsection{ Classical  "`uncertainty relations"' in  Fourier  analysis}
     In classical  mathematical harmonic analysis it is known a relation (often named  theorem) which, in terms of here used notations, is similar  with the quantum UR depicted by relation \eqref{eq:2}.  Through current mathematical representations the respective relation   can be introduced as follows. 
	
       Let be a pair of variables \textit{x} and $\xi$ , with domains $x\in(-\infty,+\infty)$  and $\xi\in(-\infty,+\infty)$, regarded as arguments of  a function $f(x)$  respectively of its Fourier transform
	\begin{equation}\label{eq:C1}
\tilde f\left( \xi  \right) = \int\limits_{ - \infty }^{ + \infty } {\exp \left( { - 2i\pi \xi } \right)}  \cdot f\left( x \right) \cdot dx
\end{equation}
If  the norm   $\left\| f \right\|$ of $f\left( x \right)$  has the property $\left\| f \right\| = 1$ ,
both $\left| {f\left( x \right)} \right|^2 $ and $\left| {\widetilde{f}\left( \xi  \right)} \right|^2 $ are probability density functions for \textit{x }and $\xi$  regarded  as real  random (stochastic) variables. The variances of such  variables, evaluated through the  corresponding probabilities, can be noted as $\left\langle {\left( {x - \left\langle x \right\rangle ^2 } \right)} \right\rangle $    and  $\left\langle {\left( {\xi  - \left\langle \xi  \right\rangle ^2 } \right)} \right\rangle $ . The respective variances  express the effective widths of  functions $f(x)$   and $\widetilde{f}(\xi)$. Then 
\cite{66} the aforesaid relation/theorem is given by the  formula		
\begin{equation}\label{eq:C2}
\left\langle {\left( {x - \left\langle x \right\rangle ^2 } \right)} \right\rangle  \cdot \left\langle {\left( {\xi  - \left\langle \xi  \right\rangle ^2 } \right)} \right\rangle  \ge \frac{1}{{16\pi ^2 }}
\end{equation}
In mathematics this formula express the fact that : \textit{ "`A nonzero function and its Fourier transform cannot both be sharply localized "`} \cite{66} .

Often formula  \eqref{eq:C2}  is transcribed in a equivalent variant as follows
\begin{equation}\label{eq:C3}
\Delta x \cdot \Delta \xi  \ge \frac{1}{{4\pi }}
\end{equation}
where $\Delta x$   and $\Delta \xi$  denote the corresponding  standard deviations of \textit{x} 
and $\xi$,   defined  through conventions  like 
$\Delta x = \sqrt {\left\langle {\left( {x - \left\langle x \right\rangle ^2 } \right)} \right\rangle }$. 
\\
     In non-quantum physics a version of relation \eqref{eq:C3}    appears  in  studies of classical signals (waves of acoustic or electromagnetic nature)  where    $\textit{ x = t }= time$ and 
		$\xi=\nu = frequency$. The  respective version is written as
\begin{equation}\label{eq:C4}
\Delta t \cdot \Delta \nu  \ge \frac{1}{{4\pi }}
\end{equation}
and  it is known \cite{67}  as '\textit{Gabor's uncertainty relation}'.  This last relation \eqref{eq:C4} means the fact that, for a classical signal (regarded as a wave packet), the product of the 'uncertainties' ('irresolutions') $\Delta t$   and $\Delta \nu$  in  time and frequency domains cannot be smaller than a specific constant.

Formally the classical relation \eqref{eq:C3} can be transposed to the case of  'quantum wave packets'
often discussed in introductory/intuitive texts about QM. Such a transposition focuses on the pairs of conjugated observables \textit{q - p} (coordinate - momentum) respectively    \textit{t - E } (time - energy) .  The corresponding  transpositions can be obtained by setting in \eqref{eq:C4}  the substitutions \textit{x= q }and $\xi=p(2\pi\hbar)^{-1}$ respectively \textit{x = t}  
 and  $\xi=E(2\pi\hbar)^{-1}$  . The substitutions of variable  $\xi$ are nothing but the so called duality relations (regarding the wave-particle  connections).  By means of the mentioned substitutions from \eqref{eq:C4} one finds the following two relations
\begin{equation}\label{eq:C5}
\Delta q \cdot \Delta p \ge \frac{\hbar }{2}\quad \quad respectively\quad \quad \Delta t \cdot \Delta E \ge \frac{\hbar }{2}
\end{equation}
These last formulas are similar  with the conventional UR \eqref{eq:2} for the pairs of observables 
\textit{q - p} respectively \textit{t -E} . Note that the mentioned similarity is admissible iff  (if and only if) one  accepts  the conventions 
$\left| {\left\langle {\left[ {\hat q,\hat p} \right]} \right\rangle _\Psi  } \right| = \hbar $  
 and $\left| {\left\langle {\left[ {\hat t,\hat E} \right]} \right\rangle _\Psi  } \right| = \hbar $. But attention, the last convention has no more than a 'metaphoric' value. This because in usual QM framework the time \textit{t}  is a deterministic but not random (stochastic) variable and, genuinely, for the respective framework a \textit{time operator $\hat t$}  is  nothing but a senseless and fictitious concept (see also the  discussions from the deficiency  $\textit{\textbf{D}}_8$).

Note that the classical relation \eqref{eq:C3} can be transposed also in another  quantum formula regarding the ground state of a  Quantum Torsion Pendulum (QTP) (see Subsection 3.6.2). For respective transposition   in \eqref{eq:C3} it should to take $f(x) = \Psi(\varphi)$,  $x = \varphi$ and $\xi=L_z\cdot(2\pi\hbar)^{-1}$. So one obtains the formula
\begin{equation}\label{eq:C6}
\Delta \varphi  \cdot \Delta L_z  \ge \frac{\hbar }{2}
\end{equation}
which is nothing but the   lowest level  version of the last of formulas \eqref{eq:13}

\begin{flushleft}
\textbf{Addendum} :
\end{flushleft}
It is worth to mention here the fact that, in the Fourier analysis, the x-unlimited relations  \eqref{eq:C3} and \eqref{eq:C4} have correspondent formulas  
in  x-limited cases (when the variable x has a finite domain  of existence). The respective fact can be evidenced as follows. 

Let be $x\in\left[0, b\right)$,  with $b$ a finite quantity and  function $f(x)$ having  the property 
$f\left( 0 \right) = f \left( {b  - 0} \right): = \mathop {\lim }\limits_{x  \to b  - \;0} \;f \left( x  \right)$. Then the quantities 
\begin{equation}\label{eq:C7}
c_n  = \frac{1}{{\sqrt b  }}\int\limits_0^b  {\exp \left( { - ik_n x} \right) \cdot f\left( x \right)}  \cdot dx
\end{equation}
represent the Fourier coefficients of $f(x)$, with $k _n  = n\cdot\frac{2\pi}{b }$ and $n$ denoting integers i.e. $n\in \mathbb{Z}$.
 
Moreover if the measure   $\left| {f\left( x \right)} \right|^2 dx$ denotes the infinitesimal probability for $x\in (x, x+dx)$ the quantity $|c_n|^2$ signify the discrete probability associated to the value $k_n$. Then for functions $A = A(x)$ and $B = B(k_n)$, depending on $x$ respectively on $k_n$, the mean (expected) values
 $\left\langle A\right\rangle$ and $\left\langle B\right\rangle$ are writen as follows
\begin{equation}\label{eq:C8}
\begin{array}{l}
 \left\langle A \right\rangle  = \int\limits_0^b {A\left( x \right)}  \cdot \left| {f\left( x \right)} \right|^2 dx \\ 
  \\ 
 \quad \left\langle B \right\rangle  = \sum\limits_n {B\left( {k_n } \right)}  \cdot \left| {c_n } \right|^2  \\ 
 \end{array}
\end{equation}
 As the most used such mean (expected) values can be quoted the following ones: first order moments $\left\langle x\right\rangle$ and $\left\langle k_n\right\rangle = \left\langle k\right\rangle$, variances $\left\langle {\left( {x - \left\langle x \right\rangle } \right)^2 } \right\rangle $ and $\left\langle {\left( {k_n  - \left\langle k \right\rangle } \right)^2 } \right\rangle $ respectively  standard deviations $\Delta x = \sqrt {\left\langle {\left( {x - \left\langle x \right\rangle } \right)^2 } \right\rangle } $ and $\Delta k = \sqrt {\left\langle {\left( {k_n  - \left\langle k \right\rangle } \right)^2 } \right\rangle } $.

In order to find the announced $x$-limited correspondents of $x$-unlimited relations \eqref{eq:C3} and \eqref{eq:C4} we take into account the following obvious formula 
\begin{equation}\label{eq:C9}
\int\limits_0^b  {\left| {\lambda \left( {x - \left\langle x \right\rangle } \right) \cdot f\left( x \right) + \left( {\frac{d}{{dx}} - i\left\langle k \right\rangle } \right) \cdot f\left( x \right)} \right|} ^2  \cdot dx \ge 0
\end{equation}
where $\lambda$ is a real, finite and arbitrary parameter. By using the above noted  probabilistic properties of function $f(x)$ and coefficients $c_n$ from \eqref{eq:C9} one obtains the relation
\begin{equation}\label{eq:C10}
\lambda ^2 \left\langle {\left( {x - \left\langle x \right\rangle } \right)^2 } \right\rangle  + \lambda \left( {b \left| {f\left( 0 \right)} \right|^2  - 1} \right) + \left\langle {\left( {k - \left\langle k \right\rangle } \right)^2 } \right\rangle  \ge 0
\end{equation}
Due to the mentioned characteristics of $\lambda$, from this last relation  one finds the next formulas for variances of $x$ and $k_n$
\begin{equation}\label{eq:C11}
\left\langle {\left( {x - \left\langle x \right\rangle } \right)^2 } \right\rangle  \cdot \left\langle {\left( {k_n  - \left\langle {k } \right\rangle } \right)^2 } \right\rangle  \ge \frac{1}{4}\left( {b \left| {f\left( 0 \right)} \right|^2  - 1} \right)^2 
\end{equation}
respectively for standard deviations of $x$ and $k_n$
\begin{equation}\label{eq:C12}
\Delta {\rm x} \cdot \Delta {\rm k} \ge \frac{{\rm 1}}{{\rm 2}}\left| {\left( {b \left| {f\left( 0 \right)} \right|^2  - 1} \right)} \right|
\end{equation}

The formulas \eqref{eq:C11} and \eqref{eq:C12} are x-limited analogues of the 
x-unlimited relations \eqref{eq:C2} and \eqref{eq:C3}.

In the end we note that formula \eqref{eq:C12} is applicable in cases of  wave functions \eqref{eq:4} regarding non-degenerate circular rotations. For such cases 
the application of \eqref{eq:C12} is obtained through the following substitutions:
$x\rightarrow\varphi$, $b \rightarrow 2\pi$, 
   $f\left( x \right) \rightarrow \Psi\left( \varphi \right)$ and $k_n\rightarrow \frac{L_z}{\hbar}$. So from \eqref{eq:C12} it results
\begin{equation}\label{eq:C13}
\Delta \varphi  \cdot \Delta L_z  \ge \frac{\hbar }{2}\left| {\left( {2\pi \left| {\Psi \left( 0 \right)} \right|^2  - 1} \right)} \right|
\end{equation}
This last formula in case of wave functions \eqref{eq:4} degenerates into trivial equality $0 = 0$ 
\newpage
			\renewcommand{\theequation}{D\arabic{equation}}
\setcounter{equation}{0}
 \subsection{ On the fluctuations  of thermodynamic observables}
  Thermodynamic systems are macroscopic bodies composed by huge  numbers of microscopic constituents (molecules and atoms). As whole bodies or through by  their macroscopic parts such systems are described by so-called thermodynamic observables.
	 The alluded observables are viewed as deterministic variables (in usual thermodynamics) respectively as stochastic quantities (in statistical physics). In the last view they are characterized by fluctuations (deviations from their deterministic values studied within  usual thermodynamics). The mentioned fluctuations are investigated within the next conceptual frameworks  : (a) phenomenological approach, (b) classical statistical mechanics, respectively (c) quantum statistical physics.	
				
       In phenomenological approach \cite{68,69,70,71,72}, proposed for the first time by Einstein, the respective fluctuations can be depicted briefly as follows. Let be a system  of the mentioned kind,  whose  properties  are described by a set  of  thermodynamic  observables $\mathbb{A}_j$ (j=1,2,3,...,n).  Each such observable $\mathbb{A}_j$  is characterized by a global fixed value 
			$ \overline{\mathbb{A}}_j$  , evaluable through the methods of  deterministic usual thermodynamics. Then the fluctuations of observables ${\mathbb{A}}_j$ should be discussed in terms of  random variables
				$\mathcal{A}_j = \mathbb{A}_j -\overline{\mathbb{A}}_j$ (j=1,2,...,n),  endowed  with continuous spectra of values such are $\mathcal{A}_j\in(-\infty,+\infty)$. The random characteristics of  variables $\mathcal{A}_j$, i.e. the fluctuations of observables $\mathbb{A}_j$, are depicted in phenomenological approach through the probability density
				$W = W\left( {\vec {\mathcal A}} \right)$, where the vector ${\vec {\mathcal A}}$  signifies the set of all 
				variables $\mathcal{A}_j $. Commonly for 	$W = W\left( {\vec {\mathcal A}} \right)$ one uses distributions of Gaussian type. The mean value (expected) value 
				$\left\langle {\mathbb{A}_j} \right\rangle _W $  and the random deviation $\delta_W \mathbb{A}_j$  of the observable 
				$\mathbb{A}_j$ are
  \begin{equation}\label{eq:D1}
\left\langle {\mathbb{A}_j } \right\rangle _W  = \int\limits_{ - \infty }^{ + \infty } {\mathbb{A}_j }  \cdot W\left( {\vec {\mathcal A}} \right) \cdot d\vec {\mathcal A}\quad  ,\quad\ \delta _W \mathbb{A}_j  = \mathbb{A}_j  - \left\langle {\overline{\mathbb{A}}_j } \right\rangle _W  = {\mathcal A}_{j} 
\end{equation}
Usually, the  fluctuations  of observables $\mathbb{A}_j$  (j=1,2,3,...,n)  are characterized by a small number of numerical parameters evaluable through the random deviations  
$\delta_W \mathbb{A}_j$ . Examples of such parameters are:   dispersions 
$\left\langle {\left( {\delta _W\mathbb{A} _j } \right)^2 } \right\rangle_W  = \left\langle {\left( {{\mathcal A}_{j} } \right)^2 } \right\rangle_W $  and their equivalents the standard deviations 
$\Delta _W \mathbb{A}_j  = \sqrt {\left\langle {\left( {\delta _W \mathbb{A} _j } \right)^2 } \right\rangle _W} $,  second order moments (correlations)   $ \left\langle \delta_W \mathbb{A}_j\cdot \delta_W \mathbb{A}_k \right\rangle_W $ ($j\neq k $) or 
even \cite{72} higher order moments (correlations) $\left\langle \left(\delta_W \mathbb{A}_j\right)^r\cdot \left(\delta_W \mathbb{A}_k\right)^s \right\rangle _ W $ ( $r + s \geq 3$).  
 
The  correlations   $ \left\langle \delta_W \mathbb{A}_j\cdot \delta_W \mathbb{A}_k \right\rangle_W $ 
($j, k =1,2,. . . ,n$)  constitute the components of a positive semi definite matrix. The respective components  satisfy \cite{70,71} the following correlation formulas
\begin{equation}\label{eq:D2}
\det \left[ {\left\langle {\delta _W \mathbb{A}_j  \cdot \delta _W \mathbb {A}_k } \right\rangle _W } \right] \ge 0
\end{equation}
where $\det \left[ {\alpha _{jk} } \right]$   denote the determinant whose components are the quantities $\alpha _{jk}$ . Particularly for two  thermodynamic observables $\mathbb{A}_1 = \mathbb{A} $ and $\mathbb{A}_2 = \mathbb{B} $  from  \eqref{eq:D2} one obtains
\begin{equation}\label{eq:D3}
\Delta _W \mathbb{A} \cdot \Delta _W \mathbb{B} \ge \left| {\left\langle {\delta _W \mathbb{A} \cdot \delta _W \mathbb{B}} \right\rangle _W } \right|
\end{equation}
where   $\Delta _W A = \sqrt {\left\langle {\left( {\delta _W A} \right)^2 } \right\rangle _W } $    denotes the standard deviation of observable $\mathbb{A}$.  Mathematically (in sense of probability theory) this last classical formula is completely analogous with the quantum UR \eqref{eq:2} . 

Regarded in their detailed expressions the standard deviations like is $\Delta_W \mathbb{A}$ 
(introduced above) have an interesting generic property. Namely they appear as being in a direct and factorized dependence of Boltzmann's constant $k_B$. The respective dependence has the following physical significance. It is known the fact that, mathematically, for a given quantity the standard deviation indicates its randomness. This in the sense that the respective quantity is a random or, alternatively, a deterministic (non-random) variable 	according as the alluded deviation has a positive or null value. Therefore $\Delta_W \mathbb{A}$ can be regarded as an indicator of randomness for the  thermodynamic observable $\mathbb{A}$.  But, for diverse cases
(of observables, systems and states),	the deviation $\Delta_W \mathbb{A}$ has various expressions in which, apparently, no common element seems to be implied.
Nevertheless such an element can be found out \cite{20,73} as being materialized by the Boltzmann's constant $k_B$. So, in  Gaussian approximation within the framework of phenomenological theory of fluctuations  one finds \cite{20,73}
\begin{equation}\label{eq:D4}
\left( {\Delta _W \mathbb{A}} \right)^2  = k_B  \cdot \sum\limits_\alpha  {\sum\limits_\beta  {\frac{{\partial \bar {\mathbb{A}}}}{{\partial \bar {\mathbb{X}}_\alpha  }}} }  \cdot \frac{{\partial \bar {\mathbb{A}}}}{{\partial \bar {\mathbb{X}}_\beta  }} \cdot \left( {\frac{{\partial ^2 \bar {\mathbb{S}}}}{{\partial \bar {\mathbb{X}}_\alpha  \partial \bar {\mathbb{X}}_\beta  }}} \right)^{ - \;1} 
\end{equation}
In this relation are used the following notations: (i) $\bar {\mathbb{A}} = \left\langle \mathbb{A} \right\rangle _W $ regarded as a  variable from usual (deterministic) thermodynamics , 
(ii) $\bar {\mathbb{X}}_\alpha$  ( $\alpha$= 1,2,...,r) denote the thermodynamic independent variables of the system , (iii) $\bar {\mathbb{S}} = \bar {\mathbb{S}} (\bar {\mathbb{X}}_\alpha)$   denotes the usual (deterministic) thermodynamic entropy of the system written as a function of  variables  $\bar {\mathbb{X}}_\alpha$  , (iv) $(\mathbb{G}_{\alpha\beta})^{-1}$  denote the inverse of matrix $(\mathbb{G}_{\alpha\beta})$.

     As a first significant aspect of the relation \eqref{eq:D4} is the fact that its right hand side  gives a generic expression of the  fluctuations indicator $(\Delta_W\mathbb{A})^2$  regarding an arbitrary thermodynamic observable $\mathbb{A}$. One can see that the mentioned expression consist in a product of  Botzmann's constant  $k_B$ (as a factorization term) with factors which are independent of $k_B$. The respective independence is evidenced by the fact that the alluded factors must coincide with deterministic (non-random) quantities from usual thermodynamics (where the fluctuations are neglected). Or it is known that such deterministic quantities do not imply $k_B$ at all. Then from \eqref{eq:D4} it results that the fluctuations indicator
		$(\Delta_W\mathbb{A})^2$ is directly proportional to $k_B$ and, consequently, it can be considered as a  non-null respectively a null quantity if one regards  $k_B\neq 0$   or
		$k_B \rightarrow 0$. (Note that because $k_B$ is a physical constant the limit 	
		$k_B \rightarrow 0$ means that the quantities directly proportional with  $k_B$ are negligible comparatively with other quantities of same dimensionality but independent of  $k_B$). 
		
		 On the other hand, a second aspect (mentioned also above) of real significance is
the fact that $(\Delta_W\mathbb{A})^2$ is a direct indicator for the  classical stochasticity (randomness) of observable $\mathbb{A}$.

    Conjointly the two mentioned aspects show that the Boltzmann  constant $k_B$ has the qualities of an authentic generic indicator of  stochasticity (randomness) associated to classical macroscopic (thermodynamic) systems. 

Now note that, a kind of non-quantum formulas completely similar with  \eqref{eq:D2} and \eqref{eq:D3},  can be reported also  for the fluctuations of thermodynamic observables  described in terms of  classical statistical mechanics.  In the respective terms the above phenomenological notations and relations  can be   transcribed formally as follows. Instead of   random variables $\mathcal{A}_j$   should to operate with the phase space  ensemble denoted as $\mu$  of all coordinates and momenta of molecules/atoms which compose the  thermodynamic  system. Also instead of   observables 
$ \mathbb{A}_j = \overline{ \mathbb{A}_j} + \mathcal{A}_j$  needs to be use the  random  functions of the form $\mathbb{A}_j =\mathbb{A}_j(\mu)$. Therewith the probability   density
 $W= W ( \vec{\mathcal{A}})$ should to be replaced with the statistical distribution function  
$\textit{w}= \textit{w} (\mu) $. Then, in terms of aforesaid description of considered fluctuations, as example, can be written the relation
\begin{equation}\label{eq:D5}
\Delta _w \mathbb{A} \cdot \Delta _w \mathbb{B} \ge \left| {\left\langle {\delta _w \mathbb{A}} \cdot \delta _w \mathbb{B} \right\rangle _w } \right|
\end{equation}
which is completely similar with (D3)  .

Add here the observation that the standard deviations $\Delta_w \mathbb{A}$   and 
$\Delta_w \mathbb{B}$  from \eqref{eq:D5}  have a factorization dependence  on $k_B$  of type  \eqref{eq:D4}, similarly  with the case of quantities $\Delta_W \mathbb{A}$   and 
$\Delta_W \mathbb{B}$ from  \eqref{eq:D3}.

For describing the  fluctuations of thermodynamic observables $\mathbb{A}_j$  in framework of quantum statistical physics as probabilities carrier instead of phenomenological  density $W = W\left( {\vec {A}} \right)$ should to use \cite{73,74,75,76}  the quantum density operator ${\hat \rho }$ :
		\begin{equation}\label{eq:D6}
\hat \rho  = \sum\limits_k {\mathfrak{p}_k } \left| {\psi _k } \right\rangle \left\langle {\psi _k } \right|
\end{equation}
Here  $ \left| {\psi _k } \right\rangle $   (k=1,2,...) denote the wave functions of pure states of system and $\mathfrak{p}_k $  are the corresponding probabilities of the respective states. In the same framework the above mentioned  random variables 
$\mathcal{A}_j$   are substituted with the  thermo-quantum  operators 
$\hat{\mathbb{A}}_j$  (j=1,2,...,n). In framework of quantum statistical physics  the mean value $ \left\langle {\mathbb{A}_j } \right\rangle _\rho $  and random deviation $\delta _\rho  \hat{\mathbb{ A}}_j $  of observable $\mathbb{ A}_j$  are 
\begin{equation}\label{eq:D7}
\begin{array}{l}
 \left\langle {\mathbb{A}_j } \right\rangle _\rho   = \sum\limits_k {\mathfrak{p}_k } \left\langle {\psi _k } \right|\hat{ \mathbb{A}}_j \left| {\psi _k } \right\rangle  = tr\left( {\sum\limits_k {\mathfrak{p}_k } \left| {\psi _k } \right\rangle \left\langle {\psi _k } \right|\hat{ \mathbb{A}}_j } \right) = tr\left( {\hat \rho  \cdot \hat{\mathbb{ A}}_j } \right) \\ 
  \\ 
 \quad \quad \quad \quad \quad \quad \quad \quad 
\delta _\rho \hat {\mathbb{ A}}_j  = \hat{ \mathbb{A}}_j  - \left\langle {\mathbb{A}_j } \right\rangle _\rho   \\ 
 \end{array}
\end{equation}
The  deviations $\delta _\rho  \hat{\mathbb{ A}}_j $ can be used in description of  numerical parameters of  fluctuations for observables $\mathbb{A}_j $ 
 in the mentioned framework. As  such parameters can be quoted: dispersions
$\left\langle {\left( {\delta _\rho  \hat{\mathbb{ A}}_j } \right)^2 } \right\rangle _\rho  $  and their equivalents standard deviations $\Delta _\rho \mathbb{ A}_j  = \sqrt {\left\langle {\left( {\delta _\rho  \hat { \mathbb{A}}_j } \right)^2 } \right\rangle _\rho  } $  ,  second order moments (correlations) $\left\langle {\delta _\rho  \hat A_j  \cdot \delta _\rho  \hat A_k } \right\rangle _\rho $ ($j\neq k$ )
or even higher order moments 
 $\left\langle {\left( {\delta _\rho  \hat A_j } \right)^r  \cdot \left( {\delta _\rho  \hat A_k } \right)^s } \right\rangle _\rho  $ ( $r + s \geq 3$).  

    In case of  two thermodynamic observables $\mathbb{A}$  and  $\mathbb{B}$ , regarded in framework of  quantum statistical physics, can be introduced also a correlation relation similar with
		\eqref{eq:D3}  and \eqref{eq:D5}. Such a relation can be introduced as follows. For the corresponding thermo-quantum operators  $\hat{\mathbb{A}}$  and  $\hat{\mathbb{B}}$ it is evidently true the relation
\begin{equation}\label{eq:D8}
\sum\limits_k {\mathfrak{p}_k } \left\langle {{\left( {\delta _\rho  \hat {\mathbb{A}} + i\lambda \,\delta _\rho  \hat {\mathbb{B}}} \right)\psi _k }}
 \mathrel{\left | {\vphantom {{\left( {\delta _\rho  \hat {\mathbb{A}} + i\lambda \,\delta _\rho  
\hat {\mathbb{B}}} \right)\psi _k } {\left( {\delta _\rho  \hat {\mathbb{A}} + i\lambda \,\delta _\rho  \hat {\mathbb{B}}} \right)\psi _k }}}
 \right. \kern-\nulldelimiterspace}
 {{\left( {\delta _\rho  \hat {\mathbb{A}}} + i\lambda \,\delta _\rho  \hat {\mathbb{B}} \right)\psi _k }} \right\rangle  \ge 0
\end{equation}
where  $\lambda$ is an arbitrary real parameter. If  in respect with the functions $\psi_k$ the operators 
$\hat{\mathbb{A}}$  and  $\hat{\mathbb{B}}$   satisfy the conditions of type \eqref{eq:A3} one can   write
\begin{equation}\label{eq:D9}
\begin{array}{l}
 \quad \;\sum\limits_k \mathfrak{p}_k  \left\langle {{\left( {\delta _\rho  \hat {\mathbb{A}} + i\lambda \,\delta _\rho  \hat {\mathbb{B}}} \right)\psi _k }}
 \mathrel{\left | {\vphantom {{\left( {\delta _ \rho  \hat {\mathbb{A}} + i\lambda \,\delta _\rho  \hat {\mathbb{B}}} \right)\psi _k } {\left( {\delta _\rho  \hat {\mathbb{A}} + i\lambda \,\delta _\rho  \hat {\mathbb{B}}} \right)\psi _k }}}
 \right. \kern-\nulldelimiterspace}
 {{\left( {\delta _\rho  \hat {\mathbb{A}} + i\lambda \,\delta _\rho  \hat {\mathbb{B}}} \right)\psi _k }} \right\rangle  =  \\ 
  \\ 
  = \sum\limits_k {\mathfrak{p}_k } \left\langle {{\psi _k \left| {\left( {\delta _\rho  \hat {\mathbb{A}}} \right)^2 } \right.}}
 \mathrel{\left | {\vphantom {{\psi _k \left| {\left( {\delta _\rho  \hat A} \right)^2 } \right.} {\psi _k }}}
 \right. \kern-\nulldelimiterspace}
 {{\psi _k }} \right\rangle  + \lambda ^2 \sum\limits_k {\mathfrak{p}_k } \left\langle {{\psi _k \left| {\left( {\delta _\rho  \hat {\mathbb{B}}} \right)^2 } \right.}}
 \mathrel{\left | {\vphantom {{\psi _k \left| {\left( {\delta _\rho  \hat {\mathbb{B}}} \right)^2 } \right.} {\psi _k }}}
 \right. \kern-\nulldelimiterspace}
 {{\psi _k }} \right\rangle  +  \\ 
  \\ 
 \quad \quad  + i\lambda \sum\limits_k {\mathfrak{p}_k } \left\langle {{\psi _k \left| {\left( {\delta _\rho  \hat {\mathbb{A}} \cdot \delta _\rho  \hat {\mathbb{B}} - \delta _\rho  \hat {\mathbb{B}} \cdot \delta _\rho  
\hat {\mathbb{A}}} \right)} \right.}}
 \mathrel{\left | {\vphantom {{\psi _k \left| {\left( {\delta _\rho  \hat {\mathbb{A}} \cdot \delta _\rho  \hat {\mathbb{B}} - \delta _\rho  \hat {\mathbb{B}} \cdot \delta _\rho  
\hat {\mathbb{A}}} \right)} \right.} {\psi _k }}}
 \right. \kern-\nulldelimiterspace}
 {{\psi _k }} \right\rangle  \\ 
 \end{array}
\end{equation}
Then from \eqref{eq:D8} it results the relation
\begin{equation}\label{eq:D10}
\left\langle {\left( {\delta _\rho  \hat {\mathbb{A}}} \right)^2 } \right\rangle _\rho   + \lambda ^2 \left\langle {\left( {\delta _\rho  \hat {\mathbb{B}}} \right)^2 } \right\rangle _\rho   + \lambda \left\langle {i\left[ {\hat {\mathbb{A}},\hat {\mathbb{B}}} \right]} \right\rangle _\rho   \ge 0
\end{equation}
where $\left[ {\hat {\mathbb{A}},\hat {\mathbb{B}}} \right]$    denotes the commutator of  operators  $\hat{\mathbb{A}}$  and  $\hat{\mathbb{B}}$ . 

Because  $\lambda$  is an arbitrary real parameter from  \eqref{eq:D10} one obtains the relation
\begin{equation}\label{eq:D11}
\left\langle {\left( {\delta _\rho  \hat {\mathbb{A}}} \right)^2 } \right\rangle _\rho   \cdot \left\langle {\left( {\delta _\rho  \hat {\mathbb{B}}} \right)^2 } \right\rangle _\rho   \ge \frac{1}{4}\left\langle {i\left[ {\hat {\mathbb{A}},\hat {\mathbb{B}}} \right]} \right\rangle ^2 _\rho  
\end{equation}
or the equivalent formula
\begin{equation}\label{eq:D12}
\Delta _\rho  \mathbb{A} \cdot \Delta _\rho  \mathbb{B} \ge \frac{1}{2}\left| {\left\langle 
{\left[ {\hat {\mathbb{A}},\hat {\mathbb{B}}} \right]} \right\rangle _\rho  } \right|
\end{equation}
\\
Now let us remind the fact  that in quantum statistics  the above discussed thermo-quantum quantities  
$\left\langle {\left( {\delta _\rho  \hat {\mathbb{A}}_j } \right)^2 } \right\rangle _\rho $ and
 $\Delta_\rho A$ are proved to be connected directly  with a quantity from deterministic (simple thermodynamic) description of thermodynamic observables. The respective connection is due by the known fluctuation-dissipation theorem \cite{76}  which is expressed by the relation 
\begin{equation}\label{eq:D13}
\left\langle {\left( {\delta _\rho  \hat {\mathbb{A}}_j } \right)^2 } \right\rangle _\rho = \left( {\Delta _\rho  \mathbb{A}_{\,j} } \right)^2  = \frac{\hbar }{{2\pi }}\int\limits_{ - \infty }^{ + \infty } {\coth \left( {\frac{{\hbar \omega }}{{2k_B T}}} \right)}  \cdot \mathcal{X}''\left( \omega  \right) \cdot d\omega 
\end{equation}
Here  $k_B$  = the Boltzmann's  constant, $\hbar$ = Planck’s constant and T = temperature of the considered system. Also in
\eqref{eq:D13} the quantity  
$\mathcal{X}''(\omega)$ denote the imaginary part of the susceptibility associated with the observable
$\mathbb{A}$. Note that $\mathcal{X}''(\omega)$  is  a deterministic quantity which is defined primarily in  non-stochastic framework of macroscopic physics \cite{77}. Due to the respective definition it is completely independent of both $k_B$  and $\hbar$.

        In the end of this Appendix  the following conclusion may be recorded. All the relations \eqref{eq:D2}, \eqref{eq:D3}, \eqref{eq:D4}(D4), \eqref{eq:D10}
				and\eqref{eq:D11}  are  formulas regarding macroscopic fluctuations  but not pieces which  should be adapted to the UR - QMS philosophy requirements.
				\renewcommand{\theequation}{E\arabic{equation}}
\setcounter{equation}{0}			
				\subsection{ On the measurements of macroscopic fluctuations}
			The fluctuations parameters, defined above Appendix D, refer to the characteristics of  intrinsic nature for the considered macroscopic systems. But in practical actions, for the same systems, one operates with global parameters, of double source (origin). A first source is given by the  intrinsic properties of  systems. A second source is provided by the actions of measuring devices. In such a vision  a measurement can be regarded as an  transmission process of information (refering to stochastic data).  Consequently the data about the intrinsic properties of measured system appear as \textit{input} (\textit{in}) information  while the global  results of the corresponding  measurement represent the\textit{ output} (\textit{out}) information.
			
			 Here below we will appeal to the aforesaid vision for giving (as in 
				\cite{91,107})  a theoretical model regarding the  measurement of thermal fluctuations.  The respective fluctuations will be considered in   a phenomenological approach (see Appendix D). For simplicity let us consider a system characterized by a single macroscopic observable $\mathbb{A} = \mathbb{\overline{A}} - \mathcal{A}$   , whose thermal fluctuations are impacted within  the random variable $\mathcal{A}$ having the spectrum $\mathcal{A}\in (-\infty,+\infty )$.  The intrinsic fluctuations of $\mathcal{A}$ is supposed to be described by the probability distribution 
		$W_{in} = W_{in}(\mathcal{A})$ regarded as carrier of input-information. The results of measurements are depicted by the distribution
		$W_{out} = W_{out}(\mathcal{A})$ regarded as bearer of out-information. Then the measuring process  may  be symbolized as a transformation of the form
	$W_{in}(\mathcal{A})\rightarrow W_{out}(\mathcal{A})$. If the measuring device is supposed to have stationary and linear characteristics, the mentioned transformation can be described as follows: 
	\begin{equation}\label{eq:E1}
W_{out} \left( \mathcal{A} \right) = \int\limits_{ - \infty }^{ + \infty } {K\left( {\mathcal{A},\mathcal{A}'} \right)}  \cdot W_{in} \left( {\mathcal{A}'} \right) \cdot d\mathcal{A}'
\end{equation}

where  $K(\mathcal{A}, \mathcal{A}')$ appears as  a doubly stochastic kernel (in sense defined in \cite{108}). This means that
 $K(\mathcal{A}, \mathcal{A}')$ satisfy the relations 
$\int\limits_{ - \infty }^{ + \infty } {K\left( {\mathcal{A},\mathcal{A'}} \right)} \;d\mathcal{A} = \int\limits_{ - \infty }^{ + \infty }
{K\left( {\mathcal{A},\mathcal{A'}}\right)}  \;d\mathcal{A'} = 1$.
    
			Add here the fact that, from a physical perspective, the  kernel $K(\mathcal{A}, \mathcal{A}')$  incorporates the theoretical description of all the characteristics of the measuring device. Particularly, for an ideal device  which ensure  
			$W_{out}(\mathcal{A}) = W_{in}(\mathcal{A})$, it must to have  the expression $K(\mathcal{A}, \mathcal{A}')= \delta(\mathcal{A}- \mathcal{A}')$ , where $\delta(\mathcal{X})$  denote the Dirac's $\delta$-function of argument $\mathcal{X}$.    

 By means of distributions $W_\eta = W_\eta(\mathcal{A})$ ($\eta = \textit{in; out}) $ can be introduced the corresponding $\eta$-numerical-characteristics of thermal fluctuations of observable 
$\mathbb{A} = \mathbb{\overline{A}}+\mathcal{A}$. Such are the   $\eta$ - mean (expected) 
value $\left\langle{\mathbb{A}} \right\rangle_\eta$ and  $\eta$ - standard deviation $\Delta_\eta \mathbb{A}$  defined through the relations 
\begin{equation}\label{eq:E2}
\left\langle \mathcal{A} \right\rangle _\eta   = \int\limits_{ - \infty }^{ + \infty } {\mathcal{A} \cdot W_\eta  } \left( \mathcal{A} \right) \cdot d\mathcal{A}\quad ,\quad \quad \left( {\Delta _\eta  \mathcal{A}} \right)^2  = \left\langle {\left( {\mathcal{A} - \left\langle \mathcal{A} \right\rangle _\eta  } \right)^2 } \right\rangle _\eta  
\end{equation}

        The above considerations allow to note some observations about the measuring  uncertainties (errors) regarding the fluctuating macroscopic observable $\mathbb{A}$ .
Firstly the $\eta = in$ - versions  of the parameters \eqref{eq:E2}  describe only the 'intrinsic' properties of the measured system. Secondly the $\eta = out$ -variants of the same parameters incorporate composite information about the respective system and considered measuring device. That is why one can say that, in terms of the above discussions, the measuring uncertainties of observable $\mathbb{A}$  should be described by the following error indicators (characteristics)
	\begin{equation}\label{eq:E3}
\begin{array}{l}
 \mathcal{E}\left\{ {\left\langle \mathbb{A} \right\rangle } \right\} =
 \left\langle \mathbb{A}\right\rangle _{out}  - \left\langle\mathbb{A} \right\rangle _{in}  \\ 
  \\ 
\mathcal{E}\left\{ {\Delta \;\mathbb{A}} \right\} = \Delta _{out} \;\mathbb{A}- \Delta _{in} \;\mathbb{A} \\ 
 \end{array}
\end{equation}
					
Observe here that because $\mathbb{A}$ has stochastic characteristics for a relevant description of its measuring uncertainties it is completely insufficient the single indicator 
$\mathcal{E}\left\{ {\left\langle \mathbb{A} \right\rangle } \right\}$ .
An  adequate minimal such description  requires at least the couple
 $\mathcal{E}\left\{ {\left\langle \mathbb{A} \right\rangle } \right\}$  and 
$\mathcal{E}\left\{ {\Delta \;\mathbb{A}} \right\}$. For further approximations of errors caused by measurements  can be taken into account \cite{111} the higher order moments like the next ones
\begin{equation}\label{eq:E4}
\mathcal{E}\left\{ {\left\langle {\left( {\delta \mathbb{A}} \right)^n } \right\rangle } \right\} = \left\langle {\left( {\delta _{out} \mathbb{A}} \right)^n } \right\rangle _{out}  - \left\langle {\left( {\delta _{in} \mathbb{A}} \right)^n } 
\right\rangle _{in} 
\end{equation}
where $\delta_\eta\mathbb{A} = \mathbb{A}-\left\langle \mathbb{A}\right\rangle$,
$\eta  = in, out$ and $n\geq3$.	

 	\renewcommand{\theequation}{F\arabic{equation}}
\setcounter{equation}{0}	    
\subsection{ An  exemplification for subsection 5.2 } 

For presenting  the announced exemplification  we will refer to  QMS of the energy for a particle of mass $m$, located in an infinite square well potential of width $L$ \cite{29}. The intrinsic state of the microparticle will be considered as being described  by the \textit{in}-wave function
$\Psi _{in} \left( x \right) = \sum\limits_{j = 1}^n {\alpha _j  \cdot \varphi _j \left( x \right)} $. Here $\varphi _j \left( x \right)$ denote the   eigenfunctions  associated to the energetic
eigenvalues $a_j  = E_j  = \Im  \cdot j^2 $ where 
$ \Im  = \left( {{\raise0.7ex\hbox{${\hbar ^2 \pi ^2 }$} \!\mathord{\left/ {\vphantom {{\hbar ^2 \pi ^2 } {2mL^2 }}}\right.\kern-\nulldelimiterspace}\!\lower0.7ex\hbox{${2mL^2 }$}}} \right)$ and
 $j = 1,2,3, ...$. In the considered  \textit{in}-wave function the quantities  $\alpha _j$ are probability amplitudes corresponding to the eigenvalues $E_j$.

We will restrict our exemplification by taking into account only the following circumstances. So we take $n = 3$ as the upper value of the inner energy of the particle while for the  amplitudes $\alpha _j$ we will consider the values which give
$\left( {\left| {\alpha _j } \right|^2 } \right)\; = \;\left( {\begin{array}{*{20}c} {0.5} & {0.4} & {0.1}  \\ \end{array}} \right)$. 

Then the intrinsic characteristics of the particle energy are described by the next mean   value  and the standard deviation 
\begin{equation}\label{eq:F1}
\left\langle E \right\rangle _{in} \; = \;3 \cdot \Im \quad ,\quad \quad \quad \Delta _{in} E\; = \;2.45 \cdot \Im 
\end{equation}

Accordingly with discussions from Subsection 5.2, for a particle in the mentioned intrinsic state, the measurement of energy can be described as  follows. We need to define a model-expression for the matrix 
$(M_{kj})$ from \eqref{eq:29}. As a  first example, we will consider a measurement done with a device endowed with flawed (FL) characteristics.  Such devices, for instance,  can be associated with a matrix $(M_{kj})$ having  the form
\begin{equation}\label{eq:F2}
\left( {M_{kj} } \right)_{FL} \; = \;\left( {\begin{array}{*{20}c}
   {0.5} & {0.3} & {0.2}  \\
   {0.4} & {0.4} & {0.2}  \\
   {0.1} & {0.3} & {0.6}  \\
\end{array}} \right)
\end{equation}
Thus the outcomes of measurement  will be characterized by probabilities
$\left( {\left| {\beta _k } \right|^2 } \right)_{FL} \; = \;\left( {\begin{array}{*{20}c}{0.34} & {0.38} & {0.23}  \\\end{array}} \right)$. 
With such probabilities, the measurement outcomes  for energy will be characterized by the next FL-expected-value and FL-standard-deviation
\begin{equation}\label{eq:F3}
\left\langle E \right\rangle _{FL} \; = \;3.98 \cdot \Im \quad ,\quad \quad \quad \Delta _{FL} \,E\; = \;3.04 \cdot \Im 
\end{equation}
Consequently, for the measurement described by \eqref{eq:F2}, the error indicators \eqref{eq:29} and \eqref{eq:31} acquire the following FL-values
\begin{equation}\label{eq:F4}
\mathcal{E}_{FL} \left\{ {\left\langle E \right\rangle } \right\}\; = \;0.9 \cdot \Im \quad ,\quad \quad \quad \mathcal{E}_{FL} \left\{ {\Delta \,E} \right\}\; = \;0.59 \cdot \Im 
\end{equation}
If,  for the above mentioned energy/particle, we want to describe a  measurement  done with a device having larger characteristics of accuracy (ACC) one can proceed as follows. In the spirit of the relations \eqref{eq:31}, for the matrix $(M_{kj})$ instead of the formula  \eqref{eq:F2} we appeal, for example, to the following expression
\begin{equation}\label{eq:F5}
\left( {M_{kj} } \right)_{ACC} \; = \;\left( {\begin{array}{*{20}c}
   {0.95} & {0.03} & {0.02}  \\
   { - 0.03} & {1.04} & { - 0.01}  \\
   {0.08} & { - 0.07} & {0.99}  \\
\end{array}} \right)
\end{equation}
So, for the probabilities  associated to the outcomes of ACC-measurement, one obtains 
$\left( {\left| {\beta _k } \right|^2 } \right)_{ACC} \; = \;\left( {\begin{array}{*{20}c}{0.489} & {0.4} & {0.11} \\ \end{array}} \right)$. Associated to the respective   probabilities, the considered ACC-measurement  of energy is characterized by the next ACC-expected value and ACC-standard-deviation
\begin{equation}\label{eq:F6}
\left\langle E \right\rangle _{ACC} \; = \;3.088 \cdot \Im \quad ,\quad \quad \quad \Delta _{ACC} \,E\; = \;2.52 \cdot \Im 
\end{equation}
By comparing values from \eqref{eq:F6} with those from \eqref{eq:F1} one sees that the referred ACC-measurement is characterized by the following error indicators
\begin{equation}\label{eq:F7}
\mathcal{E}_{ACC} \left\{ {\left\langle E \right\rangle } \right\}\; = \;0.08 \cdot \Im \quad ,\quad \quad \quad \mathcal{E}_{ACC} \left\{ {\Delta \,E} \right\}\; = \;0.07 \cdot \Im 
\end{equation}

Finally, by comparing the results reported in relations \eqref{eq:F4} and \eqref{eq:F7}, we can note the following remark.  Within the above theoretical description of measurement, the error indicators (for both expected value and standard deviation) are much smaller in the 
case dealing with higher accuracy  characteristics  comparatively with the one regarding flawed  features.

 	\renewcommand{\theequation}{G\arabic{equation}}
\setcounter{equation}{0}	    
\subsection{  Illustrations for subsection 5.4} 
     In order to illustrate   the   model discussed  in Subsection 5.4, in connection with the description of QMS,  let us present here an exercise taken by abbreviation  from our article  \cite{20} (more computational details can be found in the respective article). We will  refer to a micro-particle of mass $m$ having an one-dimensional motion along the x-axis.	Its \textit{in}-wave-function $\Psi_{in}$ is taken of the form 
$\Psi_{in}(x)= \left|\Psi_{in}(x)\right|\cdot exp\left\{i\Phi_{in}(x)\right\}$ where 
\begin{equation}\label{eq:G1}
\left| {\Psi _{in} \left( x \right)} \right| \propto \exp \left\{ { - \frac{{\left( {x - x_0} \right)^2}}{{4\sigma ^2 }}} \right\}\quad ,\quad \quad \Phi _{in} \left( x \right) = kx
\end{equation}
Here as well as below in other relations from this Appendix the  explicit notations of normalization constants are omitted ( they can be added easy by the interested readers).
According to  the wave function \eqref{eq:G1} the intrinsic features of the considered  microparticle are described by the parameters $x_0$, $\sigma$  and $k$.

Through expressions \eqref{eq:G1}, by means of formulas \eqref{eq:38}, it is simple to find the analytical expresions for probability density $\rho_{in}$ and current 
$j_{in}$.  
 As doubly stochastic kernels  suggested in \eqref{eq:40} we propose here the next two formulas 
\begin{equation}\label{eq:G2}
\Gamma \left( {x,x'} \right) \propto \exp \left\{ { - \frac{{\left( {x - x'} \right)^2 }}{{2\gamma ^2 }}} \right\}
\end{equation}
\begin{equation}\label{eq:G3}
\Lambda \left( {x,x'} \right) \propto \exp \left\{ { - \frac{{\left( {x - x'} \right)^2 }}{{2\lambda ^2 }}} \right\}
\end{equation}
Here parameters $\gamma$ and $\lambda$ depict the characteristics of measuring devices/procedures. The values of the respective parameters are associated with an ideal measurement (when both $\gamma$ and $\lambda$ tend toward zero) , respectively with a nonideal measurement (in cases when at least one of the two parameters is not-null).

Then, by using the procedures presented within Subsection 5.4, it is easy to find the \textit{out}-entities $\rho_{out}$, $j_{out}$ and $\Psi_{out}$. By using the respective entities together with the functions from \eqref{eq:G1} one can  evaluate 
the  \textit{out } and \textit{in} versions   of mean (expected) values and standard 
deviations for observables of interest. The respective evaluations ensure estimations of the corresponding error indicators.  So, for $\hat{x} = x\cdot$ = coordinate and 
$\hat{p}= -i\hbar \nabla_x$ = momentum as operators (observables) of interest, one obtains \cite{20} the following error indicators
\begin{equation}\label{eq:G4}
\mathcal{E}\left\{ {\left\langle x \right\rangle } \right\} = 0\quad ,\quad 
\mathcal{E}\left\{ {\Delta \,x} \right\} = \sqrt {\sigma ^2  + \gamma ^2 }  - \sigma 
\end{equation}
\begin{equation}\label{eq:G5}
\begin{array}{l}
 \mathcal{E}\left\{ {\left\langle p \right\rangle } \right\} = 0\quad ,\quad 
\mathcal{E}\left\{ {\Delta \,p} \right\} = \hbar \left| {\left[ {\frac{{k^2 \left( {\sigma ^2  + \gamma ^2 } \right)}}{{\sqrt {\left( {\sigma ^2  + \lambda ^2 } \right)\left( {\sigma ^2  + 2\gamma ^2  - \lambda ^2 } \right)} }} - } \right.} \right. \\ 
  \\ 
 \quad \quad \quad \quad \quad \quad  - k^2  + \left. {\left. {\frac{1}{{4\left( {\sigma ^2  + \gamma ^2 } \right)}}} \right]^{\frac{1}{2}}  - k} \right| \\ 
 \end{array}
\end{equation}
Let us now restrict in the wave function \eqref{eq:G1} to the situation when $x_0 = 0$ $k=0$ and $\sigma  = \sqrt {\frac{\hbar }{{2m\omega }}}$. Then \eqref{eq:G1} describe the ground state of a harmonic oscillator with $m$ = mass and $\omega$ = angular frequency. 
 As observable of interest of such an oscillator we
consider the energy described by the Hamiltonian $\hat H =  - \frac{{\hbar ^2 }}{{2m}}\frac{{d^2 }}{{dx^2 }} + \frac{{m\,\omega ^2 }}{2}x^2 $. For the respective observable
 one finds
\begin{equation}\label{eq:G6}
\left\langle H \right\rangle _{in}  = \frac{{\hbar \omega }}{2}\quad ,\quad \quad \Delta _{in} H = 0
\end{equation}
\begin{equation}\label{eq:G7}
\left\langle H \right\rangle _{out}  = \frac{{\omega \left[ {\hbar ^2  + \left( {\hbar  + 2m\,\omega \,\gamma ^2 } \right)^2 } \right]}}{{4\left( {\hbar  + 2m\,\omega \,\gamma ^2 } \right)}}
\end{equation}
\begin{equation}\label{eq:G8}
\Delta _{out} H = \frac{{\sqrt 2 \,m\,\omega ^2 \,\gamma ^2 \left( {\hbar  + m\,\omega \,\gamma ^2 } \right)}}{{\left( {\hbar  + 2m\,\omega \,\gamma ^2 } \right)}}
\end{equation}
\begin{equation}\label{eq:G9}
\mathcal{E}\left\{ {\left\langle H \right\rangle } \right\} = \frac{{m^2 \omega ^3 \gamma ^4 }}{{\hbar  + 2m\,\omega \,\gamma ^2 }}
\end{equation}
\begin{equation}\label{eq:G10}
\mathcal{E}\left\{ {\Delta H} \right\} = \Delta _{out} H = \frac{{\sqrt 2 \,m\,\omega ^2 \,\gamma ^2 \left( {\hbar  + m\,\omega \,\gamma ^2 } \right)}}{{\left( {\hbar  + 2\,m\,\omega \,\gamma ^2 } \right)}}
\end{equation}
\renewcommand{\theequation}{H\arabic{equation}}
\setcounter{equation}{0}	
\subsection{ A more comprehensive description of measuring errors for random observables}

      In  Subsections 5.2 and 5.4 or Appendices E, F and G, we have  discussed the measuring errors for  random observables of quantum respectively macroscopic nature. For description of that errors, were used as indicators only the lower order probabilistic parameters (moments and correlations). But those indicators give only first sequences, of limited value, for a global picture of the considered errors. A more comprehensive such a picture can be done in terms of  informational entropies. Shortly, for the above discussed observables and errors, the suggested depiction can be illustrated  as follows.
			
			     Firstly let us refer to the case of a macroscopic random observable  $\mathcal {A}$ whose measurements are outlined in Appendix E. The intrinsic characteristics (fluctuations) of $\mathcal {A}$ are considered as being described by the probability distribution
					$W_{in} = W_{in}(\mathcal {A} )  $  regarded as carrier of input-information for  measurements. The results of measurements are depicted by the distribution 
$W_{out} = W_{out}(\mathcal {A} )  $ associated with  the out-information of  measurements. The informational entropies  $\mathcal{H}_\eta$ ( $ \eta = in, out$) connected with the above noted distributions  are defined through the formulas
\begin{equation}\label{eq:H1}
\mathcal{H}_\eta  \left( \mathcal{A} \right) =  - \int\limits_{ - \infty }^{ + \infty } {W_\eta  } \left( \mathcal{A}  \right) \cdot \ln \left[ {W\left(\mathcal{A}  \right)} \right] \cdot d\mathcal{A} 
\end{equation}
By taking into account the transformation \eqref{eq:E1}, the  main properties of the  doubly stochastic kernel $K(\mathcal{A},\mathcal{A}’)$, as well  the formula  $
\ln \left( X \right)\; \le \;X - 1$  one can  write 
\begin{equation}\label{eq:H2}
\begin{array}{l}
 \mathcal{H}_{out} \left(\mathcal{A}\right) - \mathcal{H}_{in} \left( \mathcal{A} \right) =  - \int\limits_{ - \infty }^{ + \infty } {\int\limits_{ - \infty }^{ + \infty } {d\mathcal{A} \cdot d\mathcal{A}'} }  \cdot K\left( {\mathcal{A},\mathcal{A}'} \right) \cdot W_{in} \left( {\mathcal{A}'} \right) \cdot \ln \left[ {\frac{{W_{out} \left( \mathcal{A} \right)}}{{W_{in} \left( {\mathcal{A}'} \right)}}} \right] \ge  \\ 
  \\ 
 \quad \quad \quad  \ge  - \int\limits_{ - \infty }^{ + \infty } {\int\limits_{ - \infty }^{ + \infty } {d\mathcal{A} \cdot d\mathcal{A}'} }  \cdot K\left( {\mathcal{A},\mathcal{A}'} \right) \cdot W_{in} \left( {\mathcal{A}'} \right) \cdot \left[ {\frac{{W_{out} \left( \mathcal{A} \right)}}{{W_{in} \left( {\mathcal{A}'} \right)}} - 1} \right] = 0 \\ 
 \end{array}
\end{equation}
Therefore the errors specific of measurements for $\mathcal{A}$ in its wholeness can be described through the comprehensive error  indicator 
  \begin{equation}\label{eq:H3} 
\mathcal{E}\left\{ {\mathcal{H}\left( \mathcal{A} \right)} \right\}\; = \;\mathcal{H}_{out} \left( \mathcal{A} \right)\; - \;\mathcal{H}_{in} \left( \mathcal{A} \right)\; \ge \;0
\end{equation}

This relationship shows that the measuring process can be described by a non-negative change in the  informational  entropy associated with the investigated observable. The situation when the respective change is null corresponds to the case of an ideal measurement (free of errors), mentioned otherwise in connection with the relationship \eqref{eq:E1}. 
   
	Mostly, the macroscopic fluctuations described by the here used observable $\mathcal{A}$  are investigated in the so-called Gaussian approximations. Then the entities $W_{in} (\mathcal{A}) $ and $K (\mathcal{A},\mathcal{A}’)$ which appear in \eqref{eq:E1}  are given by the following formulas 
 \begin{equation}\label{eq:H4} 
W_{in} \left(\mathcal{A} \right) \propto \exp \left\{ { - {\raise0.7ex\hbox{${\mathcal{A}^2 }$} \!\mathord{\left/
 {\vphantom {{\mathcal{A}^2 } {2a^2 }}}\right.\kern-\nulldelimiterspace}
\!\lower0.7ex\hbox{${2a^2 }$}}} \right\}\;,\quad  K\left( {\mathcal{A},\;\mathcal{A}'} \right) \propto \exp \left\{ { - {\raise0.7ex\hbox{${\left( {\mathcal{A} - \mathcal{A}'} \right)^2 }$} \!\mathord{\left/
 {\vphantom {{\left( {\mathcal{A} - \mathcal{A}'} \right)^2 } {2b^2 }}}\right.\kern-\nulldelimiterspace}
\!\lower0.7ex\hbox{${2b^2 }$}}} \right\}
\end{equation}
where the explicit indication of  normalization constants are omitted (the omission can be filled easily  by interested readers).  In the first formula from \eqref{eq:H4}  $a$ denote the standard deviation of intrinsic fluctuations within the measured system. The symbol $b$  in the second expression from \eqref{eq:H4} depicts the precision parameter of measuring device.  Of course,  for a scientifically acceptable measuring process, it must be considered that $b \ll a$.

     In the alluded cases with Gaussian approximations  the output distribution
		$W_{out}(\mathcal{A})$ has the form $W_{out} \left(\mathcal{A}\right) \propto \exp \left\{ { - \left[ {{\raise0.7ex\hbox{${\mathcal{A}^2 }$} \!\mathord{\left/{\vphantom {{\mathcal{A}^2 } {2\left( {a^2  + b^2 } \right)}}}\right.\kern-\nulldelimiterspace}\!\lower0.7ex\hbox{${2\left( {a^2  + b^2 } \right)}$}}} \right]} \right\}$. Then the comprehensive error indicator \eqref{eq:H3} becomes       
\begin{equation}\label{eq:H5}
\mathcal{E}\left\{ {\mathcal{H}\left( \mathcal{A} \right)} \right\} = \frac{1}{2}\ln \left( {1 + \frac{{b^2 }}{{a^2 }}} \right) \approx \frac{1}{2} \cdot \frac{{b^2 }}{{a^2 }}
\end{equation}
Now let us refer to the comprehensive informational depiction for measuring errors in cases of  random quantum observables. We start the announced reference by discussing the case presented in Subsection 5.2, regarding the measurement of a quantum observable endowed with a discrete spectrum of eigenvalues. In the respective case the input and output data characterizing the measurement are depicted by the following corresponding probabilities
\begin{equation}\label{eq:H6}
\mathcal{P}_{in}^j  = \left| {\alpha _j } \right|^2 \;,\quad \mathcal{P}_{out}^j  = \left| {\beta _j } \right|^2 \;,\quad \;\left( {j = 1,2, \ldots ,n} \right)
\end{equation}
These probabilities can be associated with the next information entropies
\begin{equation}\label{eq:H7}
\mathcal{H}\left( {\mathcal{P}_\eta  } \right) =  - \sum\limits_{j = 1}^n {P_\eta ^j }  \cdot \ln \left( {\mathcal{P}_\eta ^j } \right)\;,\quad \left( {\eta  = in,\;out} \right)
\end{equation}
Consequently, for an extensive  description  of measuring errors for the  specified quantum observable, can be used the below comprehensive indicator
\begin{equation}\label{eq:H8}
\mathcal{E}\left\{ {\mathcal{H}\left( \mathcal{P} \right)} \right\} = \mathcal{H}\left( {\mathcal{P}_{out} } \right) - \mathcal{H}\left( {\mathcal{P}_{in} } \right)
\end{equation}
By taking into account the transformation \eqref{eq:27},  the basic properties of  doubly stochastic matrix $M_{jk}$, plus the   evident formula  $ ln (X) \leq X-1$ , through some simple calculations (similar to those appealed in \eqref{eq:H2}  and 
\eqref{eq:H3},  one finds:
\begin{equation}\label{eq:H9}
\mathcal{E}\left\{ {\mathcal{H}\left( \mathcal{P} \right)} \right\} \ge 0
\end{equation}
This formula corresponds to ideal or non-ideal measurements, in cases of equality  respectively of inequality.

Note that, in  cases of  examples presented in Appendix F related with Subsection 5.2, the relation  \eqref{eq:H8} takes the expresions 
\begin{equation}\label{eq:H10}  
\begin{array}{l}
 \mathcal{E}\left\{ {\mathcal{H}\left( \mathcal{P} \right)_{FL} } \right\}\; = \;\mathcal{H}\left( {\left( {\left| {\beta _k } \right|^2 } \right)_{FL} } \right)\; - \;\mathcal{H}\left( {\left( {\left| {\alpha _j } \right|^2 } \right)} \right)\; = \;0.131 \\ 
  \\ 
 \mathcal{E}\left\{ {\mathcal{H}\left( \mathcal{P} \right)_{ACC} } \right\}\; = \;\mathcal{H}\left( {\left( {\left| {\beta _k } \right|^2 } \right)_{ACC} } \right)\; - \;\mathcal{H}\left( {\left( {\left| {\alpha _j } \right|^2 } \right)} \right)\; = \;0.018 \\ 
 \end{array}
\end{equation}
The above expressions correspond to measurements with characteristics of flawed respectively accurate types.  Same expressions show that, even in informational-entropic approach, the measuring errors are higher in cases with flawed characteristics comparatively with the ones having accurate features.

        Now let us note some things about the comprehensive description of measuring errors in cases approached in Subsection 5.4 and in Appendix G, regarding of  quantum observables with continuous spectra. The corresponding measurements, depicted through the transformations \eqref{eq:40}, can be associated with the following informational entopies

\begin{equation}\label{eq:H11}
\begin{array}{l}
 \mathcal{H}_\eta  \left( \rho  \right) =  - \int\limits_{ - \infty }^{ + \infty } {\rho _\eta  } \left( x \right) \cdot \ln \left( {\rho _\eta  \left( x \right)} \right) \cdot dx \\ 
  \\ 
 \mathcal{H}_\eta  \left( {\left| j \right|} \right) =  - \int\limits_{ - \infty }^{ + \infty } {\left| {j_\eta  \left( x \right)} \right|}  \cdot \ln \left( {\left| {j_\eta  \left( x \right)} \right|} \right) \cdot dx \\ 
 \end{array}
\end{equation}
where  $\eta = in, out$.  Related with the above entropies can be introduced the next comprehensive error indicators
\begin{equation}\label{eq:H12}
\mathcal{E}\left\{ {H\left( \rho  \right)} \right\} = \mathcal{H}_{out} \left( \rho  \right) - \mathcal{H}_{in} \left( \rho  \right)\;,\quad \mathcal{E}\left\{ {\mathcal{H}\left( {\left| j \right|} \right)} \right\} = \mathcal{H}_{out} \left( {\left| j \right|} \right) - \mathcal{H}_{in} \left( {\left| j \right|} \right)
\end{equation}
Through some simple calculations (completely similar to the ones used in \eqref{eq:H2} and \eqref{eq:H3}) one finds that the error indicators \eqref{eq:H11}  satisfy the relations
\begin{equation}\label{eq:H13}
\mathcal{E}\left\{ {\mathcal{H}\left( \rho  \right)} \right\} \ge 0\;,\quad \mathcal{E}\left\{ {\mathcal{H}\left( {\left| j \right|} \right)} \right\} \ge 0
\end{equation}
These relations with equalities or inequalities refer to the cases of ideal respectively non-ideal measurements.
   
	In particular case of measurement illustrated  in  Appendix G, associated with  the doubly stochastic kernels \eqref{eq:G2} and \eqref{eq:G3}, the error indicators  \eqref{eq:H12} become
\begin{equation}\label{eq:H14}
\begin{array}{l}
\mathcal{E}\left\{ {\mathcal{H}\left( \rho  \right)} \right\} = \ln \sqrt {\frac{{\sigma ^2  + \gamma ^2 }}{{\sigma ^2 }}} \; \approx \;\frac{1}{2}\left( {\frac{\gamma }{\sigma }} \right)^2  \\ 
  \\ 
\mathcal{E}\left\{ {\mathcal{H}\left( {\left| j \right|} \right)} \right\} = \ln \sqrt {\frac{{\sigma ^2  + \lambda ^2 }}{{\sigma ^2 }}} \; \approx \;\frac{1}{2}\left( {\frac{\lambda }{\sigma }} \right)^2  \\ 
 \end{array}
\end{equation}
The last expressions of these indicators  imply the approximations 
$\gamma\ll\sigma$  and $\lambda\ll\sigma$,  specific to the supposition that measuring devices  have high accuracies. Of course that the cases with 
$\gamma = 0$ γ=0 and $\lambda = 0$ depict the ideal measurements.
 
 In the case of  a harmonic oscillator, mentioned in the end of Appendix G, the first error indicator from \eqref{eq:H12} get the expression
\begin{equation}\label{eq:H15}
\mathcal{E}\left\{ {\mathcal{H}\left( \rho  \right)} \right\} = \ln \sqrt {\frac{{\hbar  + 2m\omega \gamma ^2 }}{\hbar }} \; \approx \;\frac{{m\omega }}{\hbar }\gamma ^2 
\end{equation}
\newpage
\renewcommand{\theequation}{I\arabic{equation}}
\setcounter{equation}{0}	
\subsection{ A private letter from the late scientist  J.S. Bell \\to the present author}
 
\begin{center}
\includegraphics[width=0.9\textwidth]{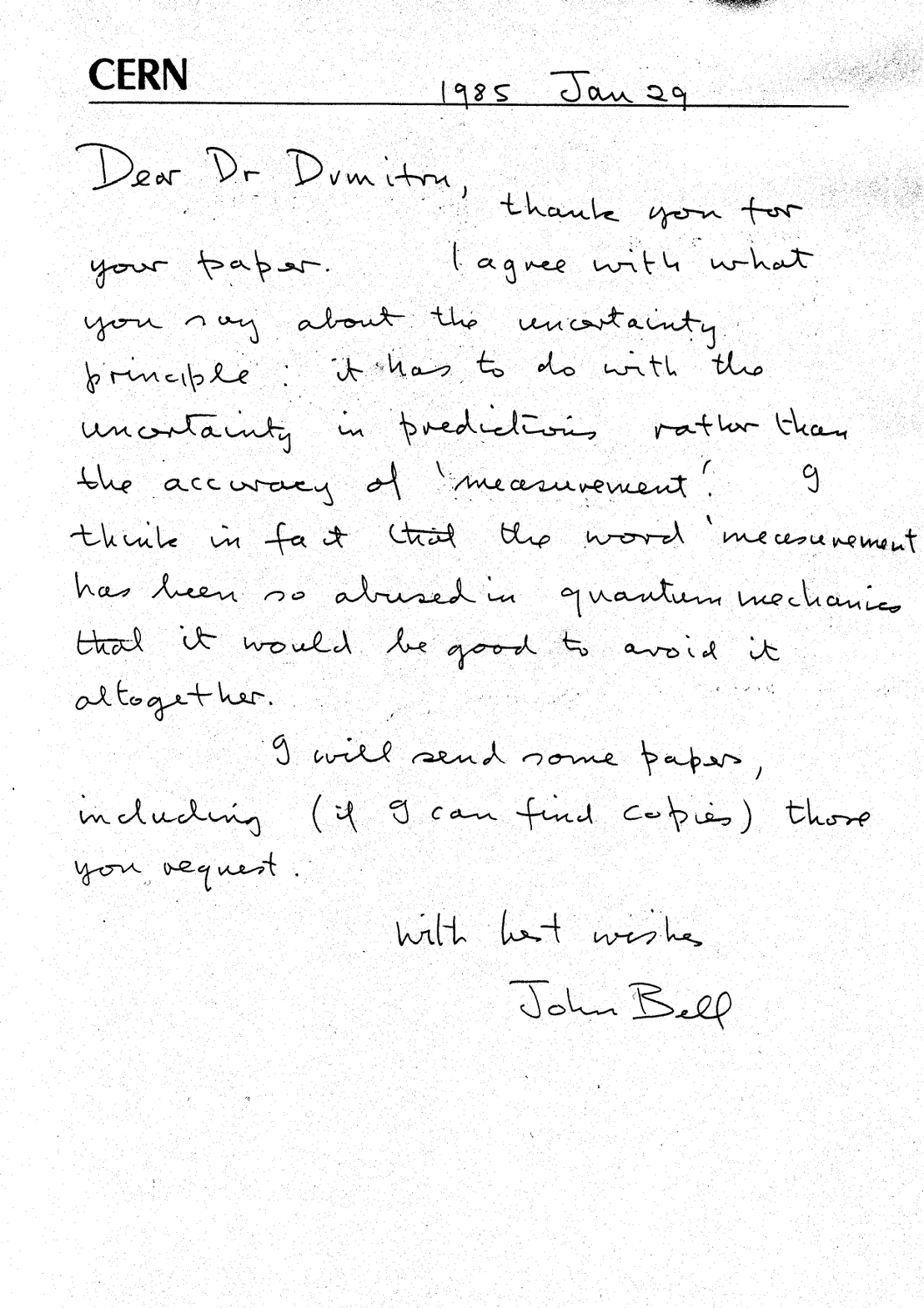}
\end{center}
\end{subappendices}

\end{document}